\documentstyle[aps,amssymb]{revtex}

\begin{document}
\title{Rigorous relativistic equation for quark-antiquark bound states at finite
temperature derived from the thermal QCD formulated in the coherent-state
representation}
\author{Jun-Chen Su}
\address{Center for Theoretical Physics, School of Physics, Jilin University,\\
Changchun 130023, People's Republic of China}
\date{}
\maketitle

\begin{abstract}
A rigorous three-dimensional relativistic equation for quark-antiquark bound
states at finite temperature is derived from the thermal QCD generating
functional which is formulated in the coherent-state representation. The
generating functional is derived newly and given a correct path-integral
expression. The perturbative expansion of the generating functional is
specifically given by means of the stationary-phase method. Especially, the
interaction kernel in the three-dimensional equation is derived by virtue of
the equations of motion satisfied by some quark-antiquark Green functions
and given a closed form which is expressed in terms of only a few types of
Green functions. This kernel is much suitable to use for exploring the
deconfinement of quarks. To demonstrate the applicability of the equation
derived , the one-gluon exchange kernel is derived and described in detail.

\end{abstract}

\section{Introduction}

Quantum Chromodynamics (QCD), as a strong interaction theory of quarks and
gluons, has a distinctive property of asymptotic freedom and infrared
slavery which makes the quarks and gluons to be confined in hadrons. It is
widely believed that in extreme conditions, i.e., at high temperature and/or
high density, the quarks and gluons would be deconfined from hadrons and
form a new matter, the quark-gluon plasma (QGP). It is highly expected that
the QCD phase transition from hadrons to QGP would take place and be
observed in the high energy heavy ion collisions at RHIC [1-3].
Theoretically, to predict the QCD phase transition, many efforts have been
made by using different approaches such as the lattice simulation, the
effective field theory, the hydrodynamic model and etc. [4-14]. According to
the prediction of the lattice QCD calculations, the phase transition could
occur when the colliding system reaches the temperatures 150-170 MeV [2].
Since the quarks and gluons are confined in hadrons which exist as the bound
states of quarks and/or gluons, it is obvious that a proper approach of
investigating the quark deconfinement is to start from an exact relativistic
equation for quark and/or antiquark bound states at finite temperature. In
this paper, we are devoted to deriving a rigorous three-dimensional
relativistic equation of Dirac-Schr\"odinger type for quark-antiquark ($q%
\overline{q})$ bound states at finite temperature. In particular, the
interaction kernel in the equation will be given a closed and explicit
expression which will be derived by following the procedure described in
Refs. [15-17]. This constitutes the main purpose of this paper. Clearly, if
the interaction kernel and the equation can be calculated by a suitable
nonperturbation method, one can exactly determine at which temperature the
quark and antiquark will be deconfined from mesons.

In this paper, we intend to derive the aforementioned equation and kernel
from the thermal QCD generating functional formulated in the coherent-state
representation. For this derivation, we need first to give a correct
expression of the generating functional in the coherent-state
representation. This constitutes another purpose of this paper. Here it is
necessary to mention that the corresponding path integral expressions for
the partition functions and generating functionals given in the previous
literature are not correct [18-22]. The incorrectness is due to that in the
previous path integral expressions, the integral representing the trace is
not separated out on the one hand and the time-dependence of the integrand
in the remaining part of the path integral is given incorrectly on the other
hand. Such path integral expressions can only be viewed as a formal
symbolism because in practical calculations, one has to return to the
original discretized forms which lead to the path integral expressions. If
one tries to perform an analytical calculation of the path integrals by
employing the general methods and formulas of computing functional
integrals, one would get a wrong result. The partition functions and
generating functionals for many-body systems discussed in quantum statistics
were rederived in the coherent-state representation and given correct
functional-integral expressions in the author's previous paper [23]. These
expressions are consistent with the corresponding coherent-state
representations of the transition amplitude and the generating functional in
the zero-temperature quantum theory [24-26]. Particularly, when the
functional integrals are of Gaussian type, the partition functions and the
generating functionals can exactly be calculated by means of the
stationary-phase method [25-27]. For the case of interacting systems, the
partition functions and finite-temperature Green functions can be
conveniently calculated from the generating functionals by the perturbation
method. The coherent-state representation of the partition function and the
generating functional given in quantum statistics is now extended to the
thermal QCD in this paper, giving a correct formulation for the quantization
of the thermal QCD in the coherent-state representation.

The remainder of this paper is arranged as follows. In Sect. 2, we quote the
main results given in our previous paper for the quantum statistical
mechanics. These results may straightforwardly be extended to the quantum
field theory. In Sect. 3, we describe the coherent-state representation of
the thermal QCD in the first order (or say, Hamiltonian) formalism. In Sect.
4, the quantization of the thermal QCD is performed in the coherent-state
representation by writing out explicitly the generating functional of
thermal Green functions. To demonstrate the applicability and correctness of
the generating functional, we pay our attention to deriving the perturbative
expansion of the generating functional in the coherent-state representation.
Section 5 will be used to establish the three-dimensional equation obeyed by
the $q\overline{q}$ bound states at finite temperature. Section 6 serves to
derive the closed expression of the interaction kernel appearing in the
three-dimensional equation. In Sect.7, the one-gluon exchange kernel and
Hamiltonian will be discussed in detail. In the last section, some
concluding remarks will be made. In Appendix, the perturbative expansion of
the generating functional given in Sect. 4 will be transformed to the
corresponding one represented in the position space.

\section{Path integral formulation of quantum statistics in the
coherent-state representation}

First, we start from the partition function for a grand canonical ensemble
which usually is written in the form [19--22] 
\begin{equation}
Z=Tre^{-\beta \widehat{K}}  \eqnum{1}
\end{equation}
where $\beta =\frac 1{kT}$ with $k$ and $T$ being the Boltzmann constant and
the temperature and 
\begin{equation}
\widehat{K}=\widehat{H}-\mu \widehat{N}  \eqnum{2}
\end{equation}
here $\mu $ is the chemical potential, $\widehat{H}$ and $\widehat{N}$ are
the Hamiltonian and particle-number operators respectively. In the
coherent-state representation, the trace in Eq. (1) will be represented by
an integral over the coherent states. To determine the concrete form of the
integral, for simplicity, let us start from an one-dimensional system. Its
partition function given in the particle-number representation is 
\begin{equation}
Z=\sum_{n=0}^\infty \left\langle n\right| e^{-\beta \widehat{K}}\left|
n\right\rangle .  \eqnum{3}
\end{equation}
Then, we use the completeness relation of the coherent states [18-28] 
\begin{equation}
\int D(a^{*}a)\mid a><a^{*}\mid =1  \eqnum{4}
\end{equation}
where $\mid a>$ denotes a normalized coherent state, i.e., the eigenstate of
the annihilation operator $\hat a$ with a complex eigenvalue $a$ [18-28] 
\begin{equation}
\widehat{a}\left| a\right\rangle =a\left| a\right\rangle  \eqnum{5}
\end{equation}
whose Hermitian conjugate is 
\begin{equation}
\left\langle a^{*}\right| \widehat{a}^{+}=a^{*}\left\langle a^{*}\right| 
\eqnum{6}
\end{equation}
and $D(a^{*}a)$ symbolizes the integration measure defined by [18-28] 
\begin{equation}
D(a^{*}a)=\{ 
\begin{array}{cc}
\frac 1\pi da^{*}da, & \text{for bosons;} \\ 
da^{*}da, & \text{for fermions.}
\end{array}
\eqnum{7}
\end{equation}
In the above, we have used the eigenvalues $a$ and $a^{*}$ to designate the
eigenstates $\left| a\right\rangle $ and $\left\langle a^{*}\right| $,
respectively. It is emphasized that since we use the normalized
eigenfunction of the coherent state whose expression in its own
representation will be shown in Eq. (15), the completeness relation in Eq.
(4) has the ordinary form as we are familiar with in quantum mechanics.
Inserting Eq. (4) into Eq. (3), we have 
\begin{equation}
Z=\sum_{n=0}^\infty \int D(a^{*}a)D(a^{\prime *}a^{\prime })\left\langle
n\mid a^{\prime }\right\rangle \left\langle a^{\prime *}\right| e^{-\beta 
\widehat{K}}\left| a\right\rangle \left\langle a^{*}\mid n\right\rangle 
\eqnum{8}
\end{equation}
where 
\begin{equation}
\begin{array}{c}
\left\langle a^{*}\mid n\right\rangle =\frac 1{\sqrt{n!}}%
(a^{*})^ne^{-a^{*}a}, \\ 
\left\langle n\mid a^{\prime }\right\rangle =\frac 1{\sqrt{n!}}(a^{\prime
})^ne^{-a^{\prime *}a^{\prime }}
\end{array}
\eqnum{9}
\end{equation}
are the energy eigenfunctions given in the coherent-state representation
(Note: for fermions, $n=0,1$) [20-26]. The both eigenfunctions commute with
the matrix element $\left\langle a^{\prime *}\mid e^{-\beta \widehat{K}}\mid
a\right\rangle $ because the operator $\widehat{K}(\widehat{a}^{+},\widehat{a%
})$ generally is a polynomial of the operator $\widehat{a}^{+}\widehat{a}$
for fermion systems. In view of the expressions in Eq. (9) and the
commutation relation [20- 26] 
\begin{equation}
a^{*}a^{\prime }=\pm a^{\prime }a^{*}  \eqnum{10}
\end{equation}
where the signs ''$+$'' and ''$-$'' are attributed to bosons and fermions
respectively, it is easy to see 
\begin{equation}
\left\langle n\mid a^{\prime }\right\rangle \left\langle a^{*}\mid
n\right\rangle =\left\langle \pm a^{*}\mid n\right\rangle \left\langle n\mid
a^{\prime }\right\rangle .  \eqnum{11}
\end{equation}
Substituting Eq. (11) in Eq. (8) and applying the completeness relations for
the particle-number states and coherent ones, one may find 
\begin{equation}
Z=\int D(a^{*}a)\left\langle \pm a^{*}\right| e^{-\beta \widehat{K}}\left|
a\right\rangle  \eqnum{12}
\end{equation}
where the plus and minus signs in front of $a^{*}$ belong to bosons and
fermions respectively.

To evaluate the matrix element in Eq. (12), we may, as usual, divide the
''time'' interval $[0,\beta ]$ into $n$ equal and infinitesimal parts, $%
\beta =n\varepsilon $. and then insert a completeness relation shown in Eq.
(4) at each dividing point. In this way, Eq. (12) may be represented as
[19-27] 
\begin{equation}
\begin{array}{c}
Z=\int D(a^{*}a)\prod\limits_{i=1}^{n-1}D(a_i^{*}a_i)\left\langle \pm
a^{*}\right| e^{-\varepsilon \widehat{K}}\left| a_{n-1}\right\rangle
\left\langle a_{n-1}^{*}\right| e^{-\varepsilon \widehat{K}}\left|
a_{n-2}\right\rangle \cdot \cdot \cdot \\ 
\times \left\langle a_{i+1}^{*}\right| e^{-\varepsilon \widehat{K}}\left|
a_i\right\rangle \left\langle a_i^{*}\right| e^{-\varepsilon \widehat{K}%
}\left| a_{i-1}\right\rangle \cdot \cdot \cdot \left\langle a_1^{*}\right|
e^{-\varepsilon \widehat{K}}\left| a\right\rangle
\end{array}
\eqnum{13}
\end{equation}
Since $\varepsilon $ is infinitesimal, we can write 
\begin{equation}
e^{-\varepsilon \widehat{K}(\widehat{a}^{+},\widehat{a})}\approx
1-\varepsilon \widehat{K}(\widehat{a}^{+},\widehat{a})  \eqnum{14}
\end{equation}
where $\widehat{K}(\widehat{a}^{+},\widehat{a})$ is assumed to be normally
ordered. Noticing this fact, when applying the equations (5) and (6) and the
inner product of two coherent states [19-27] 
\begin{equation}
\left\langle a_i^{*}\mid a_{i-1}\right\rangle =e^{-\frac 12a_i^{*}a_i-\frac 1%
2a_{i-1}^{*}a_{i-1}+a_i^{*}a_{i-1}}  \eqnum{15}
\end{equation}
which suits to the both of bosons and fermions, one can get from Eq. (13)
that 
\begin{equation}
\begin{array}{c}
Z=\int D(a^{*}a)e^{-a^{*}a}\int \prod\limits_{i=1}^{n-1}D(a_i^{*}a_i)\exp
\{-\varepsilon \sum\limits_{i=1}^nK(a_i^{*},a_{i-1}) \\ 
+\sum\limits_{i=1}^na_i^{*}a_{i-1}-\sum\limits_{i=1}^{n-1}a_i^{*}a_i\}
\end{array}
\eqnum{16}
\end{equation}
where we have set 
\begin{equation}
\pm a^{*}=a_n^{*}\text{ , }a=a_0.  \eqnum{17}
\end{equation}
It is noted that the factor $e^{-a^{*}a}$ in the first integrand comes from
the matrix elements $\left\langle \pm a^{*}\right| a_{n-1}\rangle $ and $%
\left\langle a_1^{*}\right| a\rangle $ and the last sum in the above
exponent is obtained by summing up the common terms $-\frac 12a_i^{*}a_i$
and $-\frac 12a_{i-1}^{*}a_{i-1}$ appearing in the exponents of the matrix
element $\langle a_i^{*}\mid a_{i-1}\rangle $ and its adjacent ones $\langle
a_{i+1}^{*}\mid a_i\rangle $ and $\langle a_{i-1}^{*}\mid a_{i-2}\rangle $.
As will be seen in Eq. (21), such a summation is essential to give a correct
time-dependence of the functional integrand in the partition function. The
last two sums in Eq. (16) can be rewritten in the form 
\begin{equation}
\begin{array}{c}
\sum\limits_{i=1}^na_i^{*}a_{i-1}-\sum\limits_{i=1}^{n-1}a_i^{*}a_i \\ 
=\frac 12a_n^{*}a_{n-1}+\frac 12a_1^{*}a_0+\frac \varepsilon 2%
\sum\limits_{i=1}^{n-1}[(\frac{a_{i+1}^{*}-a_i^{*}}\varepsilon )a_i-a_i^{*}(%
\frac{a_i-a_{i-1}}\varepsilon )].
\end{array}
\eqnum{18}
\end{equation}
Upon substituting Eq. (18) in Eq. (16) and taking the limit $\varepsilon
\rightarrow 0$, we obtain the path-integral expression of the partition
functions as follows:

\begin{equation}
Z=\int D(a^{*}a)e^{-a^{*}a}\int {\frak D}(a^{*}a)e^{I(a^{*},a)}  \eqnum{19}
\end{equation}
where 
\begin{equation}
{\frak D}(a^{*}a)=\{ 
\begin{array}{cc}
\prod\limits_\tau \frac 1\pi da^{*}(\tau )da(\tau ), & \text{for bosons;} \\ 
\prod\limits_\tau da^{*}(\tau )da(\tau ), & \text{for fermions}
\end{array}
\eqnum{20}
\end{equation}
and 
\begin{equation}
\begin{array}{c}
I(a^{*},a)=\frac 12a^{*}(\beta )a(\beta )+\frac 12a^{*}(0)a(0)-\int_0^\beta
d\tau [\frac 12a^{*}(\tau )\dot a(\tau ) \\ 
-\frac 12\dot a^{*}(\tau )a(\tau )+K(a^{*}(\tau ),a(\tau ))] \\ 
=a^{*}(\beta )a(\beta )-\int_0^\beta d\tau [a^{*}(\tau )\dot a(\tau
)+K(a^{*}(\tau ),a(\tau ))
\end{array}
\eqnum{21}
\end{equation}
where the last equality is obtained from the first one by a partial
integration. In accordance with the definition given in Eq. (17), we see,
the path-integral is subject to the following boundary conditions 
\begin{equation}
a^{*}(\beta )=\pm a^{*},a(0)=a  \eqnum{22}
\end{equation}
where the signs ''$+$'' and ''$-$'' are written respectively for bosons and
fermions. Here it is noted that Eq. (22) does not implies $a(\beta )=\pm a$
and $a^{*}(0)=a^{*}$. Actually, we have no such boundary conditions.

For the systems with many degrees of freedom, the functional-integral
representation of the partition functions may directly be written out from
the results given in Eqs. (19) -(22) as long as the eigenvalues $a$ and $%
a^{*}$ are understood as column matrices $a=(a_1,a_2,\cdots ,a_k,\cdots )$
and $a^{*}=(a_1^{*},a_2^{*},\cdot \cdot \cdot ,a_k^{*},\cdots )$. Written
explicitly, we have 
\begin{equation}
Z=\int D(a^{*}a)e^{-a_k^{*}a_k}\int {\frak D}(a^{*}a)e^{I(a^{*},a)} 
\eqnum{23}
\end{equation}
where 
\begin{equation}
D(a^{*}a)=\{ 
\begin{array}{cc}
\prod\limits_k\frac 1\pi da_k^{*}da_k & \text{, for bosons;} \\ 
\prod\limits_kda_k^{*}da_k & \text{, for fermions,}
\end{array}
\eqnum{24}
\end{equation}
\begin{equation}
{\frak D}(a^{*}a)=\{ 
\begin{array}{cc}
\prod\limits_{k\tau }\frac 1\pi da_k^{*}(\tau )da_k(\tau ) & \text{, for
bosons;} \\ 
\prod\limits_{k\tau }da_k^{*}(\tau )da_k(\tau ) & \text{, for fermions}
\end{array}
\eqnum{25}
\end{equation}
and 
\begin{equation}
I(a^{*},a)=a_k^{*}(\beta )a_k(\beta )-\int_0^\beta d\tau [a_k^{*}(\tau )\dot 
a_k(\tau )+K(a_k^{*}(\tau ),a_k(\tau ))].  \eqnum{26}
\end{equation}
The boundary conditions in Eq. (22) now become 
\begin{equation}
a_k^{*}(\beta )=\pm a_k^{*}\text{ , }a_k(0)=a_k.  \eqnum{27}
\end{equation}
In Eqs. (23) and (26), the repeated indices imply the summations over $k$.
If the $k$ stands for a continuous index as in the case of quantum field
theory, the summations will be replaced by integrations over $k$.

It should be pointed out that in the previous derivation of the
coherent-state representation of the partition functions, the authors did
not use the expressions given in Eqs. (16) and (18). Instead, the matrix
element in Eq. (15) was directly chosen to be the starting point and recast
in the form [18-22]

\begin{equation}
\langle a_i^{*}\mid a_{i-1}\rangle =\exp \{-\frac \varepsilon 2[a_i^{*}(%
\frac{a_i-a_{i-1}}\varepsilon )-(\frac{a_i^{*}-a_{i-1}^{*}}\varepsilon
)a_{i-1}]\}.  \eqnum{28}
\end{equation}
Substituting the above expression into Eq. (13) and taking the limit $%
\varepsilon \rightarrow 0$, it follows [18-22]

\begin{equation}
Z=\int {\frak D}(a^{*}a)\exp \{-\int_0^\beta d\tau [\frac 12a^{*}(\tau )\dot 
a(\tau )-\frac 12\dot a^{*}(\tau )a(\tau )+K(a^{*}(\tau ),a(\tau ))]\}. 
\eqnum{29}
\end{equation}
Clearly, in the above derivation, the common terms appearing in the
exponents of adjacent matrix elements were not combined together. As a
result, the time-dependence of the integrand in Eq. (29) could not be given
correctly. In comparison with the previous result shown in Eq. (29), the
expression written in Eqs. (19)-(21) has two functional integrals. The first
integral which represents the trace in Eq. (1) is absent in Eq. (29). The
second integral is defined as the same as the integral in Eq. (29); but the
integrand are different from each other. In Eq. (19), there occur two
additional factors in the integrand : one is $e^{-a^{*}a}$ which comes from
the initial and final states in Eq. (13), another is $e^{\frac 12%
[a^{*}(\beta )a(\beta )+a^{*}(0)a(0)]}$ in which $a^{*}(\beta )$ and $a(0)$
are related to the boundary conditions shown in Eq. (22). These additional
factors are also absent in Eq. (29). As will be seen soon later, the
occurrence of these factors in the functional-integral expression is
essential to give correct calculated results.

To demonstrate the correctness of the expression given in Eqs. (23)-(27),
let us compute the partition function for the system whose Hamiltonian is of
harmonic oscillator-type as we meet in the cases of ideal gases and free
fields. In this case, 
\begin{equation}
K(a^{*}a)=\omega _ka_k^{*}a_k  \eqnum{30}
\end{equation}
where $\omega _k=\varepsilon _k-\mu $ with $\varepsilon _k$ being the
particle energy and therefore Eq. (26) becomes 
\begin{equation}
I(a^{*},a)=a_k^{*}(\beta )a_k(\beta )-\int_0^\beta d\tau [a_k^{*}(\tau )\dot 
a_k(\tau )+\omega _ka_k^{*}(\tau )a_k(\tau )].  \eqnum{31}
\end{equation}
By the stationary-phase method which is established based on the property of
the Gaussian integral that the integral is equal to the extremum of the
integrand which is an exponential function [25-27], we may write 
\begin{equation}
\int {\frak D}(a^{*}a)e^{I(a^{*},a)}=e^{I_0(a^{*},a)}  \eqnum{32}
\end{equation}
where $I_0(a^{*},a)$ is obtained from $I(a^{*},a)$ by replacing the
variables $a_k^{*}(\tau )$ and $a_k(\tau )$ in $I(a^{*},a)$ with those
values which are determined from the stationary condition $\delta
I(a^{*},a)=0$. From this condition and the boundary conditions in Eq. (27)
which implies $\delta a_k^{*}(\beta )=0$ and $\delta a_k(0)=0$, it is easy
to derive the following equations of motion [24-26] 
\begin{equation}
\dot a_k(\tau )+\omega _ka_k(\tau )=0,\text{ }\dot a_k^{*}(\tau )-\omega
_ka_k^{*}(\tau )=0.  \eqnum{33}
\end{equation}
Their solutions satisfying the boundary condition are 
\begin{equation}
a_k(\tau )=a_ke^{-\omega _k\tau }\text{ , }a_k^{*}(\tau )=\pm
a_k^{*}e^{\omega _k(\tau -\beta )}.  \eqnum{34}
\end{equation}
On substituting the above solutions into Eq. (31), we obtain 
\begin{equation}
I_0(a^{*},a)=\pm a_k^{*}a_ke^{-\omega _k\beta }  \eqnum{35}
\end{equation}
With the functional integral given in Eqs. (32) and (35), the partition
functions in Eq. (23) become 
\begin{equation}
Z_0=\{ 
\begin{array}{cc}
\int D(a^{*}a)e^{-a_k^{*}a_k(1-e^{-\beta \omega _k})} & \text{ , for bosons;}
\\ 
\int D(a^{*}a)e^{-a_k^{*}a_k(1+e^{-\beta \omega _k})} & \text{ , for
fermions.}
\end{array}
\eqnum{36}
\end{equation}
For the boson case, the above integral can directly be calculated by
employing the integration formula [18]: 
\begin{equation}
\int D(a^{*}a)e^{-a^{*}(\lambda a-b)}f(a)=\frac 1\lambda f(\lambda ^{-1}b) 
\eqnum{37}
\end{equation}
The result is well-known, as shown in the following [20-22,29] 
\begin{equation}
Z_0=\prod\limits_k\frac 1{1-e^{-\beta \omega _k}}  \eqnum{38}
\end{equation}
For the fermion case, by using the property of Grassmann algebra and the
integration formulas [21-26]: 
\begin{equation}
\int da=\int da^{*}=0\text{ , }\int da^{*}a^{*}=\int daa=1  \eqnum{39}
\end{equation}
it is easy to compute the integral in Eq. (36) and get the familiar result
[21-23,29] 
\begin{equation}
Z_0=\prod\limits_k(1+e^{-\beta \omega _k})  \eqnum{40}
\end{equation}
It is noted that if the stationary-phase method is applied to the functional
integral in Eq. (29), one could not get the results as written in Eqs. (38)
and (40), showing the incorrectness of the previous functional-integral
representation for the partition functions.

Now let us turn to discuss the general case where the Hamiltonian can be
split into a free part and an interaction part. Correspondingly, we can
write 
\begin{equation}
K(a^{*},a)=K_0(a^{*},a)+H_I(a^{*},a)  \eqnum{41}
\end{equation}
where $K_0(a^{*},a)$ is the same as given in Eq. (30) and $H_I(a^{*},a)$ is
the interaction Hamiltonian. In this case, to evaluate the partition
function, it is convenient to define a generating functional through
introducing external sources $j_k^{*}(\tau )$ and $j_k(\tau )$ such that
[21-23] 
\begin{equation}
\begin{array}{c}
Z[j^{*},j]=\int D(a^{*}a)e^{-a_{^{*}k}a_k}\int {\frak D}(a^{*}a)\exp
\{a_k^{*}(\beta )a_k(\beta ) \\ 
-\int_0^\beta d\tau [a_k^{*}(\tau )\dot a_k(\tau )+K(a^{*}a)-j_k^{*}(\tau
)a_k(\tau )-a_k^{*}(\tau )j_k(\tau )]\} \\ 
=e^{-\int_0^\beta d\tau H_I(\frac \delta {\delta j_k^{*}(\tau )},\pm \frac 
\delta {\delta j_k(\tau )})}Z_0[j^{*},j]
\end{array}
\eqnum{42}
\end{equation}
where the sings ''$+$'' and ''$-$'' in front of $\frac \delta {\delta
j_k(\tau )}$ refer to bosons and fermions respectively and $Z_0[j^{*},j]$ is
defined by 
\begin{equation}
Z_0[j^{*},j]=\int D(a^{*}a)e^{-a_k^{*}a_k}\int {\frak D}%
(a^{*}a)e^{I(a^{*},a;j^{*},j)}  \eqnum{43}
\end{equation}
in which 
\begin{equation}
\begin{array}{c}
I(a^{*},a;j^{*},j)=a_k^{*}(\beta )a_k(\beta )-\int_0^\beta d\tau
[a_k^{*}(\tau )\dot a_k(\tau ) \\ 
+\omega _ka_k^{*}(\tau )a_k(\tau )-j_k^{*}(\tau )a_k(\tau )-a_k^{*}(\tau
)j_k(\tau )]
\end{array}
\eqnum{44}
\end{equation}
Obviously, the integral in Eq. (43) is of Gaussian-type. Therefore, it can
be calculated by means of the stationary-phase method as will be shown in
detail in Sect. 4.

The exact partition functions can be obtained from the generating functional
in Eq. (42) by setting the external sources to be zero 
\begin{equation}
Z=Z[j^{*},j]\mid _{j^{*}=j=0}.  \eqnum{45}
\end{equation}
In particular, the generating functional is much useful to compute the
finite-temperature Green functions. For simplicity, we take the two-point
Green function as an example to show this point. In many-body theory, the
Green function usually is defined in the operator formalism by [21,29] 
\begin{equation}
G_{kl}(\tau _1,\tau _2)=\frac 1ZTr\{e^{-\beta \widehat{K}}T[\widehat{a}%
_k(\tau _1)\widehat{a}_l^{+}(\tau _2)]\}=Tr\{e^{\beta (\Omega -\widehat{K}%
})T[\widehat{a}_k(\tau _1)\widehat{a}_l^{+}(\tau _2)]\}  \eqnum{46}
\end{equation}
where $0<\tau _1,\tau _2<\beta $, $\Omega =-\frac 1\beta \ln Z$ is the grand
canonical potential, $T$ denotes the ''time'' ordering operator, $\widehat{a}%
_k(\tau _1)$ and $\widehat{a}_l^{+}(\tau _2)$ represent the annihilation and
creation operators respectively. According to the procedure described in
Eqs. (12)-(22). it is clear to see that when taking $\tau _1$ and $\tau _2$
at two dividing points and applying the equations (5) and (6), the Green
function may be expressed as a functional integral in the coherent-state
representation as follows 
\begin{equation}
G_{kl}(\tau _1,\tau _2)=\frac 1Z\int D(a^{*}a)e^{-a_k^{*}a_k}\int {\frak D}%
(a^{*}a)a_k(\tau _1)a_l^{*}(\tau _2)e^{I(a^{*},a)}.  \eqnum{47}
\end{equation}
With the aid of the generating functional defined in Eq. (42), the above
Green function may be represented as 
\begin{equation}
G_{kl}(\tau _1,\tau _2)=\pm \frac 1Z\frac{\delta ^2Z[j^{*},j]}{\delta
j_k^{*}(\tau _1)\delta j_l(\tau _2)}\mid _{j^{*}=j=0}  \eqnum{48}
\end{equation}
where the sings ''$+$'' and ''$-$'' belong to bosons and fermions
respectively.

\section{The coherent-state representation of thermal QCD Hamiltonian and
action}

To write out explicitly a path-integral expression of thermal QCD in the
coherent-state representation, we first need to formulate the QCD in the
coherent-state representation, namely, to give exact expressions of the QCD
Hamiltonian and action in the coherent-state representation. For this
purpose, we only need to work with the classical fields by using some
skilful treatments. Let us start from the effective Lagrangian density of
QCD which appears in the path-integral of the zero-temperature QCD [22, 25,
26] 
\begin{equation}
{\cal L}=\bar \psi \{i\gamma ^\mu (\partial _\mu -igT^aA_\mu ^a)-m\}\psi -%
\frac 14F^{a\mu \nu }F_{\mu \nu }^a-\frac 1{2\alpha }(\partial ^\mu A_\mu
^a)^2-\partial ^\mu \bar C^aD_\mu ^{ab}C^b  \eqnum{49}
\end{equation}
where $T^a=\lambda ^a/2$ is the color matrix, $\psi $ and $\bar \psi $
represent the quark fields, $A_\mu ^a$ are the vector potentials of gluon
fields, $C^a$ and $\bar C^a$ designate the ghost fields, 
\begin{equation}
F_{\mu \nu }^a=\partial _\mu A_\nu ^a-\partial _\nu A_\mu ^a+gf^{abc}A_\mu
^bA_\nu ^c  \eqnum{50}
\end{equation}
and 
\begin{equation}
D_\mu ^{ab}=\delta ^{ab}\partial _\mu -gf^{abc}A_\mu ^c  \eqnum{51}
\end{equation}
For the sake of simplicity, we work in the Feynman gauge ($\alpha =1$). It
is well-known that in this gauge, the results obtained from the above
Lagrangian are equivalent to those derived from the following Lagrangian
which is given by applying the Lorentz condition $\partial ^\mu A_\mu ^a=0$
to the Lagrangian in Eq. (49), 
\begin{equation}
\begin{array}{c}
{\cal L}=\bar \psi \{i\gamma ^\mu (\partial _\mu -igT^aA_\mu ^a)-m\}\psi -%
\frac 12\partial _\mu A_\nu ^a\partial ^\mu A^{a\nu }-gf^{abc}\partial _\mu
A_\nu ^aA^{b\mu }A^{c\nu } \\ 
-\frac 14g^2f^{abc}f^{ade}A^{b\mu }A^{c\nu }A_\mu ^dA_\nu ^e-\partial ^\mu 
\bar C^a\partial _\mu C^b+gf^{abc}\partial ^\mu \bar C^aC^bA_\mu ^c
\end{array}
\eqnum{52}
\end{equation}
Here it is noted that the application of the Lorentz condition only changes
the form of free part of the gluon Lagrangian, remaining the interaction
part of the Lagrangian in Eq. (49) formally unchanged. The above Lagrangian
is written in the Minkowski metric where the $\gamma -$matrix is defined as $%
\gamma _0=\beta $ and $\vec \gamma =\beta \vec \alpha $ [26]. In the
following, it is convenient to represent the Lagrangian in the Euclidean
metric with the imaginary time $\tau =it$ where $t$ is the real time.

Since the path-integral in Eq. (42) is established in the first order (or
say, Hamiltonian) formalism, to perform the path-integral quantization of
thermal QCD in the coherent-state representation, we need to recast the
above Lagrangian in the first order form. In doing this, it is necessary to
introduce canonical conjugate momentum densities which are defined by [26,
30] 
\begin{equation}
\begin{array}{c}
\Pi _\psi =\frac{\partial {\cal L}}{\partial \partial _t\psi }=i\overline{%
\psi }\gamma ^0=i\psi ^{+}, \\ 
\Pi _{\overline{\psi }}=\frac{\partial {\cal L}}{\partial \partial _t%
\overline{\psi }}=0, \\ 
\Pi _\mu ^a=\frac{\partial {\cal L}}{\partial \partial _tA^{a\mu }}%
=-\partial _tA_\mu ^a+gf^{abc}A_\mu ^bA_0^c, \\ 
\Pi ^a=(\frac{\partial {\cal L}}{\partial \partial _tC^a})_R=-\partial _t%
\overline{C}^a, \\ 
\overline{\Pi }^a=(\frac{\partial {\cal L}}{\partial \partial _t\overline{C}%
^a})_L=-\partial _tC^a+gf^{abc}C^bA_0^c
\end{array}
\eqnum{53}
\end{equation}
where the subscripts $R$ and $L$ mark the right and left-derivatives with
respect to the real time respectively. With the above momentum densities,
the Lagrangian in Eq. (52) can be represented as 
\begin{equation}
{\cal L}=\Pi _\psi \stackrel{\cdot }{\partial _t\psi +}\Pi ^{a\mu }\partial
_tA_\mu ^a+\Pi ^a\partial _tC^a+\partial _t\overline{C}^a\overline{\Pi }-%
{\cal H}  \eqnum{54}
\end{equation}
where 
\begin{equation}
{\cal H}={\cal H}_0+{\cal H}_I  \eqnum{55}
\end{equation}
is the Hamiltonian density in which 
\begin{equation}
{\cal H}_0=\bar \psi (\vec \gamma \cdot \triangledown +m)\psi +\frac 12(\Pi
_\mu ^a)^2-\frac 12A_\mu ^a\nabla ^2A_\mu ^a-\Pi ^a\overline{\Pi }^a+\bar C%
^a\nabla ^2C^a  \eqnum{56}
\end{equation}
is the free Hamiltonian density and 
\begin{equation}
\begin{array}{c}
{\cal H}_I=ig\bar \psi T^a\gamma _\mu A_\mu ^a\psi +gf^{abc}(i\Pi _\mu
^aA_4^c+\partial _iA_\mu ^aA_i^c)A_\mu ^b-\frac 14g^2f^{abc}f^{ade}A_\mu
^bA_\mu ^d \\ 
\times (A_4^cA_4^e-A_i^cA_i^e)+gf^{abc}(i\Pi ^aA_4^c-\partial _i\bar C%
^aA_i^c)C^b
\end{array}
\eqnum{57}
\end{equation}
is the interaction Hamiltonian density here the Latin letter $i$ denotes the
spatial index. The above Hamiltonian density is written in the Euclidean
metric for later convenience. The matrix $\gamma _\mu $ in this metric is
defined by $\gamma _4=\beta $ and $\vec \gamma =-i\beta \vec \alpha $ [30].
It should be noted that the conjugate quantities $\Pi ^a$ and $\overline{\Pi 
}^a$ for the ghost fields are respectively defined by the right-derivative
and the left one as shown in Eq. (53) because only in this way one can get
correct results. This unusual definition originates from the peculiar
property of the ghost fields which are scalar fields, but subject to the
commutation rule of Grassmann algebra.

In order to derive an expression of the thermal QCD in the coherent-state
representation, one should employ the Fourier transformations for the
canonical variables of the QCD which are listed below. For the quark field
[26, 30], 
\begin{equation}
\psi (\vec x,\tau )=\int \frac{d^3p}{(2\pi )^{3/2}}[u^s(\vec p)b_s(\vec p%
,\tau )e^{i\vec p\cdot \vec x}+v^s(\vec p)d_s^{*}(\vec p,\tau )e^{-i\vec p%
\cdot \vec x}]  \eqnum{58}
\end{equation}
\begin{equation}
\overline{\psi }(\vec x,\tau )=\int \frac{d^3p}{(2\pi )^{3/2}}[\overline{u}%
^s(\vec p)b_s^{*}(\vec p,\tau )e^{-i\vec p\cdot \vec x}+\overline{v}^s(\vec p%
)d_s(\vec p,\tau )e^{i\vec p\cdot \vec x}]  \eqnum{59}
\end{equation}
where $u^s(\vec p)$ and $v^s(\vec p)$ are the spinor wave functions
satisfying the normalization conditions $u^{s+}(\vec p)u^s(\vec p)=v^{s+}(%
\vec p)v^s(\vec p)=1$, $b_s(\vec p,\tau )$ and $b_s^{*}(\vec p,\tau )$ are
the eigenvalues of the quark annihilation and creation operators $\widehat{b}%
_s(\vec p,\tau )$ and $\widehat{b}_s^{+}(\vec p,\tau )$ which are defined in
the Heisenberg picture, $d_s(\vec p,\tau )$ and $d_s^{*}(\vec p,\tau )$ are
the corresponding ones for antiquarks. For the gluon field [26, 30], 
\begin{equation}
A_\mu ^c(\vec x,\tau )=\int \frac{d^3k}{(2\pi )^{3/2}}\frac 1{\sqrt{2\omega (%
\vec k)}}\varepsilon _\mu ^\lambda (\vec k)[a_\lambda ^c(\vec k,\tau )e^{i%
\vec k\cdot \vec x}+a_\lambda ^{c*}(\vec k,\tau )e^{-i\vec k\cdot \vec x}] 
\eqnum{60}
\end{equation}
where $\varepsilon _\mu ^\lambda (\overrightarrow{k})$ is the polarization
vector and 
\begin{equation}
\Pi _\mu ^c(\vec x,\tau )=i\int \frac{d^3k}{(2\pi )^{3/2}}\sqrt{\frac{\omega
(\vec k)}2}\varepsilon _\mu ^\lambda (\vec k)[a_\lambda ^c(\vec k,\tau )e^{i%
\vec k\cdot \vec x}-a_\lambda ^{c*}(\vec k,\tau )e^{-i\vec k\cdot \vec x}] 
\eqnum{61}
\end{equation}
which follows from the definition in Eq. (53) and is consistent with the
Fourier representation of free fields. In the above, $a_\lambda ^c(\vec k%
,\tau )$ and $a_\lambda ^{c*}(\vec k,\tau )$ are the eigenvalues of the
gluon annihilation and creation operators $\widehat{a}_\lambda ^c(\vec k%
,\tau )$ and $\widehat{a}_\lambda ^{c+}(\vec k,\tau ).$ For the ghost field,
we have 
\begin{equation}
\overline{C}^a(\vec x,\tau )=\int \frac{d^3q}{(2\pi )^{3/2}}\frac 1{\sqrt{%
2\omega (\vec q)}}[\overline{c}_a(\vec q,\tau )e^{i\vec q\cdot \vec x%
}+c_a^{*}(\vec q,\tau )e^{-i\vec q\cdot \vec x}],  \eqnum{62}
\end{equation}
\begin{equation}
C^a(\vec x,\tau )=\int \frac{d^3q}{(2\pi )^{3/2}}\frac 1{\sqrt{2\omega (\vec 
q)}}[c_a(\vec q,\tau )e^{i\vec q\cdot \vec x}+\overline{c}_a^{*}(\vec q,\tau
)e^{-i\vec q\cdot \vec x}],  \eqnum{63}
\end{equation}
\begin{equation}
\Pi ^a(\vec x,\tau )=i\int \frac{d^3q}{(2\pi )^{3/2}}\sqrt{\frac{\omega (%
\vec q)}2}[\overline{c}_a(\vec q,\tau )e^{i\vec q\cdot \vec x}-c_a^{*}(\vec q%
,\tau )e^{-i\vec q\cdot \vec x}],  \eqnum{64}
\end{equation}
and 
\begin{equation}
\overline{\Pi }^a(\vec x,\tau )=i\int \frac{d^3q}{(2\pi )^{3/2}}\sqrt{\frac{%
\omega (\vec q)}2}[c_a(\vec q,\tau )e^{i\vec q\cdot \vec x}-\overline{c}%
_a^{*}(\vec q,\tau )e^{-i\vec q\cdot \vec x}].  \eqnum{65}
\end{equation}
where $c_a(\vec q,\tau )$ and $c_a^{*}(\vec q,\tau )$ are the eigenvalues of
the ghost particle annihilation and creation operators $\widehat{c}_a(\vec q%
,\tau )$ and $\widehat{c}_a^{+}(\vec q,\tau )$ and $\overline{c}_a(\vec q%
,\tau )$ and $\overline{c}_a^{*}(\vec q,\tau )$ are the ones for antighost
particles.

For simplifying the expressions of the Hamiltonian and action of the thermal
QCD, it is convenient to use abbreviation notations. Define 
\begin{equation}
b_s^\theta (\vec p,\tau )= 
{b_s(\vec p,\tau ),\text{ }if\text{ }\theta =+, \atopwithdelims\{\} d_s^{*}(\vec p,\tau ),\text{ }if\text{ }\theta =-,}
\eqnum{66}
\end{equation}
\begin{equation}
W_s^\theta (\vec p)= 
{(2\pi )^{-3/2}u^s(\vec p),\text{ }if\text{ }\theta =+, \atopwithdelims\{\} (2\pi )^{-3/2}v^s(\vec p),\text{ }if\text{ }\theta =-}
\eqnum{67}
\end{equation}
and furthermore, set $\alpha =(\vec p,s,\theta )$ and 
\begin{equation}
\sum\limits_\alpha =\sum\limits_{s\theta }\int d^3p,  \eqnum{68}
\end{equation}
Eqs. (58) and (59) may be represented as 
\begin{equation}
\begin{array}{c}
\psi (\vec x,\tau )=\sum\limits_\alpha W_\alpha b_\alpha (\tau )e^{i\theta 
\vec p\cdot \vec x}, \\ 
\overline{\psi }(\vec x,\tau )=\sum\limits_\alpha \overline{W}_\alpha
b_\alpha ^{*}(\tau )e^{-i\theta \vec p\cdot \vec x}.
\end{array}
\eqnum{69}
\end{equation}
Similarly, when we define 
\begin{equation}
a_{\lambda \theta }^c(\vec k,\tau )= 
{a_\lambda ^c(\vec k,\tau ),\text{ }if\text{ }\theta =+, \atopwithdelims\{\} a_\lambda ^{c*}(\vec k,\tau ),\text{ }if\text{ }\theta =-,}
\eqnum{70}
\end{equation}
\begin{equation}
\begin{array}{c}
A_{\mu \theta }^{c\lambda }(\vec k)=(2\pi )^{-3/2}(2\omega (\vec k%
))^{-1/2}\epsilon _\mu ^\lambda (\vec k), \\ 
\Pi _{\mu \theta }^{c\lambda }(\vec k)=i^\theta (2\pi )^{-3/2}[\omega (\vec q%
)/2]^{1/2}\epsilon _\mu ^\lambda (\vec k)
\end{array}
\eqnum{71}
\end{equation}
and furthermore, set $\alpha =(\vec k,c,\lambda ,\theta )$ and 
\begin{equation}
\sum\limits_\alpha =\sum\limits_{c\lambda \theta }\int d^3k,  \eqnum{72}
\end{equation}
Eqs. (60) and (61) can be written as 
\begin{equation}
\begin{array}{c}
A_\mu ^c(\vec x,\tau )=\sum\limits_\alpha A_\mu ^\alpha a_\alpha (\tau
)e^{i\theta \vec k\cdot \vec x}, \\ 
\Pi _\mu ^c(\vec x,\tau )=\sum\limits_\alpha \Pi _\mu ^\alpha a_\alpha (\tau
)e^{i\theta \vec k\cdot \vec x}
\end{array}
\eqnum{73}
\end{equation}
For the ghost fields, if we define 
\begin{equation}
c_\alpha ^\theta (\vec q,\tau )= 
{\overline{c}_a(\vec q,\tau ),\text{ }if\text{ }\theta =+, \atopwithdelims\{\} c_a^{*}(\vec q,\tau ),\text{ }if\text{ }\theta =-,}
\eqnum{74}
\end{equation}
\begin{equation}
\begin{array}{c}
G_\theta (\vec q)=(2\pi )^{-3/2}[2\omega (\vec q)]^{-1/2}, \\ 
\Pi _\theta (\vec q)=i^\theta (2\pi )^{-3/2}[\omega (\vec q)/2]^{1/2},
\end{array}
\eqnum{75}
\end{equation}
and furthermore set $\alpha =(\vec q,a,\theta )$ and 
\begin{equation}
\sum\limits_\alpha =\sum_{a\theta }\int d^3q  \eqnum{76}
\end{equation}
then, Eqs. (62)-(65) will be expressed as 
\begin{equation}
\begin{array}{c}
\overline{C}^a(\vec x,\tau )=\sum\limits_\alpha G_\alpha c_\alpha (\tau
)e^{i\theta \vec q\cdot \vec x} \\ 
C^a(\vec x,\tau )=\sum\limits_\alpha G_\alpha c_\alpha ^{*}(\tau
)e^{-i\theta \vec q\cdot \vec x} \\ 
\Pi ^a(\vec x,\tau )=\sum\limits_\alpha \Pi _\alpha c_\alpha (\tau
)e^{i\theta \vec q\cdot \vec x} \\ 
\overline{\Pi }^a(\vec x,\tau )=\sum\limits_\alpha \Pi _\alpha c_\alpha
^{*}(\tau )e^{-i\theta \vec q\cdot \vec x}
\end{array}
\eqnum{77}
\end{equation}

Upon substituting Eqs. (69), (73) and (77) into Eqs. (56) and (57), it is
not difficult to get 
\begin{equation}
\begin{array}{c}
H_0(\tau )=\int d^3x{\cal H}_0(x)=\sum\limits_\alpha \theta _\alpha
\varepsilon _\alpha b_\alpha ^{*}(\tau )b_\alpha (\tau ) \\ 
+\frac 12\sum\limits_\alpha \omega _\alpha a_\alpha ^{*}(\tau )a_\alpha
(\tau )+\sum\limits_\alpha \omega _\alpha c_\alpha ^{*}(\tau )c_\alpha (\tau
)
\end{array}
\eqnum{78}
\end{equation}
and 
\begin{equation}
\begin{array}{c}
H_I(\tau )=\int d^3x{\cal H}_I(x)=\sum\limits_{\alpha \beta \gamma }A(\alpha
\beta \gamma )b_\alpha ^{*}(\tau )b_\beta (\tau )a_\gamma (\tau
)+\sum\limits_{\alpha \beta \gamma }B(\alpha \beta \gamma )a_\alpha (\tau
)a_\beta (\tau )a_\gamma (\tau ) \\ 
+\sum\limits_{\alpha \beta \gamma \delta }C(\alpha \beta \gamma \delta
)a_\alpha (\tau )a_\beta (\tau )a_\gamma (\tau )a_\delta (\tau
)+\sum\limits_{\alpha \beta \gamma }D(\alpha \beta \gamma )c_\alpha
^{*}(\tau )c_\beta (\tau )a_\gamma (\tau )
\end{array}
\eqnum{79}
\end{equation}
which are the QCD Hamiltonian given in the coherent state representation. In
Eq. (78), the first, second and third terms are the free Hamiltonians for
quarks, gluons and ghost particles respectively where $\theta _\alpha \equiv
\theta $, $\varepsilon _\alpha =(\vec p^2+m^2)^{1/2}$ is the quark energy, $%
\omega _\alpha =\left| \vec k\right| $ is the energy for a gluon or a ghost
particle. In Eq. (79), the first term is the interaction Hamiltonian between
quarks and gluons, the second and third terms are the interaction
Hamiltonian among gluons and the fourth term represents the interaction
Hamiltonian between ghost particles and gluons. The coefficient functions in
Eq. (79) are defined as follows: 
\begin{equation}
A(\alpha \beta \gamma )=ig(2\pi )^3\delta ^3(\theta _\alpha \vec p_\alpha
-\theta _\beta \vec p_\beta -\theta _\gamma \vec k_\gamma )\overline{W}%
_{s_\alpha }^{\theta _\alpha }(\vec p_\alpha )T^a\gamma _\mu W_{s_\alpha
}^{\theta _\beta }(\vec p_\beta )A_{\mu \theta _\gamma }^{a\lambda _\gamma }(%
\vec k_\gamma ),  \eqnum{80}
\end{equation}
\begin{equation}
\begin{array}{c}
B(\alpha \beta \gamma )=ig(2\pi )^3\delta ^3(\theta _\alpha \vec k_\alpha
+\theta _\beta \vec k_\beta +\theta _\gamma \vec k_\gamma )f^{abc}[\Pi _{\mu
\theta _\alpha }^{a\lambda _\alpha }(\vec k_\alpha ) \\ 
\times A_{4\theta _\gamma }^{c\lambda _\gamma }(\vec k_\gamma )+\theta
_\alpha k_i^\alpha A_{\mu \theta \alpha }^{a\lambda _\alpha }(\vec k_\alpha
)A_{i\theta _\gamma }^{c\lambda _\gamma }(\vec k_\gamma )]A_{\mu \theta
_\beta }^{b\lambda _\beta }(\vec k_\beta ),
\end{array}
\eqnum{81}
\end{equation}
\begin{equation}
\begin{array}{c}
C(\alpha \beta \gamma \delta )=-\frac 14g^2(2\pi )^3\delta ^3(\theta _\alpha 
\vec k_\alpha +\theta _\beta \vec k_\beta +\theta _\rho \vec k_\rho +\theta
_\sigma \vec k_\sigma )f^{abc}f^{ade} \\ 
\times A_{\mu \theta \alpha }^{b\lambda _\alpha }(\vec k_\alpha )A_{\mu
\theta _\beta }^{d\lambda _\beta }(\vec k_\beta )[A_{4\theta _\rho
}^{c\lambda _\rho }(\vec k_\rho )A_{4\theta _\sigma }^{e\lambda _\sigma }(%
\vec k_\sigma )-A_{i\theta _\rho }^{c\lambda _\rho }(\vec k_\rho )A_{i\theta
_\sigma }^{e\lambda _\sigma }(\vec k_\sigma )]
\end{array}
\eqnum{82}
\end{equation}
and 
\begin{equation}
\begin{array}{c}
D(\alpha \beta \gamma )=ig(2\pi )^3\delta ^3(\theta _\alpha \vec q_\alpha
-\theta _\beta \vec q_\beta -\theta _\gamma \vec k_\gamma )f^{abc}G_{\theta
_\alpha }^a(\vec q_\alpha ) \\ 
\times [\Pi _{\theta \beta }^b(\vec q_\beta )A_{4\theta _\gamma }^{c\lambda
_\gamma }(\vec k_\gamma )-\theta _\alpha k_i^\alpha G_{\theta _\beta }^b(%
\vec q_\beta )A_{i\theta _\gamma }^{c\lambda _\gamma }(\vec k_\gamma )].
\end{array}
\eqnum{83}
\end{equation}
It is emphasized that the expressions in Eqs. (78) and (79) are just the
Hamiltonian of QCD appearing in the path-integral as shown in Eq. (42) where
all the creation and annihilation operators in the Hamiltonian (which are
written in a normal product) are replaced by their eigenvalues.

To write the path-integral of thermal QCD, we need also an expression of
action $S\ $given in the coherent state representation. This action can be
obtained by using the Lagrangian density shown in Eq. (54). By partial
integration and considering the following boundary conditions of the fields
[20-22]: 
\begin{equation}
\begin{array}{c}
\psi (\vec x,0)=\psi (\vec x),\text{ }\overline{\psi }(\vec x,0)=\overline{%
\psi }(\vec x), \\ 
\psi (\vec x,\beta )=-\psi (\vec x),\text{ }\overline{\psi }(\vec x,\beta )=-%
\overline{\psi }(\vec x),
\end{array}
\eqnum{84}
\end{equation}
\begin{equation}
\begin{array}{c}
A_\mu ^c(\vec x,0)=A_\mu ^c(\vec x,\beta )=A_\mu ^c(\vec x), \\ 
\Pi _\mu ^c(\vec x,0)=\Pi _\mu ^c(\vec x,\beta )=\Pi _\mu ^c(\vec x)
\end{array}
\eqnum{85}
\end{equation}
and 
\begin{equation}
\begin{array}{c}
\overline{C}^a(\vec x,0)=\overline{C}^a(\vec x,\beta )=\overline{C}^a(\vec x%
),\text{ }C^a(\vec x,0)=C^a(\vec x,\beta )=C^a(\vec x), \\ 
\overline{\Pi }^a(\vec x,0)=\overline{\Pi }^a(\vec x,\beta )=\overline{\Pi }%
^a(\vec x),\text{ }\Pi ^a(\vec x,0)=\Pi ^a(\vec x,\beta )=\Pi ^a(\vec x),
\end{array}
\eqnum{86}
\end{equation}
the action given by the Lagrangian density in Eq. (54) can be represented in
the form 
\begin{equation}
\begin{array}{c}
S=\int_0^\beta d\tau \int d^3x\{\frac 12[\psi ^{+}(\vec x,\tau )\dot \psi (%
\vec x,\tau )-\dot \psi ^{+}(\vec x,\tau )\psi (\vec x,\tau )] \\ 
+\frac i2[\Pi _\mu ^c(\vec x,\tau )\dot A_\mu ^c(\vec x,\tau )-\dot \Pi _\mu
^c(\vec x,\tau )A_\mu ^c(\vec x,\tau )] \\ 
+\frac i2[\Pi _a(\vec x,\tau )\dot C_a(\vec x,\tau )-\dot \Pi _a(\vec x,\tau
)C_a(\vec x,\tau ) \\ 
+\stackrel{\cdot }{\overline{C}}_a(\vec x,\tau )\overline{\Pi }_a(\vec x%
,\tau )-\overline{C}_a(\vec x,\tau )\stackrel{\cdot }{\overline{\Pi }}_a(%
\vec x,\tau )]-{\cal H}(\vec x,\tau )\}
\end{array}
\eqnum{87}
\end{equation}
where the first relation in Eq. (53) has been used and the symbol $"\cdot "$
in $\dot \psi (\vec x,\tau ),\dot A_\mu ^c(\vec x,\tau )\cdot \cdot \cdot
\cdot \cdot \cdot $ now denotes the derivatives of the fields with respect
to the imaginary time $\tau $. It is stressed here that only the above
expression is appropriate to use for deriving the coherent-state
representation of the action by making use of the Fourier expansions written
in Eqs. (58)-(65). On inserting Eqs. (58)-(65) into Eq. (87), it is not
difficult to get 
\begin{equation}
\begin{array}{c}
S=-\int_0^\beta d\tau \{\int d^3k\{\frac 12[b_s^{*}(\vec k,\tau )\dot b_s(%
\vec k,\tau )-\dot b_s^{*}(\vec k,\tau )b_s(\vec k,\tau )]+\frac 12[d_s^{*}(%
\vec k,\tau )\dot d_s(\vec k,\tau ) \\ 
-\dot d_s^{*}(\vec k,\tau )d_s(\vec k,\tau )]+\frac 12[a_\lambda ^{c*}(\vec k%
,\tau )\dot a_\lambda ^c(\vec k,\tau )-\dot a_\lambda ^{c*}(\vec k,\tau
)a_\lambda ^c(\vec k,\tau )]+\frac 12[\overline{c}_a^{*}(\vec k,\tau )%
\stackrel{\cdot }{\overline{c}_a}(\vec k,\tau ) \\ 
-\stackrel{\cdot }{\overline{c}_a^{*}}(\vec k,\tau )\overline{c}_a(\vec k%
,\tau )-c_a^{*}(\vec k,\tau )\dot c_a(\vec k,\tau )+\dot c_a^{*}(\vec k,\tau
)c_a(\vec k,\tau )]\}+H(\tau )\} \\ 
=-S_E
\end{array}
\eqnum{88}
\end{equation}
where $H(\tau )$ is given by the sum of the Hamiltonians in Eqs. (78) and
(79) and $S_E$ is the action defined in the Euclidean metric. It is noted
that if one considers a grand canonical ensemble of QCD, the Hamiltonian in
Eq. (88) should be replaced by $K(\tau )$ defined in Eq. (2). Employing the
abbreviation notation as denoted in Eqs. (66), (70) and (74) and letting $%
q_\alpha $ stand for $(a_\alpha ,b_\alpha ,c_\alpha )$, the action may be
compactly represented as 
\begin{equation}
S_E=\int_0^\beta d\tau \{\sum\limits_\alpha \frac 12[q_\alpha ^{*}(\tau
)\circ \dot q_\alpha (\tau )-\dot q_\alpha ^{*}(\tau )\circ q_\alpha (\tau
)]+H(\tau )\}  \eqnum{89}
\end{equation}
where we have defined 
\begin{equation}
q_\alpha ^{*}\circ q_\alpha =a_{\alpha ^{-}}a_{\alpha ^{+}}+b_\alpha
^{*}b_\alpha +\theta _\alpha c_\alpha ^{*}c_\alpha  \eqnum{90}
\end{equation}
It is emphasized that the $\theta _\alpha =\pm $ is now contained in the
subscript $\alpha .$ Therefore, each $\alpha $ may takes $\alpha ^{+}$
and/or $\alpha ^{-}$ as the first term in Eq. (90) does.

\section{Generating functional of Green functions for thermal QCD}

With the action $S_{E\text{ }}$given in the preceding section, the
quantization of the thermal QCD in the coherent-state representation is
easily implemented by writing out its generating functional of thermal Green
functions. According to the general formula shown in Eq. (42), the QCD
generating functional can be formulated as 
\begin{equation}
\begin{array}{c}
Z[j]=\int D(q^{*}q)e^{-q^{*}\cdot q}\int {\frak D}(q^{*}q)\exp \{\frac 12%
[q^{*}(\beta )\cdot q(\beta ) \\ 
-q^{*}(0)\cdot q(0)]-S_E+\int_0^\beta d\tau j^{*}(\tau )\cdot q(\tau )\}
\end{array}
\eqnum{91}
\end{equation}
where we have defined 
\begin{equation}
q^{*}\cdot q=\frac 12a_\alpha ^{*}a_\alpha +\theta _\alpha b_\alpha
^{*}b_\alpha +c_\alpha ^{*}c_\alpha  \eqnum{92}
\end{equation}
and 
\begin{equation}
j^{*}\cdot q=\xi _\alpha ^{*}a_\alpha +\theta _\alpha (\eta _\alpha
^{*}b_\alpha +b_\alpha ^{*}\eta _\alpha +\zeta _\alpha ^{*}c_\alpha
+c_\alpha ^{*}\zeta _\alpha )  \eqnum{93}
\end{equation}
here $\xi _\alpha ,\eta _\alpha $ and $\zeta _\alpha $ are the sources for
gluons, quarks and ghost particles respectively and the repeated index
implies summation. It is noted that the product $q^{*}\cdot q$ defined above
is different from the $q_\alpha ^{*}\circ q_\alpha $ defined in Eq. (90) in
the terms for quarks and ghost particles and the subscript $\alpha $ in Eqs.
(92) and (93) is also defined by containing $\theta _\alpha =\pm $ . In what
follows, we assign $\alpha ^{\pm }$ to represent the $\alpha $ with $\theta
_\alpha =\pm $. According to this notation, the sources in Eq. (93) are
specifically defined as follows: 
\begin{equation}
\begin{array}{c}
\xi _{\alpha ^{+}}=\xi _\alpha ,\text{ }\xi _{\alpha ^{-}}=\xi _\alpha ^{*}
\\ 
\eta _{\alpha ^{+}}=\eta _\alpha ,\text{ }\eta _{\alpha ^{-}}=\overline{\eta 
}_\alpha ^{*} \\ 
\zeta _{\alpha ^{+}}=\zeta _\alpha ,\text{ }\zeta _{\alpha ^{-}}=\overline{%
\zeta }_\alpha ^{*}
\end{array}
\eqnum{94}
\end{equation}
where the subscript $\alpha $ on the right hand side of each equality no
longer contains $\theta _\alpha $ and the gluon term in Eq. (92) $%
(1/2)a_\alpha ^{*}a_\alpha $ may be replaced by $a_{\alpha ^{-}}a_{\alpha
^{+}}$. The integration measures $D(q^{*}q)$ and ${\frak D}(q^{*}q)$ are
defined as in Eqs. (24) and (25).

The generating functional in Eq. (91) is nonperturbative. Now we are
interested in describing the perturbation method of calculating the QCD
generating functional. Since the Hamiltonian can be split into two parts $%
H_0(\tau )$ and $H_I(\tau )$ as shown in Eqs. (78) and (79), the generating
functional in Eq. (91) may be perturbatively represented in the form 
\begin{equation}
Z[j]=\exp \{-\int_0^\beta d\tau H_I(\frac \delta {\delta j(\tau )})\}Z^0[j] 
\eqnum{95}
\end{equation}
where $Z^0[j]$ is the generating functional for the free system and the
exponential may be expanded in a Taylor series. In the above, the
commutativity between $H_I$ and $Z^0[j]$ has been considered. Obviously, the 
$Z^0[j]$ can be written as 
\begin{equation}
Z^0[j]=Z_g^0[\xi ]Z_q^0[\eta ]Z_c^0[\zeta ]  \eqnum{96}
\end{equation}
where $Z_g^0[\xi ]$, $Z_q^0[\eta ]$ and $Z_c^0[\zeta ]$ are the generating
functionals contributed from the free Hamiltonians of gluons, quarks and
ghost particles respectively. They are separately and specifically described
below.

In view of the expressions in Eqs. (91), (88) and (78), the generating
functional $Z_g^0[\xi ]$ is of the form 
\begin{equation}
\begin{array}{c}
Z_g^0[\xi ]=\int D(a^{*}a)\exp \{-\int d^3ka_\lambda ^{*}(\vec k)a_\lambda (%
\vec k)\} \\ 
\times \int {\frak D}(a^{*}a)\exp \{I_g(a_\lambda ^{*},a_\lambda ;\xi
_\lambda ^{*},\xi _\lambda )\}
\end{array}
\eqnum{97}
\end{equation}
where 
\begin{equation}
\begin{array}{c}
I_g(a_\lambda ^{*},a_\lambda ;\xi _\lambda ^{*},\xi _\lambda )=\int d^3k%
\frac 12[a_\lambda ^{*}(\vec k,\beta )a_\lambda (\vec k,\beta )+a_\lambda
^{*}(\vec k,0)a_\lambda (\vec k,0)] \\ 
-\int_0^\beta d\tau \int d^3k\{\frac 12[a_\lambda ^{*}(\vec k,\tau )\dot a%
_\lambda (\vec k,\tau )-\dot a_\lambda ^{*}(\vec k,\tau )a_\lambda (\vec k%
,\tau )]+\omega (\vec k)a_\lambda ^{*}(\vec k,\tau )a_\lambda (\vec k,\tau )
\\ 
-\xi _\lambda ^{*}(\vec k,\tau )a_\lambda (\vec k,\tau )-a_\lambda ^{*}(\vec 
k,\tau )\xi _\lambda (\vec k,\tau )\}
\end{array}
\eqnum{98}
\end{equation}
and 
\begin{equation}
\begin{array}{c}
D(a^{*}a)=\prod\limits_{\vec k\lambda }\frac 1\pi da_\lambda ^{*}(\vec k%
)da_\lambda (\vec k), \\ 
{\frak D}(a^{*}a)=\prod\limits_{\vec k\lambda \tau }\frac 1\pi da_\lambda
^{*}(\vec k,\tau )da_\lambda (\vec k,\tau ).
\end{array}
\eqnum{99}
\end{equation}
The subscript $\lambda $ in the above is now assigned to denote polarization
and color. When we perform a partial integration, Eq. (98) becomes 
\begin{equation}
\begin{array}{c}
I_g(a_\lambda ^{*},a_\lambda ;\xi _\lambda ^{*},\xi _\lambda )=\int
d^3ka_\lambda ^{*}(\vec k,\beta )a_\lambda (\vec k,\beta )-\int_0^\beta
d\tau \int d^3k\{a_\lambda ^{*}(\vec k,\tau )\dot a_\lambda (\vec k,\tau )
\\ 
+\omega (\vec k)a_\lambda ^{*}(\vec k,\tau )a_\lambda (\vec k,\tau )-\xi
_\lambda ^{*}(\vec k,\tau )a_\lambda (\vec k,\tau )-a_\lambda ^{*}(\vec k%
,\tau )\xi _\lambda (\vec k,\tau )\}.
\end{array}
\eqnum{100}
\end{equation}

For the generating functional $Z_q^0[\eta ]$, we can write 
\begin{equation}
\begin{array}{c}
Z_q^0[\eta ]=\int D(b^{*}bd^{*}d)\exp \{-\int d^3k[b_s^{*}(\vec k)b_s(\vec k%
)+d_s^{*}(\vec k)d_s(\vec k)]\} \\ 
\times \int {\frak D}(b^{*}bd^{*}d)\exp \{I_q(b_s^{*},b_s,d_s^{*},d_s;\eta
_s^{*},\eta _s,\overline{\eta }_s^{*},\overline{\eta }_s)\}
\end{array}
\eqnum{101}
\end{equation}
where 
\begin{equation}
\begin{array}{c}
I_q(b_s^{*},b_s,d_s^{*},d_s;\eta _s^{*},\eta _s,\overline{\eta }_s^{*},%
\overline{\eta }_s)=\int d^3k\frac 12[b_s^{*}(\vec k,\beta )b_s(\vec k,\beta
)+d_s^{*}(\vec k,\beta )d_s(\vec k,\beta ) \\ 
+b_s^{*}(\vec k,0)b_s(\vec k,0)+d_s^{*}(\vec k,0)d_s(\vec k,0)]-\int_0^\beta
d\tau \int d^3k\{\frac 12[b_s^{*}(\vec k,\tau )\dot b_s(\vec k,\tau ) \\ 
-\dot b_s^{*}(\vec k,\tau )b_s(\vec k,\tau )]+\frac 12[d_s^{*}(\vec k,\tau )%
\dot d_s(\vec k,\tau )-\dot d_s^{*}(\vec k,\tau )d_s(\vec k,\tau )] \\ 
+\varepsilon (\vec k)[b_s^{*}(\vec k,\tau )b_s(\vec k,\tau )+d_s^{*}(\vec k%
,\tau )d_s(\vec k,\tau )]-[\eta _s^{*}(\vec k,\tau )b_s(\vec k,\tau ) \\ 
+b_s^{*}(\vec k,\tau )\eta _s(\vec k,\tau )+\overline{\eta }_s^{*}(\vec k%
,\tau )d_s(\vec k,\tau )+d_s^{*}(\vec k,\tau )\overline{\eta }_s\vec k,\tau
)]\}
\end{array}
\eqnum{102}
\end{equation}
and 
\begin{equation}
\begin{array}{c}
D(b^{*}bd^{*}d)=\prod\limits_{\vec ks}db_s^{*}(\vec k)db_s(\vec k)dd_s^{*}(%
\vec k)dd_s(\vec k), \\ 
{\frak D}(b^{*}bd^{*}d)=\prod\limits_{\vec ks\tau }db_s^{*}(\vec k,\tau
)db_s(\vec k,\tau )dd_s^{*}(\vec k,\tau )dd_s(\vec k,\tau )
\end{array}
\eqnum{103}
\end{equation}
in which the subscript $s$ stands for spin, color and flavor. By a partial
integration over $\tau $, Eq. (102) may be given a simpler expression

\begin{equation}
\begin{array}{c}
I_q(b_s^{*},b_s,d_s^{*},d_s;\eta _s^{*},\eta _s,\overline{\eta }_s^{*},%
\overline{\eta }_s)=\int d^3k[b_s^{*}(\vec k,\beta )b_s(\vec k,\beta
)+d_s^{*}(\vec k,\beta )d_s(\vec k,\beta )] \\ 
-\int_0^\beta d\tau \int d^3k\{b_s^{*}(\vec k,\tau )\dot b_s(\vec k,\tau
)+d_s^{*}(\vec k,\tau )\dot d_s(\vec k,\tau ) \\ 
+\varepsilon (\vec k)[b_s^{*}(\vec k,\tau )b_s(\vec k,\tau )+d_s^{*}(\vec k%
,\tau )d_s(\vec k,\tau )]-[\eta _s^{*}(\vec k,\tau )b_s(\vec k,\tau ) \\ 
+b_s^{*}(\vec k,\tau )\eta _s(\vec k,\tau )+\overline{\eta }_s^{*}(\vec k%
,\tau )d_s(\vec k,\tau )+d_s^{*}(\vec k,\tau )\overline{\eta }_s\vec k,\tau
)]\}.
\end{array}
\eqnum{104}
\end{equation}

As for the generating functional $Z_c^0[\zeta ]$, we have 
\begin{equation}
\begin{array}{c}
Z_c^0[\zeta ]=\int D(\overline{c}^{*}\overline{c}cc^{*})\exp \{-\int d^3k[%
\overline{c}_a^{*}(\vec k)\overline{c}_a(\vec k)-c_a^{*}(\vec k)c_a(\vec k%
)]\} \\ 
\times \int {\frak D}(\overline{c}^{*}\overline{c}cc^{*})\exp
\{I_c(c_a^{*},c_a,\overline{c}_a^{*},\overline{c}_a;\zeta _a^{*},\zeta _a,%
\overline{\zeta }_a^{*},\overline{\zeta }_a)\}
\end{array}
\eqnum{105}
\end{equation}
where 
\begin{equation}
\begin{array}{c}
I_c(c_a^{*},c_a,\overline{c}_a^{*},\overline{c}_a;\zeta _a^{*},\zeta _a,%
\overline{\zeta }_a^{*},\overline{\zeta }_a)=\int d^3k\frac 12[\overline{c}%
_a^{*}(\vec k,\beta )\overline{c}_a(\vec k,\beta )-c_a^{*}(\vec k,\beta )c_a(%
\vec k,\beta ) \\ 
+\overline{c}_a^{*}(\vec k,0)\overline{c}_a(\vec k,0)-c_a^{*}(\vec k,0)c_a(%
\vec k,0)]-\int_0^\beta d\tau \int d^3k\{\frac 12[\overline{c}_a^{*}(\vec k%
,\tau )\stackrel{\cdot }{\overline{c}_a}(\vec k,\tau ) \\ 
-\stackrel{\cdot }{\overline{c}_a^{*}}(\vec k,\tau )\overline{c}_a(\vec k%
,\tau )]-\frac 12[c_a^{*}(\vec k,\tau )\dot c_a(\vec k,\tau )-\dot c_a^{*}(%
\vec k,\tau )c_a(\vec k,\tau )] \\ 
+\omega (\vec k)[\overline{c}_a^{*}(\vec k,\tau )\overline{c}_a(\vec k,\tau
)-c_a^{*}(\vec k,\tau )c_a(\vec k,\tau )]-[\zeta _a^{*}(\vec k,\tau )c_a(%
\vec k,\tau ) \\ 
+c_a^{*}(\vec k,\tau )\zeta _a(\vec k,\tau )+\overline{\zeta }_a^{*}(\vec k%
,\tau )\overline{c}_a(\vec k,\tau )+\overline{c}_a^{*}(\vec k,\tau )%
\overline{\zeta }_a(\vec k,\tau )]\}
\end{array}
\eqnum{106}
\end{equation}
and 
\begin{equation}
\begin{array}{c}
D(\overline{c}^{*}\overline{c}cc^{*})=\prod\limits_{\vec ka}d\overline{c}%
_a^{*}(\vec k)d\overline{c}_a(\vec k)dc_a(\vec k)dc_a^{*}(\vec k), \\ 
{\frak D}(\overline{c}^{*}\overline{c}cc^{*})=\prod\limits_{\vec ka\tau }d%
\overline{c}_a^{*}(\vec k,\tau )d\overline{c}_a(\vec k,\tau )dc_a(\vec k%
,\tau )dc_a^{*}(\vec k,\tau )
\end{array}
\eqnum{107}
\end{equation}
in which the subscript $a$ is a color index. After a partial integration,
Eq. (106) is reduced to 
\begin{equation}
\begin{array}{c}
I_c(c_a^{*},c_a,\overline{c}_a^{*},\overline{c}_a;\zeta _a^{*},\zeta _a,%
\overline{\zeta }_a^{*},\overline{\zeta }_a)=\int d^3k[\overline{c}_a^{*}(%
\vec k,\beta )\overline{c}_a(\vec k,\beta )-c_a^{*}(\vec k,\beta )c_a(\vec k%
,\beta )] \\ 
-\int_0^\beta d\tau \int d^3k\{\overline{c}_a^{*}(\vec k,\tau )\stackrel{%
\cdot }{\overline{c}_a}(\vec k,\tau )-c_a^{*}(\vec k,\tau )\dot c_a(\vec k%
,\tau )+\omega (\vec k)[\overline{c}_a^{*}(\vec k,\tau )\overline{c}_a(\vec k%
,\tau ) \\ 
-c_a^{*}(\vec k,\tau )c_a(\vec k,\tau )]-[\zeta _a^{*}(\vec k,\tau )c_a(\vec 
k,\tau )+c_a^{*}(\vec k,\tau )\zeta _a(\vec k,\tau )+\overline{\zeta }_a^{*}(%
\vec k,\tau )\overline{c}_a(\vec k,\tau ) \\ 
+\overline{c}_a^{*}(\vec k,\tau )\overline{\zeta }_a(\vec k,\tau )]\}.
\end{array}
\eqnum{108}
\end{equation}
Here it is noted that all the terms related to the quantities $c_a^{*}$ and $%
c_a$ are opposite in sign to the terms related to the $\overline{c}_a^{*}$
and $\overline{c}_a$ and, correspondingly, the definitions of the
integration measures for these quantities, as shown in Eq. (107), are
different from each other in the order of the differentials.

The generating functionals in Eqs. (97), (101) and (105) are all of
Gaussian-type, therefore, they can exactly be calculated by the
stationary-phase method. First, we calculate the functional integral $%
Z_g^0[\xi ]$. According to the stationary-phase method, the functional $%
Z_g^0[\xi ]$ can be represented in the form 
\begin{equation}
Z_g^0[\xi ]=\int D(a^{*}a)\exp \{-\int d^3k[a_\lambda ^{*}(\vec k)a_\lambda (%
\vec k)+I_g^0(a_\lambda ^{*},a_\lambda ;\xi _\lambda ^{*},\xi _\lambda )]\} 
\eqnum{109}
\end{equation}
where $I_g^0(a_\lambda ^{*},a_\lambda ;\xi _\lambda ^{*},\xi _\lambda )$ is
given by the stationary condition $\delta I_g(a_\lambda ^{*},a_\lambda ;\xi
_\lambda ^{*},\xi _\lambda )=0.$ By this condition and the boundary
condition [20-22]: 
\begin{equation}
a_\lambda ^{*}(\vec k,\beta )=a_\lambda ^{*}(\vec k)\text{, }a_\lambda (\vec 
k,0)=a_\lambda (\vec k),  \eqnum{110}
\end{equation}
one may derive from Eq. (98) or (100) the following inhomogeneous equations
of motion [23-26]: 
\begin{equation}
\begin{array}{c}
\dot a_\lambda (\vec k,\tau )+\omega (\vec k)a_\lambda (\vec k,\tau )=\xi
_\lambda (\vec k,\tau ), \\ 
\dot a_\lambda ^{*}(\vec k,\tau )-\omega (\vec k)a_\lambda ^{*}(\vec k,\tau
)=-\xi _\lambda ^{*}(\vec k,\tau ).
\end{array}
\eqnum{111}
\end{equation}
In accordance with the general method of solving such a kind of equations,
one may first solve the homogeneous linear equations as written in Eq. (33).
Based on the solutions shown in Eq. (34) and the boundary condition denoted
in Eq. (110), one may assume [23-26] 
\begin{equation}
\begin{array}{c}
a_\lambda (\vec k,\tau )=[a_\lambda (\vec k)+u_\lambda (\vec k,\tau
)]e^{-\omega (\vec k)\tau }, \\ 
a_\lambda ^{*}(\vec k,\tau )=[a_\lambda ^{*}(\vec k)+u_\lambda ^{*}(\vec k%
,\tau )]e^{\omega (\vec k)(\tau -\beta )}
\end{array}
\eqnum{112}
\end{equation}
where the unknown functions $u_\lambda (\vec k,\tau )$ and $u_\lambda ^{*}(%
\vec k,\tau )$ are required to satisfy the boundary conditions [23-26]: 
\begin{equation}
u_\lambda (\vec k,0)=u_\lambda (\vec k,\beta )=u_\lambda ^{*}(\vec k%
,0)=u_\lambda ^{*}(\vec k,\beta )=0.  \eqnum{113}
\end{equation}
Inserting Eq. (112) into Eq. (111), we find 
\begin{equation}
\begin{array}{c}
\dot u_\lambda (\tau )=\xi _\lambda (\vec k,\tau )e^{\omega (\vec k)\tau }.
\\ 
\dot u_\lambda ^{*}(\tau )=-\xi _\lambda ^{*}(\vec k,\tau )e^{\omega (\vec k%
)(\beta -\tau )}.
\end{array}
\eqnum{114}
\end{equation}
Integrating these two equations and applying the boundary conditions in Eq.
(113), one can get 
\begin{equation}
\begin{array}{c}
u_\lambda (\vec k,\tau )=\int_0^\tau d\tau ^{\prime }e^{\omega (\vec k)\tau
^{\prime }}\xi _\lambda (\vec k,\tau ^{\prime }), \\ 
u_\lambda ^{*}(\vec k,\tau )=-\int_\beta ^\tau d\tau ^{\prime }e^{\omega (%
\vec k)(\beta -\tau ^{\prime })}\xi _\lambda ^{*}(\vec k,\tau ^{\prime }).
\end{array}
\eqnum{115}
\end{equation}
Substitution of these solutions in Eq. (112) yields [23-26] 
\begin{equation}
\begin{array}{c}
a_\lambda (\vec k,\tau )=a_\lambda (\vec k)e^{-\omega (\vec k)\tau
}+\int_0^\tau d\tau ^{\prime }e^{-\omega (\vec k)(\tau -\tau ^{\prime })}\xi
_\lambda (\vec k,\tau ^{\prime }), \\ 
a_\lambda ^{*}(\vec k,\tau )=a_\lambda ^{*}(\vec k)e^{\omega (\vec k)(\tau
-\beta )}+\int_\tau ^\beta d\tau ^{\prime }e^{\omega (\vec k)(\tau -\tau
^{\prime })}\xi _\lambda ^{*}(\vec k,\tau ^{\prime }).
\end{array}
\eqnum{116}
\end{equation}
When Eq. (116) is inserted into Eq. (98) or Eq. (100), one may obtain the $%
I_g^0(a_\lambda ^{*},a_\lambda ;\xi _\lambda ^{*},\xi _\lambda )$ which
leads to another expression of Eq. (109) like this 
\begin{equation}
\begin{array}{c}
Z_g^0[\xi ]=\int D(a^{*}a)\exp \{-\int d^3k[a_\lambda ^{*}(\vec k)a_\lambda (%
\vec k)(1-e^{-\beta \omega (\vec k)})-a_\lambda ^{*}(\vec k)e^{-\beta \omega
(\vec k)} \\ 
\times \int_0^\beta d\tau e^{\omega (\vec k)\tau }\xi _\lambda (\vec k,\tau
)-\int_0^\beta d\tau e^{-\omega (\vec k)\tau }\xi _\lambda ^{*}(\vec k,\tau
)a_\lambda (\vec k)] \\ 
+\int_0^\beta d\tau _1\int_0^\beta d\tau _2\int d^3k\xi _\lambda ^{*}(\vec k%
,\tau _1)\theta (\tau _1-\tau _2)e^{-\omega (\vec k)(\tau _1-\tau _2)}\xi
_\lambda (\vec k,\tau _2)\}.
\end{array}
\eqnum{117}
\end{equation}
When we set 
\begin{equation}
\begin{array}{c}
\lambda =1-e^{-\beta \omega (\vec k)}, \\ 
b=e^{-\beta \omega (\vec k)}\int_0^\beta d\tau e^{\omega (\vec k)\tau }\xi
_\lambda (\vec k,\tau ), \\ 
f(a)=\int_0^\beta d\tau e^{-\omega (\vec k)\tau }\xi _\lambda ^{*}(\vec k%
,\tau )a_\lambda (\vec k),
\end{array}
\eqnum{118}
\end{equation}
by employing the formula denoted in Eq. (37), the integral over $a_\lambda
^{*}(\vec k)$ and $a_\lambda (\vec k)$ in Eq. (117) can easily be
calculated. The result is 
\begin{equation}
Z_g^0[\xi ]=Z_g^0\exp \{-\int_0^\beta d\tau _1\int_0^\beta d\tau _2\int
d^3k\xi _a^{\lambda *}(\vec k,\tau _1)\Delta _{\lambda \lambda ^{\prime
}}^{aa^{\prime }}(\vec k,\tau _1-\tau _2)\xi _{a^{\prime }}^{\lambda
^{\prime }}(\vec k,\tau _2)\}  \eqnum{119}
\end{equation}
where 
\begin{equation}
Z_g^0=\prod\limits_{\overrightarrow{k}\lambda a}[1-e^{-\beta \omega (\vec k%
)}]^{-1}=\prod\limits_{\overrightarrow{k}a}[1-e^{-\beta \omega (\vec k%
)}]^{-4}  \eqnum{120}
\end{equation}
is precisely the partition function contributed from the free gluons [21-24,
29] and 
\begin{equation}
\Delta _{\lambda \lambda ^{\prime }}^{aa^{\prime }}(\vec k,\tau _1-\tau
_2)=g_{\lambda \lambda ^{\prime }}\delta ^{aa^{\prime }}\Delta _g(\vec k%
,\tau _1-\tau _2)  \eqnum{121}
\end{equation}
with

\begin{equation}
\Delta _g(\vec k,\tau _1-\tau _2)=\theta (\tau _1-\tau _2)-(1-e^{\beta
\omega (\vec k)})^{-1}e^{-\omega (\vec k)(\tau _1-\tau _2)}  \eqnum{122}
\end{equation}
is the free gluon propagator given in the Feynman gauge and in the Minkowski
metric \{Note: in Euclidean metric, $g_{\lambda \lambda ^{\prime
}}\rightarrow -\delta _{\lambda \lambda ^{\prime }}$). In Eq. (119), the
color index $"a"$ has been explicitly written out and the $\lambda $ now
merely designates the polarization index. In the other expressions, we still
use $\lambda $ to mark the both of color and polarization indices for
simplicity. When we interchange the integration variables $\tau _1$ and $%
\tau _2$ and make a transformation $\vec k\rightarrow -\vec k$ in Eq. (119),
by considering the relation 
\begin{equation}
\xi _\lambda ^{*}(\vec k,\tau )=\xi _\lambda (-\vec k,\tau )  \eqnum{123}
\end{equation}
which will be interpreted in the Appendix, one may find that the propagator
in Eq. (122) can be represented in the form 
\begin{equation}
\Delta _g(\vec k,\tau _1-\tau _2)=\frac 12[\overline{n}_b(\vec k)e^{-\omega (%
\vec k)\left| \tau _1-\tau _2\right| }-n_b(\vec k)e^{\omega (\vec k)\left|
\tau _1-\tau _2\right| }]  \eqnum{124}
\end{equation}
where 
\begin{equation}
\overline{n}_b(\vec k)=(1-e^{-\beta \varepsilon (\vec k)})^{-1},\text{ }n_b(%
\vec k)=(1-e^{\beta \varepsilon (\vec k)})^{-1}  \eqnum{125}
\end{equation}
are just the boson distribution functions [20-23, 29].

Let us turn to the calculation of the functional integral in Eq. (101).
Based on the stationary-phase method, we can write

\begin{equation}
\begin{array}{c}
Z_q^0[\eta ]=\int D(b^{*}bd^{*}d)\exp \{-\int d^3k[b_s^{*}(\vec k)b_s(\vec k%
)+d_s^{*}(\vec k)d_s(\vec k)]\} \\ 
\times \exp \{I_q^0(b_s^{*},b_s,d_s^{*},d_s;\eta _s^{*},\eta _s,\overline{%
\eta }_s^{*},\overline{\eta }_s)\}
\end{array}
\eqnum{126}
\end{equation}
where $I_q^0(b_s^{*},b_s,d_s^{*},d_s;\eta _s^{*},\eta _s,\overline{\eta }%
_s^{*},\overline{\eta }_s)$ will be obtained from Eq. (102) or Eq. (104) by
the stationary condition $\delta I_q(b_s^{*},b_s,d_s^{*},d_s;\eta
_s^{*},\eta _s,\overline{\eta }_s^{*},\overline{\eta }_s)=0$. From this
condition and the boundary conditions [21-26]:

\begin{equation}
\begin{array}{c}
b_s^{*}(\vec k,\beta )=-b_s^{*}(\vec k),\text{ }b_s(\vec k,0)=b_s(\vec k),
\\ 
d_s^{*}(\vec k,\beta )=-d_s^{*}(\vec k),\text{ }d_s(\vec k,0)=d_s(\vec k),
\end{array}
\eqnum{127}
\end{equation}
one may deduce from Eq. (102) or Eq. (104) the following equations [23-26]: 
\begin{equation}
\begin{array}{c}
\dot b_s(\vec k,\tau )+\varepsilon (\vec k)b_s(\vec k,\tau )=\eta _s(\vec k%
,\tau ), \\ 
\dot b_s^{*}(\vec k,\tau )-\varepsilon (\vec k)b_s^{*}(\vec k,\tau )=-\eta
_s^{*}(\vec k,\tau ), \\ 
\dot d_s(\vec k,\tau )+\varepsilon (\vec k)d_s(\vec k,\tau )=\overline{\eta }%
_s(\vec k,\tau ), \\ 
\dot d_s^{*}(\vec k,\tau )-\varepsilon (\vec k)d_s^{*}(\vec k,\tau )=-%
\overline{\eta }_s^{*}(\vec k,\tau ).
\end{array}
\eqnum{128}
\end{equation}
Following the procedure described in Eqs. (111)-(116), the solutions to the
above equations, which satisfies the boundary conditions in Eqs. (127) and
the conditions like those in Eq. (113), can be found to be [23-26] 
\begin{equation}
\begin{array}{c}
b_s(\vec k,\tau )=b_s(\vec k)e^{-\varepsilon (\vec k)\tau }+\int_0^\tau
d\tau ^{\prime }e^{-\varepsilon (\vec k)(\tau -\tau ^{\prime })}\eta _s(\vec 
k,\tau ^{\prime }), \\ 
b_s^{*}(\vec k,\tau )=-b_s^{*}(\vec k)e^{\varepsilon (\vec k)(\tau -\beta
)}+\int_\tau ^\beta d\tau ^{\prime }e^{\varepsilon (\vec k)(\tau -\tau
^{\prime })}\eta _s^{*}(\vec k,\tau ^{\prime }), \\ 
d_s(\vec k,\tau )=d_s(\vec k)e^{-\varepsilon (\vec k)\tau }+\int_0^\tau
d\tau ^{\prime }e^{-\varepsilon (\vec k)(\tau -\tau ^{\prime })}\overline{%
\eta }_s(\vec k,\tau ^{\prime }), \\ 
d_s^{*}(\vec k,\tau )=-d_s^{*}(\vec k)e^{\varepsilon (\vec k)(\tau -\beta
)}+\int_\tau ^\beta d\tau ^{\prime }e^{\varepsilon (\vec k)(\tau -\tau
^{\prime })}\overline{\eta }_s^{*}(\vec k,\tau ^{\prime }).
\end{array}
\eqnum{129}
\end{equation}
Substituting the above solutions into Eq. (102) or (104), we find 
\begin{equation}
\begin{array}{c}
I_q^0(b_s^{*},b_s,d_s^{*},d_s;\eta _s^{*},\eta _s,\overline{\eta }_s^{*},%
\overline{\eta }_s)=\int d^3k\{-e^{-\beta \varepsilon (\vec k)}[b_s^{*}(\vec 
k)b_s(\vec k)+d_s^{*}(\vec k)d_s(\vec k)] \\ 
+\int_0^\beta d\tau e^{-\varepsilon (\vec k)\tau }[\eta _s^{*}(\vec k,\tau
)b_s(\vec k)+\overline{\eta }_s^{*}(\vec k,\tau )d_s(\vec k)]-e^{-\beta
\varepsilon (\vec k)}\int_0^\beta d\tau e^{\varepsilon (\vec k)\tau } \\ 
\times [b_s^{*}(\vec k)\eta _s(\vec k,\tau )+d_s^{*}(\vec k)\overline{\eta }%
_s(\vec k,\tau )]\}+B[\eta _s^{*},\eta _s,\overline{\eta }_s^{*},\overline{%
\eta }_s]
\end{array}
\eqnum{130}
\end{equation}
where 
\begin{equation}
\begin{array}{c}
B[\eta _s^{*},\eta _s,\overline{\eta }_s^{*},\overline{\eta }_s]=\frac 12%
\int_0^\beta d\tau _1\int_0^\beta d\tau _2\int d^3k\{\theta (\tau _1-\tau
_2)e^{-\varepsilon (\vec k)(\tau _1-\tau _2)} \\ 
\times [\eta _s^{*}(\vec k,\tau _1)\eta _s(\vec k,\tau _2)+\overline{\eta }%
_s^{*}(\vec k,\tau _1)\overline{\eta }_s(\vec k,\tau _2)]+\theta (\tau
_2-\tau _1)e^{-\varepsilon (\vec k)(\tau _2-\tau _1)} \\ 
\times [\eta _s^{*}(\vec k,\tau _2)\eta _s(\vec k,\tau _1)+\overline{\eta }%
_s^{*}(\vec k,\tau _2)\overline{\eta }_s(\vec k,\tau _1)]\}
\end{array}
\eqnum{131}
\end{equation}
On inserting Eq. (130) into Eq. (126), we have 
\begin{equation}
Z_q^0[\eta ]=A[\eta _s^{*},\eta _s,\overline{\eta }_s^{*},\overline{\eta }%
_s]e^{B[\eta _s^{*},\eta _s,\overline{\eta }_s^{*},\overline{\eta }_s]} 
\eqnum{132}
\end{equation}
where 
\begin{equation}
\begin{array}{c}
A[\eta _s^{*},\eta _s,\overline{\eta }_s^{*},\overline{\eta }_s]=\int
D(b^{*}b)\exp \{-\int d^3k[b_s^{*}(\vec k)b_s(\vec k)(1+e^{-\beta
\varepsilon (\vec k)}) \\ 
+e^{-\beta \varepsilon (\overrightarrow{k})}b_s^{*}(\vec k)\int_0^\beta
d\tau e^{\varepsilon (\vec k)\tau }\eta _s(\vec k,\tau )-\int_0^\beta d\tau
e^{-\varepsilon (\vec k)\tau }\eta _s^{*}(\vec k,\tau )b_s(\vec k)]\} \\ 
\times \int D(d^{*}d)\exp \{-\int d^3k[d_s^{*}(\vec k)d_s(\vec k%
)(1+e^{-\beta \varepsilon (\vec k)})+e^{-\beta \varepsilon (\vec k)}d_s^{*}(%
\vec k) \\ 
\times \int_0^\beta d\tau e^{\varepsilon (\vec k)\tau }\overline{\eta }_s(%
\vec k,\tau )-\int_0^\beta d\tau e^{-\varepsilon (\vec k)\tau }\overline{%
\eta }_s^{*}(\vec k,\tau )d_s(\vec k)]\}
\end{array}
\eqnum{133}
\end{equation}
here the fact that the two integrals over \{$b^{*},b\}$ and \{$d^{*},d\}$
commute with each other has been noted. Obviously, each of the above
integrals can easily be calculated by applying the integration formulas
shown in Eq. (39). The result is 
\begin{equation}
\begin{array}{c}
A[\eta _s^{*},\eta _s,\overline{\eta }_s^{*},\overline{\eta }_s]=Z_q^0\exp
\{-\frac 12\int_0^\beta d\tau _1\int_0^\beta d\tau _2\int d^3k(1+e^{\beta
\varepsilon (\vec k)})^{-1} \\ 
\times \{e^{-\varepsilon (\vec k)(\tau _1-\tau _2)}[\eta _s^{*}(\vec k,\tau
_1)\eta _s(\vec k,\tau _2)+\overline{\eta }_s^{*}(\vec k,\tau _1)\overline{%
\eta }_s(\vec k,\tau _2)] \\ 
+e^{-\varepsilon (\vec k)(\tau _2-\tau _1)}[\eta _s^{*}(\vec k,\tau _2)\eta
_s(\vec k,\tau _1)+\overline{\eta }_s^{*}(\vec k,\tau _2)\overline{\eta }_s(%
\vec k,\tau _1)]\}\}
\end{array}
\eqnum{134}
\end{equation}
where

\begin{equation}
Z_q^0=\prod\limits_{\overrightarrow{k}s}[1+e^{-\beta \varepsilon (\vec k)}]^2
\eqnum{135}
\end{equation}
which just is the partition function contributed from free quarks and
antiquarks [21-23,29]. It is noted that the two terms in the exponent of Eq.
(134) are equal to one another as seen from the interchange of the
integration variables $\tau _1$ and $\tau _2$. After Eqs. (131) and 134) are
substituted in Eq. (132), we get 
\begin{equation}
\begin{array}{c}
Z_q^0[\eta ]=Z_q^0\exp \{\int_0^\beta d\tau _1\int_0^\beta d\tau _2\int
d^3k\{[\theta (\tau _1-\tau _2)-(1+e^{\beta \varepsilon (\vec k%
)})^{-1}]e^{-\varepsilon (\vec k)(\tau _1-\tau _2)} \\ 
\times [\eta _s^{*}(\vec k,\tau _1)\eta _s(\vec k,\tau _2)+\overline{\eta }%
_s^{*}(\vec k,\tau _1)\overline{\eta }_s(\vec k,\tau _2)]+[\theta (\tau
_2-\tau _1)-(1+e^{\beta \varepsilon (\vec k)})^{-1}] \\ 
\times e^{-\varepsilon (\vec k)(\tau _2-\tau _1)}[\eta _s^{*}(\vec k,\tau
_2)\eta _s(\vec k,\tau _1)+\overline{\eta }_s^{*}(\vec k,\tau _2)\overline{%
\eta }_s(\vec k,\tau _1)]\}\}.
\end{array}
\eqnum{136}
\end{equation}
When we interchange the variables $\tau _1$ and $\tau _2$ and set $\vec k%
\rightarrow -\vec k$ in the second term of the above integrals and notice
the relation 
\begin{equation}
\overline{\eta }_s^{*}(\vec k,\tau _2)\overline{\eta }_s(\vec k,\tau
_1)=\eta _s^{*}(-\vec k,\tau _1)\eta _s(-\vec k,\tau _2)  \eqnum{137}
\end{equation}
which will be proved in the Appendix , the functional $Z_q^0[\eta ]$ will
eventually be represented as 
\begin{equation}
\begin{array}{c}
Z_q^0[\eta ]=Z_q^0\exp \{\int_0^\beta d\tau _1\int_0^\beta d\tau _2\int
d^3k[\eta _s^{*}(\vec k,\tau _1)\Delta _q^{ss^{\prime }}(\vec k,\tau _1-\tau
_2)\eta _{s^{\prime }}(\vec k,\tau _2) \\ 
+\overline{\eta }_s^{*}(\vec k,\tau _1)\Delta _q^{ss^{\prime }}(\vec k,\tau
_1-\tau _2)\overline{\eta }_{s^{\prime }}(\vec k,\tau _2)]\}
\end{array}
\eqnum{138}
\end{equation}
where 
\begin{equation}
\Delta _q^{ss^{\prime }}(\vec k,\tau _1-\tau _2)=\delta ^{ss^{\prime
}}\Delta _q(\vec k,\tau _1-\tau _2)  \eqnum{139}
\end{equation}
with 
\begin{equation}
\Delta _q(\vec k,\tau _1-\tau _2)=\frac 12[\overline{n}_f(\vec k%
)e^{-\varepsilon (\vec k)\left| \tau _1-\tau _2\right| }-n_f(\vec k%
)e^{\varepsilon (\vec k)\left| \tau _1-\tau _2\right| }]  \eqnum{140}
\end{equation}
is the free quark (antiquark) propagator. In the above, 
\begin{equation}
\overline{n}_f(\vec k)=(1+e^{-\beta \varepsilon (\vec k)})^{-1},\text{ }n_f(%
\vec k)=(1+e^{\beta \varepsilon (\vec k)})^{-1}  \eqnum{141}
\end{equation}
are the fermion distribution functions [21-23, 29].

Finally, let us calculate the generating functional $Z_c^0[\zeta ]$. From
the stationary condition $\delta I_c(c_a^{*},c_a,\overline{c}_a^{*},%
\overline{c}_a;\zeta _a^{*},\zeta _a,\overline{\zeta }_a^{*},\overline{\zeta 
}_a)=0$ and the boundary conditions: 
\begin{equation}
\overline{c}_a^{*}(\vec k,\beta )=\overline{c}_a^{*}(\vec k),\text{ }c_a^{*}(%
\vec k,\beta )=c_a^{*}(\vec k)\text{, }\overline{c}_a(\vec k,0)=\overline{c}%
_a(\vec k)\text{, }c_a(\vec k,0)=c_a(\vec k)\text{,}  \eqnum{142}
\end{equation}
which are the same as those for scalar fields other than for the fermion
fields [21], it is easy to derive from Eq. (106) or (108) the following
equations of motion: 
\begin{equation}
\begin{array}{c}
\dot c_a(\vec k,\tau )+\omega (\vec k)c_a(\vec k,\tau )=-\zeta _a(\vec k%
,\tau ), \\ 
\dot c_a^{*}(\vec k,\tau )-\omega (\vec k)c_a^{*}(\vec k,\tau )=\zeta _a^{*}(%
\vec k,\tau ), \\ 
\stackrel{\cdot }{\overline{c}_a}(\vec k,\tau )+\omega (\vec k)\overline{c}%
_a(\vec k,\tau )=\overline{\zeta }_a(\vec k,\tau ), \\ 
\stackrel{\cdot }{\overline{c}_a^{*}}(\vec k,\tau )-\omega (\vec k)\overline{%
c}_a^{*}(\vec k,\tau )=-\overline{\zeta }_a^{*}(\vec k,\tau ).
\end{array}
\eqnum{143}
\end{equation}
By the same procedure as stated in Eqs. (111)-(116), the solutions to the
above equations can be found to be 
\begin{equation}
\begin{array}{c}
c_a(\vec k,\tau )=c_a(\vec k)e^{-\omega (\vec k)\tau }-\int_0^\tau d\tau
^{\prime }e^{-\omega (\vec k)(\tau -\tau ^{\prime })}\zeta _a(\vec k,\tau
^{\prime }), \\ 
c_a^{*}(\vec k,\tau )=c_a^{*}(\vec k)e^{\omega (\vec k)(\tau -\beta
)}-\int_\tau ^\beta d\tau ^{\prime }e^{\omega (\vec k)(\tau -\tau ^{\prime
})}\zeta _a^{*}(\vec k,\tau ^{\prime }), \\ 
\overline{c}_a(\vec k,\tau )=\overline{c}_a(\vec k)e^{-\omega (\vec k)\tau
}+\int_0^\tau d\tau ^{\prime }e^{-\omega (\vec k)(\tau -\tau ^{\prime })}%
\overline{\zeta }_a(\vec k,\tau ^{\prime }), \\ 
\overline{c}_a^{*}(\vec k,\tau )=\overline{c}_a^{*}(\vec k)e^{\omega (\vec k%
)(\tau -\beta )}+\int_\tau ^\beta d\tau ^{\prime }e^{\omega (\vec k)(\tau
-\tau ^{\prime })}\overline{\zeta }_a^{*}(\vec k,\tau ^{\prime }).
\end{array}
\eqnum{144}
\end{equation}
Upon substituting the above solutions into Eq. (106) or Eq. (108), we find 
\begin{equation}
\begin{array}{c}
I_c^0(c_a^{*},c_a,\overline{c}_a^{*},\overline{c}_a;\zeta _a^{*},\zeta _a,%
\overline{\zeta }_a^{*},\overline{\zeta }_a)=\int d^3k\{e^{-\beta \omega (%
\vec k)}[\overline{c}_a^{*}(\vec k)\overline{c}_a(\vec k)-c_a^{*}(\vec k)c_a(%
\vec k)] \\ 
+e^{-\beta \omega (\vec k)}\int_0^\beta d\tau e^{\omega (\vec k)\tau
}[c_a^{*}(\vec k)\zeta _a(\vec k,\tau )+\overline{c}_a^{*}(\vec k)\overline{%
\zeta }_a(\vec k,\tau )] \\ 
-\int_0^\beta d\tau e^{-\omega (\vec k)\tau }[\zeta _a^{*}(\vec k,\tau )c_a(%
\vec k)+\overline{\zeta }_a^{*}(\vec k,\tau )\overline{c}_a(\vec k)]\} \\ 
-\int_0^\beta d\tau _1\int_0^\beta d\tau _2\int d^3k\theta (\tau _1-\tau
_2)e^{-\omega (\vec k)(\tau _1-\tau _2)}[\zeta _a^{*}(\vec k,\tau _1)\zeta
_a(\vec k,\tau _2) \\ 
-\overline{\zeta }_a^{*}(\vec k,\tau _1)\overline{\zeta }_a(\vec k,\tau _2)].
\end{array}
\eqnum{145}
\end{equation}
On inserting the above expression into the following integral given by the
stationary-phase method: 
\begin{equation}
Z_c^0[\zeta ]=\int D(\overline{c}^{*}\overline{c}cc^{*})\exp \{-\int d^3k[%
\overline{c}_a^{*}(\vec k)\overline{c}_a(\vec k)-c_a^{*}(\vec k)c_a(\vec k%
)]+I_c^0(c_a^{*},c_a,\overline{c}_a^{*},\overline{c}_a;\zeta _a^{*},\zeta _a,%
\overline{\zeta }_a^{*},\overline{\zeta }_a)\},  \eqnum{146}
\end{equation}
and applying the integration formulas in Eq. (39) , one can get 
\begin{equation}
\begin{array}{c}
Z_c^0[\zeta ]=Z_c^0\exp \{\int_0^\beta d\tau _1\int_0^\beta d\tau _2\int
d^3k[\theta (\tau _1-\tau _2)-(1-e^{\beta \omega (\vec k)})^{-1}]e^{-\omega (%
\vec k)(\tau _1-\tau _2)} \\ 
\times [\overline{\zeta }_a^{*}(\vec k,\tau _1)\overline{\zeta }_a(\vec k%
,\tau _2)-\zeta _a^{*}(\vec k,\tau _1)\zeta _a(\vec k,\tau _2)]\}
\end{array}
\eqnum{147}
\end{equation}
where 
\begin{equation}
Z_c^0==\prod\limits_{\overrightarrow{k}a}[1-e^{-\beta \omega (\vec k)}]^2 
\eqnum{148}
\end{equation}
is just the partition function arising from the free ghost particles which
plays the role of cancelling out the unphysical contribution contained in
Eq. (120). If we change the integration variables in Eq. (147) and
considering the relations 
\begin{equation}
\zeta _a^{*}(\vec k,\tau )=-\overline{\zeta }_a(-\vec k,\tau )\text{, }\zeta
_a(\vec k,\tau )=-\overline{\zeta }_a^{*}(-\vec k,\tau )  \eqnum{149}
\end{equation}
which will be interpreted in the Appendix, Eq. (147) may be recast in the
form 
\begin{equation}
\begin{array}{c}
Z_c^0[\zeta ]=Z_c^0\exp \{\int_0^\beta d\tau _1\int_0^\beta d\tau _2\int
d^3k[\overline{\zeta }_a^{*}(\vec k,\tau _1)\Delta _c^{aa^{\prime }}(\vec k%
,\tau _1-\tau _2)\overline{\zeta }_{a^{\prime }}(\vec k,\tau _2) \\ 
-\zeta _a^{*}(\vec k,\tau _1)\Delta _c^{aa^{\prime }}(\vec k,\tau _1-\tau
_2)\zeta _{a^{\prime }}(\vec k,\tau _2)]\}
\end{array}
\eqnum{150}
\end{equation}
where 
\begin{equation}
\Delta _c^{aa^{\prime }}(\vec k,\tau _1-\tau _2)=\delta ^{aa^{\prime
}}\Delta _g(\vec k,\tau _1-\tau _2)  \eqnum{151}
\end{equation}
here $\Delta _g(\vec k,\tau _1-\tau _2)$ was written in Eq. (124).

Up to the present, the perturbative expansion of the thermal QCD generating
functional in the coherent-state representation has exactly been given by
the combination of Eqs. (95), (96), (119), (138) and (150). In the
derivation of the perturbation expansion, as one has seen, to obtain the
final expressions of the propagators shown in Eqs. (124), (140) and (151),
it is necessary to use the functional properties and relations for the
external sources as denoted in Eqs. (123), (137) and (149). As a result of
the derivation of the generating functional, the partition function for the
free system has simultaneously been given by the combination of Eqs. (120),
(135) and (148). The partition function for the interacting system can be
calculated in the way as shown in Eq. (45). Here it should be noted that the
differential $\delta /\delta j(\tau )$ in Eq. (95) represent the collection
of the differentials $\delta /\delta \xi _\lambda ^{a*}(\vec k,\tau )$, $%
\delta /\delta \xi _\lambda ^a(\vec k,\tau )$, $-\delta /\delta \eta _s(\vec 
k,\tau )$, $\delta /\delta \eta _s^{*}(\vec k,\tau )$, $-\delta /\delta 
\overline{\eta }_s(\vec k,\tau )$, $\delta /\delta \overline{\eta }_s^{*}(%
\vec k,\tau )$, $-\delta /\delta \zeta _a(\vec k,\tau )$, $\delta /\delta
\zeta _a^{*}(\vec k,\tau )$, $-\delta /\delta \overline{\zeta }_a(\vec k%
,\tau )$ and $\delta /\delta \overline{\zeta }_a^{*}(\vec k,\tau )$.
Ordinarily, the generating functional of thermal QCD represented in the
position space is used in the literature. In the Appendix, it will be shown
that this generating functional can readily be derived from the generating
functional described in this section.

\section{Relativistic equation for $q\overline{q\text{ }}$bound states}

With the generating functional given in the preceding section, we are ready
to derive the relativistic equation for $q\overline{q}$ bound states at
finite temperature. It is well-known that a bound state exists in the
space-like\ Minkowski space in which there always is an equal-time Lorentz
frame. Since in the equal-time frame, the relativistic equation is reduced
to a three-dimensional one without loss of any rigorism, in this section we
only pay our attention to the three-dimensional equation which may be
derived from the equations of motion satisfied by the following $q\overline{q%
}$ two-''time'' (temperature) four-point Green function [15-17] 
\begin{equation}
\begin{array}{c}
{\cal G}(\alpha \beta ;\gamma \delta ;\tau _1-\tau _2)=Tr\{e^{\beta (\Omega -%
\widehat{K})}T\{N[\widehat{b}_\alpha (\tau _1)\widehat{b}_\beta ^{+}(\tau
_1)]N[\widehat{b}_\gamma (\tau _2)\widehat{b}_\delta ^{+}(\tau _2)]\}\} \\ 
\equiv \left\langle T\{N[\widehat{b}_\alpha (\tau _1)\widehat{b}_\beta
^{+}(\tau _1)]N[\widehat{b}_\gamma (\tau _2)\widehat{b}_\delta ^{+}(\tau
_2)]\}\right\rangle _\beta
\end{array}
\eqnum{152}
\end{equation}
where the symbol $\left\langle \cdot \cdot \cdot \right\rangle _\beta $
represents the statistical average and $N$ symbolizes the normal product
whose definition can be given from the corresponding definition at
zero-temperature by replacing the vacuum average with the statistical
average [17] 
\begin{equation}
N[\widehat{b}_\alpha (\tau _1)\widehat{b}_\beta ^{+}(\tau _2)]=T[\widehat{b}%
_\alpha (\tau _1)\widehat{b}_\beta ^{+}(\tau _2)]-S_{\alpha \beta }(\tau
_1-\tau _2)  \eqnum{153}
\end{equation}
where 
\begin{equation}
S_{\alpha \beta }(\tau _1-\tau _2)=\left\langle T[\widehat{b}_\alpha (\tau
_1)\widehat{b}_\beta ^{+}(\tau _2)]\right\rangle _\beta  \eqnum{154}
\end{equation}
is the quark or antiquark thermal propagator. The normal product in Eq.
(152) plays a role of excluding the contraction between the quark and the
antiquark operators from the Green function when the quark and antiquark are
of the same flavor. Physically, this avoids the $q\overline{q}$ annihilation
that would break stability of a bound state. Substituting Eq. (153) in Eq.
(152), we have 
\begin{equation}
{\cal G}(\alpha \beta ;\gamma \delta ;\tau _1-\tau _2)=G(\alpha \beta
;\gamma \delta ;\tau _1-\tau _2)-S_{\alpha \beta }S_{\gamma \delta } 
\eqnum{155}
\end{equation}
where 
\begin{equation}
G(\alpha \beta ;\gamma \delta ;\tau _1-\tau _2)=\left\langle T\{\widehat{b}%
_\alpha (\tau _1)\widehat{b}_\beta ^{+}(\tau _1)\widehat{b}_\gamma (\tau _2)%
\widehat{b}_\delta ^{+}(\tau _2)\}\right\rangle _\beta  \eqnum{156}
\end{equation}
is the ordinary Green function and, $S_{\alpha \beta }$ and $S_{\gamma
\delta }$ are the equal-time quark (antiquark) propagators. Obviously, in
order to derive the equation of motion satisfied by the Green function $%
{\cal G}(\alpha \beta ;\gamma \delta ;\tau _1-\tau _2)$, we need first to
derive the equation of motion for the Green function $G(\alpha \beta ;\gamma
\delta ;\tau _1-\tau _2)$.

Let us start with the generating functional in Eq. (91). As shown in Sect.
4, by partial integration of the second term on the right hand side of Eq.
(89), the generating functional may be written in the form 
\begin{equation}
\begin{array}{c}
Z[j]=\int D(q^{*}q)e^{-q^{*}\cdot q}\int {\frak D}(q^{*}q)\exp \{q^{*}(\beta
)\cdot q(\beta ) \\ 
-S_E+\int_0^\beta d\tau j^{*}(\tau )\cdot q(\tau )\}
\end{array}
\eqnum{157}
\end{equation}
where 
\begin{equation}
S_E=\int_0^\beta d\tau \{\sum\limits_\alpha q_\alpha ^{*}(\tau )\circ \dot q%
_\alpha (\tau )+H(\tau )\}  \eqnum{158}
\end{equation}
here $H(\tau )$ was given in Eqs. (78) and (79). First, we derive an
equation of motion describing the variation of the $q\overline{q\text{ }}$
four-point Green function with the ''time'' variable $\tau _1$. For this
purpose, let us differentiate the generating functional in Eq. (157) with
respect to $b_\alpha ^{*}(\tau _1)$. Considering that the generating
functional is independent of $b_\alpha ^{*}(\tau _1)$ and noticing the
expressions given in Eqs. (158), (78), (79) and (93), one may obtain 
\begin{equation}
\begin{array}{c}
\frac{\delta Z[j]}{\delta b_\alpha ^{*}(\tau _1)}=\int
D(q^{*}q)e^{-q^{*}\cdot q}\int {\frak D}(q^{*}q)[-\dot b_\alpha (\tau
_1)-\varepsilon _\alpha \theta _\alpha b_\alpha (\tau _1) \\ 
-\sum\limits_{\rho \lambda }A_{\alpha \rho \lambda }b_\rho (\tau
_1)a_\lambda (\tau _1)+\theta _\alpha \eta _\alpha (\tau _1)]\exp
\{q^{*}(\beta )\cdot q(\beta )-S_E \\ 
-\int_0^\beta d\tau j^{*}(\tau )\cdot q(\tau )\}=0.
\end{array}
\eqnum{159}
\end{equation}
When the $b_\alpha (\tau _1)$ and $a_\lambda (\tau _1)$ in the above are
replaced by the functional derivatives $\theta _\alpha \delta /\delta \eta
_\alpha ^{*}(\tau _1)$ and $\delta /\delta j_\lambda ^{*}(\tau _1)$
respectively and multiplying the both sides of Eq. (159) with $\theta
_\alpha $, the above equation can be written as 
\begin{equation}
\{\frac d{d\tau _1}\frac \delta {\delta \eta _\alpha ^{*}(\tau _1)}+\theta
_\alpha \varepsilon _\alpha \frac \delta {\delta \eta _\alpha ^{*}(\tau _1)}%
+\sum\limits_{\rho \lambda }\theta _\alpha \theta _\rho A(\alpha \rho
\lambda )\frac{\delta ^2}{\delta \eta _\rho ^{*}(\tau _1)\delta j_\lambda
^{*}(\tau _1)}-\eta _\alpha (\tau _1)\}Z[j]=0.  \eqnum{160}
\end{equation}
Then, we differentiate the above equation with respect to the sources $\eta
_\beta ((\tau _1)$, giving

\begin{equation}
\begin{array}{c}
\{(\frac d{d\tau _1}\frac \delta {\delta \eta _\alpha ^{*}(\tau _1)})\frac 
\delta {\delta \eta _\beta (\tau _1)}+\theta _\alpha \varepsilon _\alpha 
\frac{\delta ^2}{\delta \eta _\alpha ^{*}(\tau _1)\delta \eta _\beta (\tau
_1)}+\sum\limits_{\rho \lambda }\theta _\alpha \theta _\rho A(\alpha \rho
\lambda )\frac{\delta ^3}{\delta \eta _\rho ^{*}(\tau _1)\delta \eta _\beta
(\tau _1)\delta j_\lambda ^{*}(\tau _1)} \\ 
+\delta _{\alpha \beta }-\eta _\alpha (\tau _1)\frac \delta {\delta \eta
_\beta (\tau _1)}\}Z[j]=0.
\end{array}
\eqnum{161}
\end{equation}
Furthermore, successive differentiations of Eq. (161) with respect sources $%
\eta _\gamma ^{*}(\tau _2)$ and $\eta _\delta (\tau _2)$ yield 
\begin{equation}
\begin{array}{c}
\{(\frac d{d\tau _1}\frac \delta {\delta \eta _\alpha ^{*}(\tau _1)})\frac{%
\delta ^3}{\delta \eta _\beta (\tau _1)\delta \eta _\gamma ^{*}(\tau
_2)\delta \eta _\delta (\tau _2)}+\theta _\alpha \varepsilon _\alpha \frac{%
\delta ^4}{\delta \eta _\alpha ^{*}(\tau _1)\delta \eta _\beta (\tau
_1)\delta \eta _\gamma ^{*}(\tau _2)\delta \eta _\delta (\tau _2)} \\ 
+\sum\limits_{\lambda \sigma }\theta _\alpha \theta _\rho A(\alpha \rho
\lambda )\frac{\delta ^5}{\delta \eta _\rho ^{*}(\tau _1)\delta \eta _\beta
(\tau _1)\delta \eta _\gamma ^{*}(\tau _2)\delta \eta _\delta (\tau
_2)\delta j_\lambda ^{*}(\tau _1)}+\delta _{\alpha \beta }\frac{\delta ^2}{%
\delta \eta _\gamma ^{*}(\tau _2)\delta \eta _\delta (\tau _2)} \\ 
-\delta _{\alpha \delta }\delta (\tau _1-\tau _2)\frac{\delta ^2}{\delta
\eta _\gamma ^{*}(\tau _2)\delta \eta _\beta (\tau _1)}-\eta _\alpha (\tau
_1)\frac{\delta ^3}{\delta \eta _\beta (\tau _1)\delta \eta _\gamma
^{*}(\tau _2)\delta \eta _\delta (\tau _2)}\}Z[j]=0.
\end{array}
\eqnum{162}
\end{equation}
Similarly, when differentiating Eq. (157) with respect $b_\beta (\tau _1)$,
one may obtain 
\begin{equation}
\{\frac d{d\tau _1}\frac \delta {\delta \eta _\beta (\tau _1)}-\theta _\beta
\varepsilon _\beta \frac \delta {\delta \eta _\beta (\tau _1)}%
-\sum\limits_{\sigma \lambda }\theta _\sigma \theta _\beta A(\sigma \beta
\lambda )\frac{\delta ^2}{\delta \eta _\sigma (\tau _1)\delta j_\lambda
^{*}(\tau _1)}-\eta _\beta ^{*}(\tau _1)\}Z[j]=0,  \eqnum{163}
\end{equation}
Subsequently, On differentiating the above equation with respect to $\eta
_\alpha ^{*}(\tau _1)$, we get 
\begin{equation}
\begin{array}{c}
\{\frac \delta {\delta \eta _\alpha ^{*}(\tau _1)}(\frac d{d\tau _1}\frac 
\delta {\delta \eta _\beta (\tau _1)})-\theta _\beta \varepsilon _\beta 
\frac{\delta ^2}{\delta \eta _\alpha ^{*}(\tau _1)\delta \eta _\beta (\tau
_1)}-\sum\limits_{\sigma \lambda }\theta _\beta \theta _\sigma A(\sigma
\beta \lambda )\frac{\delta ^3}{\delta \eta _\alpha ^{*}(\tau _1)\delta \eta
_\sigma (\tau _1)\delta j_\lambda ^{*}(\tau _1)} \\ 
-\delta _{\alpha \beta }+\eta _\beta ^{*}(\tau _1)\frac \delta {\delta \eta
_\alpha ^{*}(\tau _1)}\}Z[j]=0.
\end{array}
\eqnum{164}
\end{equation}
Finally, successive differentiations of the above equation with respect to
the sources $\eta _\gamma ^{*}(\tau _2)$ and $\eta _\delta (\tau _2)$ give
rise to 
\begin{equation}
\begin{array}{c}
\{\frac \delta {\delta \eta _\alpha ^{*}(\tau _1)}(\frac d{d\tau _1}\frac 
\delta {\delta \eta _\beta (\tau _1)})\frac{\delta ^2}{\delta \eta _\gamma
^{*}(\tau _2)\delta \eta _\delta (\tau _2)}-\theta _\beta \varepsilon _\beta 
\frac{\delta ^4}{\delta \eta _\alpha ^{*}(\tau _1)\delta \eta _\beta (\tau
_1)\delta \eta _\gamma ^{*}(\tau _2)\delta \eta _\delta (\tau _2)} \\ 
-\sum\limits_{\lambda \sigma }\theta _\beta \theta _\sigma A(\sigma \beta
\lambda )\frac{\delta ^5}{\delta \eta _\alpha ^{*}(\tau _1)\delta \eta
_\sigma (\tau _1)\delta \eta _\gamma ^{*}(\tau _2)\delta \eta _\delta (\tau
_2)\delta j_\lambda ^{*}(\tau _1)}-\delta _{\alpha \beta }\frac{\delta ^2}{%
\delta \eta _\gamma ^{*}(\tau _2)\delta \eta _\delta (\tau _2)} \\ 
+\delta _{\beta \gamma }\delta (\tau _1-\tau _2)\frac{\delta ^2}{\delta \eta
_\alpha ^{*}(\tau _1)\delta \eta _\delta (\tau _2)}+\eta _\beta ^{*}(\tau _1)%
\frac{\delta ^3}{\delta \eta _\alpha ^{*}(\tau _1)\delta \eta _\gamma
^{*}(\tau _2)\delta \eta _\delta (\tau _2)}\}Z[j]=0.
\end{array}
\eqnum{165}
\end{equation}

Adding Eq. (161) to Eq. (164), then multiplying the both sides of the
equation thus obtained with $-\theta _\alpha \theta _\beta $ and finally
setting the external sources $\eta _\alpha ^{*}=\eta _\beta =0$, but
remaining the gluon source $j_\lambda \neq 0$, we get 
\begin{equation}
(\frac d{d\tau _1}+\theta _\alpha \varepsilon _\alpha -\theta _\beta
\varepsilon _\beta )S_{\alpha \beta }^{j_\lambda }+\sum\limits_{\rho \sigma
\lambda }[A(\alpha \rho \lambda )\delta _{\beta \sigma }-A(\sigma \beta
\lambda )\delta _{\alpha \rho }]\frac \delta {\delta j_\lambda ^{*}(\tau _1)}%
S_{\rho \sigma }^{j_\lambda }=0  \eqnum{166}
\end{equation}
where 
\begin{equation}
S_{\alpha \beta }^{j_\lambda }=-\frac 1Z\theta _\alpha \theta _\beta \frac{%
\delta ^2Z[j]}{\delta \eta _\alpha ^{*}(\tau _1)\delta \eta _\beta (\tau _1)}%
\mid _{\eta _\alpha ^{*}=\eta _\beta =0}  \eqnum{167}
\end{equation}
is the quark (antiquark) equal-time propagator in the presence of source $%
j_\lambda $. If we define 
\begin{equation}
H(\alpha \beta ;\rho \sigma ;\tau _1)^{j_\lambda }=(\frac d{d\tau _1}+\theta
_\alpha \varepsilon _\alpha -\theta _\beta \varepsilon _\beta )\delta
_{\alpha \rho }\delta _{\beta \sigma }+\sum\limits_\lambda f(\alpha \beta
;\rho \sigma \lambda )\frac \delta {\delta j_\lambda ^{*}(\tau _1)} 
\eqnum{168}
\end{equation}
where 
\begin{equation}
f(\alpha \beta ;\rho \sigma \lambda )=A(\alpha \rho \lambda )\delta _{\beta
\sigma }-A(\sigma \beta \lambda )\delta _{\alpha \rho },  \eqnum{169}
\end{equation}
Eq. (166) can simply be represented as 
\begin{equation}
\sum\limits_{\rho \sigma }H(\alpha \beta ;\rho \sigma ;\tau _1)^{j_\lambda
}S_{\rho \sigma }^{j_\lambda }=0.  \eqnum{170}
\end{equation}

When summing up the both equations in Eqs. (162) and (165), then multiplying
the equation thus obtained with $\theta _\alpha \theta _\beta \theta _\gamma
\theta _\delta $ and finally setting all the sources but $j_\lambda $ to be
zero, one may get 
\begin{equation}
\begin{array}{c}
(\frac d{d\tau _1}+\theta _\alpha \varepsilon _\alpha -\theta _\beta
\varepsilon _\beta )G(\alpha \beta ;\gamma \delta ;\tau _1-\tau
_2)^{j_\lambda }+\sum\limits_{\rho \sigma \lambda }f(\alpha \beta ;\rho
\sigma \lambda )\frac \delta {\delta j_\lambda ^{*}(\tau _1)}G(\rho \sigma
;\gamma \delta ;\tau _1-\tau _2)^{j_\lambda } \\ 
=\delta (\tau _1-\tau _2)[\delta _{\beta \gamma }S_{\alpha \delta }(\tau
_1-\tau _2)^{j_\lambda }-\delta _{\alpha \delta }S_{\gamma \beta }(\tau
_2-\tau _1)^{j_\lambda }]
\end{array}
\eqnum{171}
\end{equation}
where 
\begin{equation}
G(\alpha \beta ;\gamma \delta ;\tau _1-\tau _2)^{j_\lambda }=\frac 1Z\theta
_\alpha \theta _\beta \theta _\gamma \theta _\delta \frac{\delta ^4Z[j]}{%
\delta \eta _\alpha ^{*}(\tau _1)\delta \eta _\beta (\tau _1)\delta \eta
_\gamma ^{*}(\tau _2)\delta \eta _\delta (\tau _2)}\mid _{\eta ^{*}=\eta =0}
\eqnum{172}
\end{equation}
and 
\begin{equation}
S_{\alpha \beta }(\tau _1-\tau _2)^{j_\lambda }=-\frac 1Z\theta _\alpha
\theta _\beta \frac{\delta ^2Z[j]}{\delta \eta _\alpha ^{*}(\tau _1)\delta
\eta _\beta (\tau _2)}\mid _{\eta ^{*}=\eta =0}  \eqnum{173}
\end{equation}
are respectively the $q\overline{q}$ two-''time'' four-point thermal Green
function and the quark or antiquark thermal propagator in presence of source 
$j_\lambda $. When the source $j_\lambda $ is turned off, Eqs. (172) and
(173) will respectively go over to the Green function in Eq. (156) and the
propagator in Eq. (154). It is noted that due to the restriction of the
delta function, the propagators in Eq. (171) are actually
''time''-independent. With the definition in Eq. (168), Eq. (171) may be
represented as 
\begin{equation}
\sum\limits_{\rho \sigma }H(\alpha \beta ;\rho \sigma ;\tau _1)^{j_\lambda
}G(\rho \sigma ;\gamma \delta ;\tau _1-\tau _2)^{j_\lambda }=-\delta (\tau
_1-\tau _2)S(\alpha \beta ;\gamma \delta )^{j_\lambda }  \eqnum{174}
\end{equation}
where 
\begin{equation}
S(\alpha \beta ;\gamma \delta )^{j_\lambda }=\delta _{\alpha \delta
}S_{\gamma \beta }^{j_\lambda }-\delta _{\beta \gamma }S_{\alpha \delta
}^{j_\lambda }.  \eqnum{175}
\end{equation}
Acting on the both sides of Eq. (155) with the operator $H(\alpha \beta
;\rho \sigma ;\tau _1)^{j_\lambda }$ and using the equations in Eqs. (170)
and (174), we find 
\begin{equation}
\begin{array}{c}
\sum\limits_{\rho \sigma }H(\alpha \beta ;\rho \sigma ;\tau _1)^{j_\lambda }%
{\cal G}(\rho \sigma ;\gamma \delta ;\tau _1-\tau _2)^{j_\lambda
}=\sum\limits_{\rho \sigma }H(\alpha \beta ;\rho \sigma ;\tau _1)^{j_\lambda
}G(\rho \sigma ;\gamma \delta ;\tau _1-\tau _2)^{j_\lambda } \\ 
=-\delta (\tau _1-\tau _2)S(\alpha \beta ;\gamma \delta )^{j_\lambda }
\end{array}
\eqnum{176}
\end{equation}
This indicates that the equation of motion satisfied by the Green function $%
{\cal G}(\alpha \beta ;\gamma \delta ;\tau _1-\tau _2)$ formally is the same
as the one shown in Eq. (171). Therefore, in the case that the source $%
j_{\lambda \text{ }}$vanishes, we can write 
\begin{equation}
\begin{array}{c}
(\frac d{d\tau _1}+\theta _\alpha \varepsilon _\alpha -\theta _\beta
\varepsilon _\beta ){\cal G}(\alpha \beta ;\gamma \delta ;\tau _1-\tau _2)
\\ 
=-\delta (\tau _1-\tau _2)S(\alpha \beta ;\gamma \delta )-\sum\limits_{\rho
\sigma \lambda }f(\alpha \beta ;\rho \sigma \lambda ){\cal G}(\rho \sigma
\lambda ;\gamma \delta ;\tau _1-\tau _2)
\end{array}
\eqnum{177}
\end{equation}
where

\begin{equation}
\begin{array}{c}
{\cal G}(\rho \sigma \lambda ;\gamma \delta ;\tau _1-\tau _2)=\frac \delta {%
\delta j_\lambda ^{*}(\tau _1)}{\cal G}(\rho \sigma ;\gamma \delta ;\tau
_1-\tau _2)^{j_\lambda }\mid _{j_\lambda =0} \\ 
=\left\langle T\{N[\widehat{b}_\rho (\tau _1)\widehat{b}_\sigma ^{+}(\tau _1)%
\widehat{a}_\lambda (\tau _1)]N[\widehat{b}_\gamma (\tau _2)\widehat{b}%
_\delta ^{+}(\tau _2)]\}\right\rangle _\beta
\end{array}
\eqnum{178}
\end{equation}
and 
\begin{equation}
S(\alpha \beta ;\gamma \delta )=\delta _{\alpha \delta }S_{\gamma \beta
}-\delta _{\beta \gamma }S_{\alpha \delta }=-\langle [\widehat{b}_\alpha 
\widehat{b}_\beta ^{+},\widehat{b}_\gamma \widehat{b}_\delta
^{+}]_{-}\rangle _\beta .  \eqnum{179}
\end{equation}
It is noted here that similar to the definition in Eq. (153), the normal
product $N[\widehat{b}_\rho (\tau _1)\widehat{b}_\sigma ^{+}(\tau _1)%
\widehat{a}_\lambda (\tau _1)]$ in Eq. (178) is defined as 
\begin{equation}
N[\widehat{b}_\rho (\tau _1)\widehat{b}_\sigma ^{+}(\tau _1)\widehat{a}%
_\lambda (\tau _1)]=T[\widehat{b}_\rho (\tau _1)\widehat{b}_\sigma ^{+}(\tau
_1)\widehat{a}_\lambda (\tau _1)]-\Lambda (\rho \sigma \lambda )  \eqnum{180}
\end{equation}
where 
\begin{equation}
\Lambda (\rho \sigma \lambda )=\left\langle T[\widehat{b}_\rho (\tau _1)%
\widehat{b}_\sigma ^{+}(\tau _1)\widehat{a}_\lambda (\tau _1)]\right\rangle
_\beta .  \eqnum{181}
\end{equation}
Substituting Eqs. (153) and (180) into Eq. (178), we have 
\begin{equation}
{\cal G}(\rho \sigma \lambda ;\gamma \delta ;\tau _1-\tau _2)=G(\rho \sigma
\lambda ;\gamma \delta ;\tau _1-\tau _2)-\Lambda (\rho \sigma \lambda
)S_{\gamma \delta }  \eqnum{182}
\end{equation}
where 
\begin{equation}
G(\rho \sigma \lambda ;\gamma \delta ;\tau _1-\tau _2)=\left\langle T\{%
\widehat{b}_\rho (\tau _1)\widehat{b}_\sigma ^{+}(\tau _1)\widehat{a}%
_\lambda (\tau _1)\widehat{b}_\gamma (\tau _2)\widehat{b}_\delta ^{+}(\tau
_2)\}\right\rangle _\beta  \eqnum{183}
\end{equation}
is the ordinary five-point thermal Green function including a gluon operator
in it.

By the argument as mentioned in the Appendix (see Eq. (A9) or (A18)), it is
easy to prove that the Green functions ${\cal G}(\alpha \beta ;\gamma \delta
;\tau _1-\tau _2)$ and ${\cal G}(\rho \sigma \lambda ;\gamma \delta ;\tau
_1-\tau _2)$ are periodic. Therefore, we have the following Fourier
expansions: 
\begin{equation}
\begin{array}{c}
{\cal G}(\alpha \beta ;\gamma \delta ;\tau )=\frac 1\beta \sum\limits_n{\cal %
G}(\alpha \beta ;\gamma \delta ;\omega _n)e^{-i\omega _n\tau }, \\ 
{\cal G}(\rho \sigma \lambda ;\gamma \delta ;\tau )=\frac 1\beta
\sum\limits_n{\cal G}(\rho \sigma \lambda ;\gamma \delta ;\omega
_n)e^{-i\omega _n\tau }
\end{array}
\eqnum{184}
\end{equation}
where $\tau =\tau _1-\tau _2$ and $\omega _n=\frac{2\pi n}\beta $. Upon
inserting Eq. (184) into Eq. (177) and performing the integration $\frac 12%
\int_{-\beta }^\beta d\tau e^{i\omega _n\tau }$, we arrive at 
\begin{equation}
\begin{array}{c}
(i\omega _n-\theta _\alpha \varepsilon _\alpha +\theta _\beta \varepsilon
_\beta ){\cal G}(\alpha \beta ;\gamma \delta ;\omega _n) \\ 
=-S(\alpha \beta ;\gamma \delta )+\sum\limits_{\rho \sigma \lambda }f(\alpha
\beta ;\rho \sigma \lambda ){\cal G}(\rho \sigma \lambda ;\gamma \delta
;\omega _n).
\end{array}
\eqnum{185}
\end{equation}
It is well-known that the Green function ${\cal G}(\rho \sigma \lambda
;\gamma \delta ;\omega _n)$ is B-S (two-particle) reducible [15-17].
Therefore, we can write 
\begin{equation}
\sum\limits_{\lambda \tau \rho }f(\alpha \beta ;\rho \sigma \lambda ){\cal G}%
(\rho \sigma \lambda ;\gamma \delta ;\omega _n)=\sum\limits_{\mu \nu
}K(\alpha \beta ;\mu \nu ;\omega _n){\cal G}(\mu \nu ;\gamma \delta ;\omega
_n)  \eqnum{186}
\end{equation}
where $K(\alpha \beta ;\mu \nu ;\omega _n)$ is called interaction kernel.
Thus, Eq. (185) can be written in a closed form 
\begin{equation}
(i\omega _n-\theta _\alpha \varepsilon _\alpha +\theta _\beta \varepsilon
_\beta ){\cal G}(\alpha \beta ;\gamma \delta ;\omega _n)=-S(\alpha \beta
;\gamma \delta )+\sum\limits_{\mu \nu }K(\alpha \beta ;\mu \nu ;\omega _n)%
{\cal G}(\mu \nu ;\gamma \delta ;\omega _n).  \eqnum{187}
\end{equation}

Now, let us turn to the equation satisfied by $q\overline{q}$ bound states.
This equation can be derived from Eq. (187) with the aid of the following
Lehmann representation of the four-point Green function which may be derived
by expanding the time-ordered product in Eq. (152) and then inserting the
complete set of $q\overline{q}$ bound states into Eq. (152) [15-17, 31], 
\begin{equation}
{\cal G}(\alpha \beta ;\gamma \delta ;\omega _l)=\frac 12e^{\beta \Omega
}\sum\limits_{mn}\Delta _{mn}\{\frac{\chi _{nm}(\alpha \beta )\chi
_{mn}(\gamma \delta )}{i\omega _l-E_{nm}}-\frac{\chi _{nm}(\gamma \delta
)\chi _{mn}(\alpha \beta )}{i\omega _l+E_{nm}}\}  \eqnum{188}
\end{equation}
where 
\begin{equation}
\chi _{nm}(\alpha \beta )=\left\langle m\left| N[\widehat{b}_\alpha \widehat{%
b}_\beta ^{+}]\right| n\right\rangle  \eqnum{189}
\end{equation}
which is the transition amplitude from the state with energy $E_n$ to the
state with energy $E_m$ and 
\begin{equation}
\Delta _{nm}=e^{-\beta E_n}-e^{-\beta E_m}.  \eqnum{190}
\end{equation}
Upon substituting Eq. (188) into Eq. (187) and then taking the limit: $%
\lim_{i\omega _l\rightarrow E_{nm}}(i\omega _l-E_{nm})$, we get the
following equation satisfied by the transition amplitude 
\begin{equation}
(E_{nm}-\theta _\alpha \varepsilon _\alpha +\theta _\beta \varepsilon _\beta
)\chi _{nm}(\alpha \beta )=\sum\limits_{\gamma \delta }K(\alpha \beta
;\gamma \delta ;E_{nm})\chi _{nm}(\gamma \delta )  \eqnum{191}
\end{equation}
where the fact that the function $S(\alpha \beta ;\gamma \delta )$ has no
bound state poles has been considered. If we take $\mid m\rangle $ to be the
vacuum state $\mid 0\rangle $ and set $E=E_{n0}$ and $\chi _n(\alpha \beta
)=\left\langle 0\left| N[\widehat{b}_\alpha \widehat{b}_\beta ^{+}]\right|
n\right\rangle $, we can write from the above equation that 
\begin{equation}
(E-\theta _\alpha \varepsilon _\alpha +\theta _\beta \varepsilon _\beta
)\chi _n(\alpha \beta )=\sum\limits_{\gamma \delta }K(\alpha \beta ;\gamma
\delta ;E)\chi _n(\gamma \delta ).  \eqnum{192}
\end{equation}
where the subscript $n$ in $E_n$ has been suppressed. This just is the
equation satisfied by the $q\overline{q}$ bound states at finite temperature.

Since the index $\alpha $ contains $\theta _\alpha =\pm $, Eq. (192)
actually is a set of coupled equations for the amplitudes $\chi _n(\alpha
^{+}\beta ^{-})$, $\chi _n(\alpha ^{-}\beta ^{+})$, $\chi _n(\alpha
^{+}\beta ^{+})$ and $\chi _n(\alpha ^{-}\beta ^{-})$. Following the
procedure described in Refs. (16) and (17), one may reduce the above
equation to an equivalent equation satisfied by the amplitude of positive
energy. We do not repeat the derivation here. We only show the result as
follows: 
\begin{equation}
\lbrack E-\varepsilon (\vec k_\alpha )-\varepsilon (\vec k_\beta )]\psi
(\alpha \beta ;E)=\sum\limits_{\gamma \delta }V(\alpha \beta ;\gamma \delta
;E)\psi (\gamma \delta ;E).  \eqnum{193}
\end{equation}
where $\psi (\alpha \beta ;E)=\chi _n(\alpha ^{+}\beta ^{-})$ and $V(\alpha
\beta ;\gamma \delta ;E)$ is the interaction Hamiltonian which can be
expressed as

\begin{equation}
V(\alpha \beta ;\gamma \delta ;E)=\sum\limits_{n=0}V^{(n)}(\alpha \beta
;\gamma \delta ;E),  \eqnum{194}
\end{equation}
in which 
\begin{equation}
V^{(0)}(\alpha \beta ;\gamma \delta ;E)=K_{++++}(\alpha \beta ;\gamma \delta
;E),  \eqnum{195}
\end{equation}
\begin{equation}
V^{(1)}(\alpha \beta ;\gamma \delta ;E)=\sum\limits_{ab\neq
++}\sum\limits_{\rho \sigma }\frac{K_{++ab}(\alpha \beta ;\rho \sigma
;E)K_{ab++}(\rho \sigma ;\gamma \delta ;E)}{E-a\varepsilon (\vec k_\rho
)-b\varepsilon (\vec k_\sigma )},  \eqnum{196}
\end{equation}
\begin{equation}
\begin{tabular}{l}
$V^{(2)}(\alpha \beta ;\gamma \delta ;E)$ \\ 
$=\sum\limits_{ab\neq ++}\sum\limits_{cd\neq ++}\sum\limits_{\rho \sigma
}\sum\limits_{\mu \nu }\frac{K_{++ab}(\alpha \beta ;\rho \sigma
;E)K_{abcd}(\rho \sigma ;\mu \nu ;E)K_{cd++}(\mu \nu ;\gamma \delta ;E)}{%
(E-a\varepsilon (\vec k_\rho )-b\varepsilon (\vec k_\sigma ))(E-c\varepsilon
(\vec k_\mu )-d\varepsilon (\vec k_\nu ))},$ \\ 
$\cdot \cdot \cdot \cdot \cdot \cdot .$%
\end{tabular}
\eqnum{197}
\end{equation}
here $a,b=\pm $, and 
\begin{equation}
\begin{array}{c}
K_{++++}(\alpha \beta ;\gamma \delta ;E)=K(\alpha ^{+}\beta ^{-};\gamma
^{+}\delta ^{-};E), \\ 
K_{----}(\alpha \beta ;\gamma \delta ;E)=K(\alpha ^{-}\beta ^{+};\gamma
^{-}\delta ^{+};E) \\ 
K_{+-+-}(\alpha \beta ;\gamma \delta ;E)=K(\alpha ^{+}\beta ^{+};\gamma
^{+}\delta ^{+};E), \\ 
K_{-+-+}(\alpha \beta ;\gamma \delta ;E)=K(\alpha ^{-}\beta ^{-};\gamma
^{-}\delta ^{-};E).
\end{array}
\eqnum{198}
\end{equation}

\section{Closed expression of the interaction kernel in the equation for $q%
\overline{q}$ bound states}

In this section, we are devoted to deriving a closed expression of the
interaction kernel appearing in Eq. (192) and defined in Eq. (186). For this
derivation, we need equations of motion which describe evolution of the
Green functions ${\cal G}(\alpha \beta ;\gamma \delta ;\tau _1-\tau _2)$ and 
${\cal G}(\alpha \beta \sigma ;\gamma \delta ;\tau _1-\tau _2)$ with time $%
\tau _2$. Taking the derivatives of the generating functional in Eq. (157)
with respect to $b_\gamma ^{*}(\tau _2)$ and $b_\delta (\tau _2)$
respectively, by the same procedure as described in the derivation of Eq.
(160), one may obtain 
\begin{equation}
\{\frac d{d\tau _2}\frac \delta {\delta \eta _\gamma ^{*}(\tau _2)}+\theta
_\gamma \varepsilon _\gamma \frac \delta {\delta \eta _\gamma ^{*}(\tau _2)}%
+\sum\limits_{\rho \lambda }\theta _\gamma \theta _\rho A(\gamma \rho
\lambda )\frac{\delta ^2}{\delta \eta _\rho ^{*}(\tau _2)\delta j_\lambda
^{*}(\tau _2)}-\eta _\gamma (\tau _2)\}Z[j]=0.  \eqnum{199}
\end{equation}
and

\begin{equation}
\{\frac d{d\tau _2}\frac \delta {\delta \eta _\delta (\tau _2)}-\theta
_\delta \varepsilon _\delta \frac \delta {\delta \eta _\delta (\tau _2)}%
-\sum\limits_{\sigma \lambda }\theta _\delta \theta _\sigma A(\sigma \delta
\lambda )\frac{\delta ^2}{\delta \eta _\sigma (\tau _2)\delta j_\lambda
^{*}(\tau _2)}-\eta _\delta ^{*}(\tau _2)\}Z[j]=0.  \eqnum{200}
\end{equation}
Performing differentiations of Eqs. (199) and (200) with respect to the
sources $\eta _\delta (\tau _2)$ and $\eta _\gamma ^{*}(\tau _2)$
respectively, we get 
\begin{equation}
\begin{array}{c}
\{(\frac d{d\tau _2}\frac \delta {\delta \eta _\gamma ^{*}(\tau _2)})\frac 
\delta {\delta \eta _\delta (\tau _2)}+\theta _\gamma \varepsilon _\gamma 
\frac{\delta ^2}{\delta \eta _\gamma ^{*}(\tau _2)\delta \eta _\delta (\tau
_2)}+\sum\limits_{\rho \lambda }\theta _\gamma \theta _\rho A(\gamma \rho
\lambda )\frac{\delta ^3}{\delta \eta _\rho ^{*}(\tau _2)\delta \eta _\delta
(\tau _2)\delta j_\lambda ^{*}(\tau _2)} \\ 
+\delta _{\gamma \delta }-\eta _\gamma (\tau _2)\frac \delta {\delta \eta
_\delta (\tau _2)}\}Z[j]=0
\end{array}
\eqnum{201}
\end{equation}
and 
\begin{equation}
\begin{array}{c}
\{\frac \delta {\delta \eta _\gamma ^{*}(\tau _2)}(\frac d{d\tau _2}\frac 
\delta {\delta \eta _\delta (\tau _2)})-\theta _\delta \varepsilon _\delta 
\frac \delta {\delta \eta _\gamma ^{*}(\tau _2)\delta \eta _\delta (\tau _2)}%
-\sum\limits_{\sigma \lambda }\theta _\delta \theta _\sigma A(\sigma \delta
\lambda )\frac{\delta ^3}{\delta \eta _\gamma ^{*}(\tau _2)\delta \eta
_\sigma (\tau _2)\delta j_\lambda ^{*}(\tau _2)} \\ 
-\delta _{\gamma \delta }+\eta _\delta ^{*}(\tau _2)\frac \delta {\delta
\eta _\gamma ^{*}(\tau _2)}\}Z[j]=0.
\end{array}
\eqnum{202}
\end{equation}
Furthermore, by successively differentiating Eqs. (201) and (202) with
respect to the sources $\eta _\alpha ^{*}(\tau _1)$ and $\eta _\beta (\tau
_1)$, one obtains 
\begin{equation}
\begin{array}{c}
\{\frac{\delta ^2}{\delta \eta _\alpha ^{*}(\tau _1)\delta \eta _\beta (\tau
_1)}(\frac d{d\tau _2}\frac \delta {\delta \eta _\gamma ^{*}(\tau _2)})\frac 
\delta {\delta \eta _\delta (\tau _2)}+\theta _\gamma \varepsilon _\gamma 
\frac{\delta ^4}{\delta \eta _\alpha ^{*}(\tau _1)\delta \eta _\beta (\tau
_1)\delta \eta _\gamma ^{*}(\tau _2)\delta \eta _\delta (\tau _2)} \\ 
+\sum\limits_{\lambda \sigma }\theta _\gamma \theta _\lambda A(\gamma \rho
\lambda )\frac{\delta ^5}{\delta \eta _\alpha ^{*}(\tau _1)\delta \eta
_\beta (\tau _1)\delta \eta _\rho ^{*}(\tau _2)\delta \eta _\delta (\tau
_2)\delta j_\lambda ^{*}(\tau _2)}+\delta _{\gamma \delta }\frac{\delta ^2}{%
\delta \eta _\alpha ^{*}(\tau _2)\delta \eta _\beta (\tau _2)} \\ 
-\delta _{\beta \gamma }\delta (\tau _1-\tau _2)\frac{\delta ^2}{\delta \eta
_\alpha ^{*}(\tau _1)\delta \eta _\delta (\tau _2)}-\eta _\gamma (\tau _2)%
\frac{\delta ^3}{\delta \eta _\alpha ^{*}(\tau _1)\delta \eta _\beta (\tau
_1)\delta \eta _\delta (\tau _2)}\}Z[j]=0
\end{array}
\eqnum{203}
\end{equation}
and 
\begin{equation}
\begin{array}{c}
\{\frac{\delta ^3}{\delta \eta _\alpha ^{*}(\tau _1)\delta \eta _\beta (\tau
_1)\delta \eta _\gamma ^{*}(\tau _2)}(\frac d{d\tau _2}\frac \delta {\delta
\eta _\delta (\tau _2)})-\theta _\delta \varepsilon _\delta \frac{\delta ^4}{%
\delta \eta _\alpha ^{*}(\tau _1)\delta \eta _\beta (\tau _1)\delta \eta
_\gamma ^{*}(\tau _2)\delta \eta _\delta (\tau _2)} \\ 
-\sum\limits_{\sigma \lambda }\theta _\delta \theta _\sigma A(\sigma \delta
\lambda )\frac{\delta ^5}{\delta \eta _\alpha ^{*}(\tau _1)\delta \eta
_\beta (\tau _1)\delta \eta _\gamma ^{*}(\tau _2)\delta \eta _\sigma (\tau
_2)\delta j_\lambda ^{*}(\tau _2)}-\delta _{\gamma \delta }\frac{\delta ^2}{%
\delta \eta _\alpha ^{*}(\tau _2)\delta \eta _\beta (\tau _2)} \\ 
+\delta _{\alpha \delta }\delta (\tau _1-\tau _2)\frac{\delta ^2}{\delta
\eta _\gamma ^{*}(\tau _2)\delta \eta _\beta (\tau _1)}+\eta _\delta
^{*}(\tau _2)\frac{\delta ^3}{\delta \eta _\alpha ^{*}(\tau _1)\delta \eta
_\beta (\tau _1)\delta \eta _\gamma ^{*}(\tau _2)}\}Z[j]=0.
\end{array}
\eqnum{204}
\end{equation}

Let us sum up Eqs. (201) and (202) at first, then multiply the both sides of
the equation thus obtained with $-\theta _\gamma \theta _\delta $ and
finally set all the sources but the source $j_\lambda $ to vanish. By these
operations, we get 
\begin{equation}
\sum\limits_{\rho \sigma }\overline{H}(\gamma \delta ;\rho \sigma ;\tau
_2)^{j_\lambda }S_{\rho \sigma }^{j_\lambda }=0  \eqnum{205}
\end{equation}
where 
\begin{equation}
\overline{H}(\gamma \delta ;\rho \sigma ;\tau _2)^{j_\lambda }=(\frac d{%
d\tau _2}+\theta _\gamma \varepsilon _\gamma -\theta _\delta \varepsilon
_\delta )\delta _{\gamma \rho }\delta _{\delta \sigma }-\sum\limits_\lambda
f(\rho \sigma \lambda ;\gamma \delta )\frac \delta {\delta j_\lambda
^{*}(\tau _2)}  \eqnum{206}
\end{equation}
in which 
\begin{equation}
f(\rho \sigma \lambda ;\gamma \delta )=A(\sigma \delta \lambda )\delta
_{\gamma \rho }-A(\gamma \rho \lambda ))\delta _{\delta \sigma }=-f(\gamma
\delta ;\rho \sigma \lambda )  \eqnum{207}
\end{equation}
and $S_{\rho \sigma }^{j_\lambda }$ was defined in Eq. (167).

When we sum up Eqs. (203) and (204), then multiply the both sides of the
equation thus obtained with $\theta _\alpha \theta _\beta \theta _\gamma
\theta _\delta $ and finally set all the sources but the source $j_\lambda $
to be zero, according to the definitions in Eqs. (172) and (173), it is
found that 
\begin{equation}
\sum\limits_{\rho \sigma }\overline{H}(\gamma \delta ;\rho \sigma ;\tau
_2)^{j_\lambda }G(\alpha \beta ;\gamma \delta ;\tau _1-\tau _2)^{j_\lambda
}=\delta (\tau _1-\tau _2)[\delta _{\alpha \delta }S_{\gamma \beta }(\tau
_2-\tau _1)^{j_\lambda }-\delta _{\beta \gamma }S_{\alpha \delta }(\tau
_1-\tau _2)^{j_\lambda }].  \eqnum{208}
\end{equation}

In order to derive the equation of motion satisfied by the Green function $%
G(\lambda \tau \sigma ;\gamma \delta ;\tau _1-\tau _2)$ defined in Eq.
(183), we may take the derivative of Eq. (208) with respect to $j_\lambda
^{*}(\tau _1)$. In this way, we get 
\begin{equation}
\sum\limits_{\rho \sigma }\overline{H}(\gamma \delta ;\rho \sigma ;\tau
_2)^{j_\lambda }G(\alpha \beta \lambda ;\gamma \delta ;\tau _1-\tau
_2)^{j_\lambda }=\delta (\tau _1-\tau _2)[\delta _{\alpha \delta }\Lambda
(\gamma \beta \rho ;\tau _2-\tau _1)^{j_\lambda }-\delta _{\beta \gamma
}\Lambda (\alpha \delta \rho ;\tau _1-\tau _2)^{j_\lambda }]  \eqnum{209}
\end{equation}
where 
\begin{equation}
\Lambda (\gamma \beta \rho ;\tau _2-\tau _1)^{j_\lambda }=\frac \delta {%
\delta j_\lambda ^{*}(\tau _1)}S_{\gamma \beta }(\tau _2-\tau _1)^{j_\lambda
},  \eqnum{210}
\end{equation}
\begin{equation}
\Lambda (\alpha \delta \rho ;\tau _1-\tau _2)^{j_\lambda }=\frac \delta {%
\delta j_\lambda ^{*}(\tau _1)}S_{\alpha \delta }(\tau _1-\tau
_2)^{j_\lambda }  \eqnum{211}
\end{equation}
and 
\begin{equation}
G(\alpha \beta \lambda ;\gamma \delta ;\tau _1-\tau _2)=\frac \delta {\delta
j_\lambda ^{*}(\tau _1)}G(\alpha \beta ;\gamma \delta ;\tau _1-\tau
_2)^{j_\lambda }.  \eqnum{212}
\end{equation}

Acting on Eqs. (155) and (182) with the operator $\overline{H}(\gamma \delta
;\rho \sigma ;\tau _2)$ and employing Eq. (205), we find 
\begin{equation}
\sum\limits_{\rho \sigma }\overline{H}(\gamma \delta ;\rho \sigma ;\tau
_2)^{j_\lambda }{\cal G}(\alpha \beta ;\rho \sigma ;\tau _1-\tau
_2)^{j_\lambda }=\sum\limits_{\rho \sigma }\overline{H}(\gamma \delta ;\rho
\sigma ;\tau _2)^{j_\lambda }G(\alpha \beta ;\rho \sigma ;\tau _1-\tau
_2)^{j_\lambda }  \eqnum{213}
\end{equation}
and 
\begin{equation}
\sum\limits_{\rho \sigma }\overline{H}(\gamma \delta ;\rho \sigma ;\tau
_2)^{j_\lambda }{\cal G}(\alpha \beta \lambda ;\rho \sigma ;\tau _1-\tau
_2)^{j_\lambda }=\sum\limits_{\rho \sigma }\overline{H}(\gamma \delta ;\rho
\sigma ;\tau _2)^{j_\lambda }G(\alpha \beta \lambda ;\rho \sigma ;\tau
_1-\tau _2)^{j_\lambda }.  \eqnum{214}
\end{equation}
The above two equalities further indicate that the equations of motion
satisfied by the Green functions ${\cal G}(\alpha \beta ;\rho \sigma ;\tau
_1-\tau _2)^{j_\lambda }$ and ${\cal G}(\alpha \beta \lambda ;\rho \sigma
;\tau _1-\tau _2)^{j_\lambda }$ are formally the same as those for the
ordinary Green functions $G(\alpha \beta ;\rho \sigma ;\tau _1-\tau
_2)^{j_\lambda }$ and $G(\alpha \beta \lambda ;\rho \sigma ;\tau _1-\tau
_2)^{j_\lambda }$ respectively. Upon inserting Eqs. (208) into Eq. (213) and
Eq. (209) into Eq. (214) and turning off the source $j_\lambda $, noticing
the definition in Eq. (206), we derive the following equations 
\begin{equation}
\begin{array}{c}
(\frac d{d\tau _2}+\theta _\gamma \varepsilon _\gamma -\theta _\delta
\varepsilon _\delta ){\cal G}(\alpha \beta ;\gamma \delta ;\tau _1-\tau
_2)=\delta (\tau _1-\tau _2)[\delta _{\alpha \delta }S_{\gamma \beta }(\tau
_2-\tau _1) \\ 
-\delta _{\beta \gamma }S_{\alpha \delta }(\tau _1-\tau
_2)]+\sum\limits_{\lambda \tau \sigma }{\cal G}(\alpha \beta ;\lambda \tau
\sigma ;\tau _1-\tau _2)f(\lambda \tau \sigma ;\gamma \delta )
\end{array}
\eqnum{215}
\end{equation}
and 
\begin{equation}
\begin{array}{c}
(\frac d{d\tau _2}+\theta _\gamma \varepsilon _\gamma -\theta _\delta
\varepsilon _\delta ){\cal G}(\alpha \beta \rho ;\gamma \delta ;\tau _1-\tau
_2)=\delta (\tau _1-\tau _2)[\delta _{\alpha \delta }\Lambda (\gamma \beta
\rho ;\tau _2-\tau _1) \\ 
-\delta _{\beta \gamma }\Lambda (\alpha \delta \rho ;\tau _1-\tau
_2)]+\sum\limits_{\lambda \tau \sigma }{\cal G}(\alpha \beta \rho ;\lambda
\tau \sigma ;\tau _1-\tau _2)f(\lambda \tau \sigma ;\gamma \delta )
\end{array}
\eqnum{216}
\end{equation}
where some indices have been changed for convenience, 
\begin{equation}
\begin{array}{c}
\Lambda (\gamma \beta \rho ;\tau _2-\tau _1)=\left\langle T[\widehat{b}%
_\gamma (\tau _2)\widehat{b}_\beta ^{+}(\tau _1)\widehat{a}_\rho (\tau
_1)]\right\rangle _\beta \\ 
\Lambda (\alpha \delta \rho ;\tau _1-\tau _2)=\left\langle T[\widehat{b}%
_\alpha (\tau _1)\widehat{b}_\delta ^{+}(\tau _2)\widehat{a}_\rho (\tau
_1)]\right\rangle _\beta
\end{array}
\eqnum{217}
\end{equation}
which are given by Eqs. (210) and (211) with setting $j_\lambda =0$ and 
\begin{equation}
\begin{array}{c}
{\cal G}(\lambda \tau \rho ;\gamma \delta \sigma ;\tau _1-\tau _2)=\frac{%
\delta ^2}{\delta j_\rho ^{*}(\tau _1)\delta j_\sigma ^{*}(\tau _2)}{\cal G}%
(\lambda \tau ;\gamma \delta ;\tau _1-\tau _2)^{j_\lambda }\mid _{j_\lambda
=0} \\ 
=\left\langle T\{N[\widehat{b}_\lambda (\tau _1)\widehat{b}_\tau ^{+}(\tau
_1)\widehat{a}_\rho (\tau _1)]N[\widehat{b}_\gamma (\tau _2)\widehat{b}%
_\delta ^{+}(\tau _2)\widehat{a}_\sigma (\tau _2)]\}\right\rangle _\beta
\end{array}
\eqnum{218}
\end{equation}
is the six-point Green function including two gluon operators in it.
According to the definition in Eq. (180), we have 
\begin{equation}
{\cal G}(\lambda \tau \rho ;\gamma \delta \sigma ;\tau _1-\tau _2)=G(\lambda
\tau \rho ;\gamma \delta \sigma ;\tau _1-\tau _2)-\Lambda (\lambda \tau \rho
)\Lambda (\gamma \delta \sigma )  \eqnum{219}
\end{equation}
where 
\begin{equation}
G(\lambda \tau \rho ;\gamma \delta \sigma ;\tau _1-\tau _2)=\langle T[%
\widehat{b}_\lambda (\tau _1)\widehat{b}_\tau ^{+}(\tau _1)\widehat{a}_\rho
(\tau _1)\widehat{b}_\gamma (\tau _2)\widehat{b}_\delta ^{+}(\tau _2)%
\widehat{a}_\sigma (\tau _2)]\rangle _\beta  \eqnum{220}
\end{equation}
is the ordinary six-point Green function. It should be noted that due to the
restriction of the delta function, the terms in the brackets on the right
hand sides of Eqs. (215) and (216) actually are ''time''-independent.

It is easy to see that the Green functions ${\cal G}(\alpha \beta ;\lambda
\tau \sigma ;\tau _1-\tau _2)$ and ${\cal G}(\lambda \tau \rho ;\gamma
\delta \sigma ;\tau _1-\tau _2)$, as the Green functions ${\cal G}(\alpha
\beta ;\gamma \delta ;\tau _1-\tau _2)$ and ${\cal G}(\alpha \beta \rho
;\gamma \delta ;\tau _1-\tau _2)$, are periodic. Therefore, by the Fourier
transformation, i.e. by the integration $\int_0^\beta d\tau e^{i\omega
_n\tau }$, noticing $d/d\tau _2=-d/d\tau $, Eqs. (215) and (216) will be
transformed to 
\begin{equation}
\begin{array}{c}
(i\omega _n+\theta _\gamma \varepsilon _\gamma -\theta _\delta \varepsilon
_\delta ){\cal G}(\alpha \beta ;\gamma \delta ;\omega _n)=S(\alpha \beta
;\gamma \delta ) \\ 
+\sum\limits_{\lambda \tau \sigma }{\cal G}(\alpha \beta ;\lambda \tau
\sigma ;\omega _n)f(\lambda \tau \sigma ;\gamma \delta )
\end{array}
\eqnum{221}
\end{equation}
where $S(\alpha \beta ;\gamma \delta )$ was defined in Eq. (179) and 
\begin{equation}
\begin{array}{c}
(i\omega _n+\theta _\gamma \varepsilon _\gamma -\theta _\delta \varepsilon
_\delta ){\cal G}(\alpha \beta \rho ;\gamma \delta ;\omega _n)=R(\alpha
\beta \rho ;\gamma \delta ) \\ 
+\sum\limits_{\lambda \tau \sigma }{\cal G}(\alpha \beta \rho ;\lambda \tau
\sigma ;\omega _n)f(\lambda \tau \sigma ;\gamma \delta )
\end{array}
\eqnum{222}
\end{equation}
where 
\begin{equation}
\begin{array}{c}
R(\alpha \beta \rho ;\gamma \delta )=\delta _{\alpha \delta }\Lambda (\gamma
\beta \rho )-\delta _{\beta \gamma }\Lambda (\alpha \delta \rho ) \\ 
=\langle [\widehat{b}_\alpha \widehat{b}_\beta ^{+},\widehat{b}_\gamma 
\widehat{b}_\delta ^{+}]_{-}\widehat{a}_\rho \rangle _\beta
\end{array}
\eqnum{223}
\end{equation}
which is ''time''-independent.

Now we are ready to derive the interaction kernel. Acting on the both sides
of Eq. (186) with $(i\omega _n+\theta _\gamma \varepsilon _\gamma -\theta
_\delta \varepsilon _\delta )$ and using Eqs. (221) and (222), one gets 
\begin{equation}
\begin{array}{c}
\sum\limits_{\mu \nu }K(\alpha \beta ;\mu \nu ;\omega _n)S(\mu \nu ;\gamma
\delta )=\sum\limits_{\lambda \tau \rho }f(\alpha \beta ;\lambda \tau \rho
)R(\lambda \tau \rho ;\gamma \delta )+\sum\limits_{\lambda \tau \rho
}\sum\limits_{\xi \eta \sigma }f(\alpha \beta ;\lambda \tau \rho ) \\ 
\times {\cal G}(\lambda \tau \rho ;\xi \eta \sigma ;\omega _n)f(\xi \eta
\sigma ;\gamma \delta )-\sum\limits_{\mu \nu }\sum\limits_{\xi \eta \sigma
}K(\alpha \beta ;\mu \nu ;\omega _n){\cal G}(\alpha \beta ;\xi \eta \sigma
;\omega _n)f(\lambda \tau \sigma ;\gamma \delta ).
\end{array}
\eqnum{224}
\end{equation}
Operating on the both sides of Eq. (186) with the inverse of ${\cal G}(\mu
\nu ;\gamma \delta ;\omega _n)$, we have 
\begin{equation}
K(\alpha \beta ;\gamma \delta ;\omega _n)=\sum\limits_{\gamma \delta
}\sum\limits_{\lambda \tau \sigma }f(\alpha \beta ;\lambda \tau \rho ){\cal G%
}(\lambda \tau \rho ;\mu \nu ;\omega _n){\cal G}^{-1}(\mu \nu ;\gamma \delta
;\omega _n).  \eqnum{225}
\end{equation}
Upon substituting Eq. (225) onto the right hand side of Eq. (224) and acting
on Eq. (224) with the inverse $S^{-1}(\mu \nu ;\gamma \delta ),$ we
eventually arrive at 
\begin{equation}
\begin{array}{c}
K(\alpha \beta ;\gamma \delta ;E)=\sum\limits_{\mu \nu
}\{\sum\limits_{\lambda \tau \rho }f(\alpha \beta ;\lambda \tau \rho
)R(\lambda \tau \rho ;\mu \nu )+\sum\limits_{\lambda \tau \rho
}\sum\limits_{\xi \eta \sigma }f(\alpha \beta ;\lambda \tau \rho ){\cal G}%
(\lambda \tau \rho ;\xi \eta \sigma ;E)f(\xi \eta \sigma ;\mu \nu ) \\ 
-\sum\limits_{\lambda \tau \rho }\sum\limits_{\xi \eta \sigma
}\sum\limits_{\kappa \varsigma }\sum\limits_{\pi \theta }f(\alpha \beta
;\lambda \tau \rho ){\cal G}(\lambda \tau \rho ;\kappa \varsigma ;E){\cal G}%
^{-1}(\kappa \varsigma ;\pi \theta ;E){\cal G}(\pi \theta ;\xi \eta \sigma
;E)f(\xi \eta \sigma ;\mu \nu )\}S^{-1}(\mu \nu ;\gamma \delta )
\end{array}
\eqnum{226}
\end{equation}
where $\omega _n$ has been replaced by $E$. This just is the wanted closed
expression of the interaction kernel appearing in Eq. (192). In accordance
with Eq. (186), the last term in Eq. (226) can be written in the form 
\begin{equation}
\sum\limits_{\rho \sigma }\sum\limits_{\xi \eta }\sum\limits_{\mu \nu
}K(\alpha \beta ;\rho \sigma ;E){\cal G}(\rho \sigma ;\xi \eta ;E)K(\xi \eta
;\mu \nu ;E)S^{-1}(\mu \nu ;\gamma \delta )  \eqnum{227}
\end{equation}
which exhibits a typical B-S reducible structure [17]. Therefore, the last
term in Eq. (226) plays the role of cancelling the B-S reducible part
contained in the other terms in Eq. (226) to make the kernel to be B-S
irreducible. If we use the above expression in place of the last term in Eq.
(226) and acting on Eq. (226) with $S(\gamma \delta ;\mu \nu )$, we obtain
from Eq. (226) an integral equation satisfied by the kernel $K(\alpha \beta
;\gamma \delta ;E)$. Define 
\begin{equation}
{\cal R}(\alpha \beta ;\gamma \delta )=\sum\limits_{\lambda \tau \rho
}f(\alpha \beta ;\lambda \tau \rho )R(\lambda \tau \rho ;\gamma \delta ) 
\eqnum{228}
\end{equation}
and 
\begin{equation}
{\cal Q}(\alpha \beta ;\gamma \delta )=\sum\limits_{\lambda \tau \rho
}\sum\limits_{\xi \eta \sigma }f(\alpha \beta ;\lambda \tau \rho ){\cal G}%
(\lambda \tau \rho ;\xi \eta \sigma ;E)f(\xi \eta \sigma ;\gamma \delta ), 
\eqnum{229}
\end{equation}
the integral equation can be written in the matrix form as follows 
\begin{equation}
KS={\cal R}+{\cal Q}-K{\cal G}K.  \eqnum{230}
\end{equation}

For comparison with the kernel in Eq. (226) and for convenience of
nonperturbative investigations, we would like to show the corresponding
closed expression given in the position space without giving derivation.
This kernel can be obtained from the kernel in Eq. (226) by making use of
the inverse of the Fourier transformations written in sect. 3 or derived
from the generating functional represented in the position space (see
Appendix ) by completely following the procedure as described in this
section. The kernel is represented as follows:

\begin{equation}
\begin{tabular}{l}
$K(\vec x_{1,}\vec x_2;\vec y_1,\vec y_2;E)=\int d^3z_1d^3z_2\{{\cal R}(\vec 
x_{1,}\vec x_2;\vec z_1,\vec z_2)$ \\ 
$+{\cal Q}(\vec x_{1,}\vec x_2;\vec z_1,\vec z_2;E)-{\cal D}(\vec x_{1,}\vec 
x_2;\vec z_1,\vec z_2;E)\}S^{-1}(\vec z_1,\vec z_2;\vec y_1,\vec y_2)$%
\end{tabular}
\eqnum{231}
\end{equation}
where ${\cal R}(\vec x_{1,}\vec x_2;\vec z_1,\vec z_2),{\cal Q}(\vec x_1,%
\vec x_2;\vec z,,\vec z_2;E)$ and ${\cal D}(\vec x_1,\vec x_2;\vec z_1,\vec z%
_2;E)$ are separately described below. 

The function ${\cal R}(\vec x_{1,}\vec x_2;\vec z_1,\vec z_2)$ can be
represented as  
\begin{equation}
{\cal R}(\vec x_1,\vec x_2;\vec z_1,\vec z_2)=\sum_{i=1}^2\Omega _i^{a\mu }%
{\cal R}_\mu ^{(i)a}(\vec x_{1,}\vec x_2;\vec z_1,\vec z_2)  \eqnum{232}
\end{equation}
in which 
\begin{equation}
\Omega _1^{a\mu }=ig\gamma _1^4\gamma _1^\mu T_1^a,\text{ }\Omega _2^{b\nu
}=ig\gamma _2^4\gamma _2^\nu \overline{T}_2^b  \eqnum{233}
\end{equation}
with $T_1^a=\lambda ^a/2$ and $\overline{T}_2^b=-\lambda ^{a*}/2$ being the
quark and antiquark color matrices respectively and 
\begin{equation}
{\cal R}_\mu ^{(i)a}(\vec x_1,\vec x_2;\vec z_1,\vec z_2)=\delta ^3(\vec x_1-%
\vec z_1)\gamma _1^4\Lambda _\mu ^{{\bf c}a}(\vec x_i\mid \vec x_2,\vec z%
_2)+\delta ^3(\vec x_2-\vec z_2)\gamma _2^4\Lambda _\mu ^a(\vec x_i\mid \vec 
x_1,\vec z_1)  \eqnum{234}
\end{equation}
here $\Lambda _\mu ^a(\vec x_i\mid \vec x_1,\vec y_1)$ and $\Lambda _\mu ^{%
{\bf c}a}(\vec x_i\mid \vec x_2,\vec y_2)$ are defined as 
\begin{equation}
\begin{array}{c}
\Lambda _\mu ^a(\vec x_i\mid \vec x_1,\vec y_1)=\langle T[{\bf A}_\mu ^a(%
\vec x_i,\tau _1){\bf \psi }(\vec x_1,\tau _1)\overline{{\bf \psi }}(\vec y%
_1,\tau _1)]\rangle _\beta , \\ 
\Lambda _\mu ^{{\bf c}a}(\vec x_i\mid \vec x_2,\vec y_2)=\langle T[{\bf A}%
_\mu ^a(\vec x_i,\tau _1){\bf \psi }^c(\vec x_2,\tau _1)\overline{{\bf \psi }%
}^c(\vec y_2,\tau _1)]\rangle _\beta 
\end{array}
\eqnum{235}
\end{equation}
which are time-independent due to the translation-invariance property of the
Green functions. 

The function is of the form
\begin{equation}
{\cal Q}(\vec x_1,\vec x_2;\vec z_1,\vec z_2;E)=\sum_{i,j=1}^2\Omega
_i^{a\mu }{\cal G}_{\mu \nu }^{ab}(\vec x_i,\vec z_j\mid \vec x_1,\vec x_2;%
\vec z_1,\vec z_2;E)\overline{\Omega }_j^{b\nu }  \eqnum{236}
\end{equation}
in which 
\begin{equation}
\overline{\Omega }_1^{a\mu }=ig\gamma _1^\mu \gamma _1^4T_1^a,\text{ }%
\overline{\Omega }_2^{a\mu }=ig\gamma _2^\mu \gamma _2^4\overline{T}_2^a, 
\eqnum{237}
\end{equation}
${\cal G}_{\mu \nu }^{ab}(\vec x_i,\vec z_j\mid \vec x_1,\vec x_2;\vec z_1,%
\vec z_2;E)$ is the Fourier transform of the Green function defined by 
\begin{equation}
\begin{tabular}{l}
${\cal G}_{\mu \nu }^{ab}(\vec x_i,\vec z_j\mid \vec x_1,\vec x_2;\vec z_1,%
\vec z_2;\tau _1-\tau _2)$ \\ 
$=\langle T\{N[{\bf A}_\mu ^a(\vec x_i,\tau _1){\bf \psi }(\vec x_1,\tau _1)%
{\bf \psi }^c(\vec x_2,\tau _1)]N[{\bf A}_\nu ^b(\vec z_j,\tau _2)\overline{%
{\bf \psi }}(\vec z_1,\tau _2)\overline{{\bf \psi }}^c(\vec z_2,\tau
_2)]\}\rangle _\beta $%
\end{tabular}
\eqnum{238}
\end{equation}

The function ${\cal D}(\vec x_1,\vec x_2;\vec z_1,\vec z_2;E)$ is expressed
by  
\begin{equation}
\begin{tabular}{l}
${\cal D}(\vec x_1,\vec x_2;\vec z_1,\vec z_2;E)=\int
\prod\limits_{k=1}^2d^3u_kd^3v_k\sum\limits_{i,j=1}^2\Omega _i^{a\mu }{\cal G%
}_\mu ^{(i)a}(\vec x_i\mid \vec x_1,\vec x_2;\vec u_1,\vec u_2;E)$ \\ 
$\times {\cal G}^{-1}(\vec u_1,\vec u_2;\vec v_1,\vec v_2;E){\cal G}_\nu
^{(j)b}(\vec z_j\mid \vec v_1,\vec v_2;\vec z_1,\vec z_2;E)\overline{\Omega }%
_j^{b\nu }$%
\end{tabular}
\eqnum{239}
\end{equation}
in which ${\cal G}_\mu ^{(i)a}(\vec x_i\mid \vec x_1,\vec x_2;\vec u_1,\vec u%
_2;E)$ and ${\cal G}_\nu ^{(j)b}(\vec z_j\mid \vec v_1,\vec v_2;\vec z_1,%
\vec z_2;E)$ are the Fourier transforms of the following Green functions 
\begin{equation}
\begin{tabular}{l}
${\cal G}_\mu ^{(i)a}(\vec x_i\mid \vec x_1,\vec x_2;\vec u_1,\vec u_2;\tau
_1-\tau _2)$ \\ 
$=\langle T\{N[{\bf A}_\mu ^a(\vec x_i,\tau _1){\bf \psi }(\vec x_1,\tau _1)%
{\bf \psi }^c(\vec x_2,\tau _1)]N[\overline{{\bf \psi }}(\vec u_1,\tau _2)%
\overline{{\bf \psi }}^c(\vec u_2,\tau _2)]\}\rangle _\beta $%
\end{tabular}
\eqnum{240}
\end{equation}
and 
\begin{equation}
\begin{tabular}{l}
${\cal G}_\nu ^{(j)b}(\vec z_j\mid \vec v_1,\vec v_2;\vec z_1,\vec z_2;\tau
_1-\tau _2)$ \\ 
$=\langle T\{N[{\bf \psi }(\vec v_1,\tau _1){\bf \psi }^c(\vec v_2,\tau
_1)]N[{\bf A}_\nu ^b(\vec z_j,\tau _2)\overline{{\bf \psi }}(\vec z_1,\tau
_2)\overline{{\bf \psi }}^c(\vec z_2,\tau _2)]\}\rangle _\beta $%
\end{tabular}
\eqnum{241}
\end{equation}
The $S^{-1}(\vec z_1,\vec z_2;\vec y_1,\vec y_2)$ in Eq. (231) is the
inverse of the function defined by 
\begin{equation}
S(\vec x_1,\vec x_2;\vec z_1,\vec z_2)=\delta ^3(\vec x_1-\vec z_1)\gamma
_1^4S_F^c(\vec x_2-\vec z_2)+\delta ^3(\vec x_2-\vec z_2)\gamma _2^4S_F(\vec 
x_1-\vec z_1)  \eqnum{242}
\end{equation}
in which $S_F(\vec x_1-\vec z_1)$ and $S_F^c(\vec x_2-\vec z_2)$ are the
equal-time quark and antiquark thermal propagators respectively. It is clear
that there is one-to-one correspondence between the both kernels written in
Eqs. (226) and (231). It is noted that the interaction kernel derived in
this section is nonperturbative because the Green functions included in the
kernel are defined in the Heisenberg picture. Perturbative calculations of
the kernel can easily be done by using the familiar perturbative expansions
of Green functions as illustrated in the next section.

\section{One gluon exchange kernel and Hamiltonian}

In this section, we would like to show the one-gluon exchange kernel given
by the expression in Eq. (226). In the lowest order approximation of
perturbation, only the first term of the series in Eq. (194), i.e., the
kernel $K(\alpha ^{+}\beta ^{-};\gamma ^{+}\delta ^{-};E)$ represented in
Eq. (195) is needed to be taken into account in Eqs. (192) and (193). In the
lowest order approximation, as seen from Eq. (223), the first term in Eq.
(226) vanishes because the function $R(\lambda \tau \rho ;\gamma ^{+}\delta
^{-})$ gives no contribution to the kernel owing to the expectation value $%
\left\langle n\left| \widehat{a}_\rho \right| n\right\rangle $ vanishes.
Therefore, the one-gluon exchange kernel can only arises from the second
term in Eq. (226) where the Green function ${\cal G}(\lambda \tau \rho ;\xi
\eta \sigma ;E)$ reduces to the ordinary one $G(\lambda \tau \rho ;\xi \eta
\sigma ;E)$ in the lowest order approximation because the last term in Eq.
(226) vanishes in this case for the reason as argued for the function $%
R(\lambda \tau \rho ;\gamma ^{+}\delta ^{-})$. To evaluate the inverse $%
S^{-1}(\mu \nu ;\gamma ^{+}\delta ^{-})$, we first evaluate $S(\gamma
^{+}\delta ^{-};\mu \nu )$. From the expression denoted in Eq. (179), it is
easy to find that the nonvanishing contribution of $S(\gamma ^{+}\delta
^{-};\mu \nu )$ is given by 
\begin{equation}
\begin{array}{c}
S(\gamma ^{+}\delta ^{-};\mu ^{-}\nu ^{+})=\delta _{\gamma ^{+}\nu
^{+}}S_{\mu ^{-}\delta ^{-}}-\delta _{\mu ^{-}\delta ^{-}}S_{\gamma ^{+}\nu
^{+}} \\ 
=\delta _{_{\gamma ^{+}\nu ^{+}}}\delta _{\mu ^{-}\delta ^{-}}-\delta _{\mu
^{-}\delta ^{-}}S_{\gamma ^{+}\nu ^{+}}-\delta _{\gamma ^{+}\nu ^{+}}%
\overline{S}_{\delta ^{-}\mu ^{-}}
\end{array}
\eqnum{243}
\end{equation}
where $S_{\gamma ^{+}\nu ^{+}}$ and $\overline{S}_{\delta ^{-}\mu ^{-}}$ are
the quark and antiquark equal-time propagators. Their expressions, according
to the common definition, can be read from Eqs. (139) and (140) by setting $%
\tau _1-\tau _2\rightarrow 0^{+}$, 
\begin{equation}
\begin{array}{c}
S_{\gamma ^{+}\nu ^{+}}=\delta _{\gamma ^{+}\nu ^{+}}\Delta _q(\vec q%
_1,0^{+})=\delta _{\gamma ^{+}\nu ^{+}}\frac 12[\overline{n}_f(\vec q_1)-n_f(%
\vec q_1)], \\ 
\overline{S}_{\delta ^{-}\mu ^{-}}=\delta _{\delta ^{-}\mu ^{-}}\Delta _q(%
\vec q_2,0^{+})=\delta _{\delta ^{-}\mu ^{-}}\frac 12[\overline{n}_f(\vec q%
_2)-n_f(\vec q_2)].
\end{array}
\eqnum{244}
\end{equation}
Therefore, we can write 
\begin{equation}
S^{-1}(\mu ^{-}\nu ^{+};\gamma ^{+}\delta ^{-})=\delta _{_{\gamma ^{+}\nu
^{+}}}\delta _{\mu ^{-}\delta ^{-}}{\cal S}^{-1}(\gamma \delta )  \eqnum{245}
\end{equation}
where 
\begin{equation}
{\cal S}(\gamma \delta )=1-\frac 12[\overline{n}_f(\vec q_1)-n_f(\vec q_1)+%
\overline{n}_f(\vec q_2)-n_f(\vec q_2)]  \eqnum{246}
\end{equation}
Thus, to derive the lowest order approximate kernel, we only need to
consider 
\begin{equation}
K(\alpha ^{+}\beta ^{-};\gamma ^{+}\delta ^{-};E)=\sum\limits_{\lambda \tau
\rho }\sum\limits_{\xi \zeta \sigma }f(\alpha ^{+}\beta ^{-};\lambda \tau
\rho )G(\lambda \tau \rho ;\mu \nu \sigma ;E)f(\mu \nu \sigma ;\delta
^{-}\gamma ^{+}){\cal S}^{-1}(\gamma \delta ).  \eqnum{247}
\end{equation}
From Eqs. (169), (207), (80) and (67), it is clearly seen that when $\lambda
,$ $\tau $, $\mu $ and $\nu $ take $\lambda ^{+},$ $\tau ^{-}$, $\mu ^{-}$
and $\nu ^{+},$ the functions $f(\alpha ^{+}\beta ^{-};\lambda \tau \rho )$
and $f(\mu \nu \sigma ;\delta ^{-}\gamma ^{+})$ give the quark-antiquark
interaction taking place in the t-channel scattering process; while, when $%
\lambda ,$ $\tau $, $\mu $ and $\nu $ take $\lambda ^{-},$ $\tau ^{+}$, $\mu
^{+}$ and $\nu ^{-},$ the $f(\alpha ^{+}\beta ^{-};\lambda \tau \rho )$ and $%
f(\mu \nu \sigma ;\delta ^{-}\gamma ^{+})$ will give the quark-antiquark
vertices which describe the $q\overline{q}$ annihilation process. Since the
expectation value of the $q\overline{q}$ color matrix appearing in the $q%
\overline{q}$ lowest order annihilation process is zero in the color
singlet, it is only necessary to consider the following interaction kernel 
\begin{equation}
\begin{array}{c}
K(\alpha ^{+}\beta ^{-};\gamma ^{+}\delta ^{-};E)=\sum\limits_{\mu \nu
}\sum\limits_{\lambda \tau \rho }\sum\limits_{\xi \eta \sigma }[f(\alpha
^{+}\beta ^{-};\lambda ^{+}\tau ^{-}\rho ^{+})G(\lambda ^{+}\tau ^{-}\rho
^{+};\mu ^{-}\nu ^{+}\sigma ^{-};E)f(\mu ^{-}\nu ^{+}\sigma ^{-};\delta
^{-}\gamma ^{+}) \\ 
+f(\alpha ^{+}\beta ^{-};\lambda ^{+}\tau ^{-}\rho ^{-})G(\lambda ^{+}\tau
^{-}\rho ^{-};\mu ^{-}\nu ^{+}\sigma ^{+};E)f(\mu ^{-}\nu ^{+}\sigma
^{+};\delta ^{-}\gamma ^{+})]{\cal S}^{-1}(\gamma \delta ).
\end{array}
\eqnum{248}
\end{equation}
Noting that the functions $f(\alpha ^{+}\beta ^{-};\lambda ^{+}\tau ^{-}\rho
)$ and $f(\mu ^{-}\nu ^{+}\sigma ;\delta ^{-}\gamma ^{+})$ are proportional
to the coupling constant $g$, in the approximation of order $g^2$, we only
need to consider the zero-order of the Green function $G(\lambda ^{+}\tau
^{-}\rho ^{\pm };\mu ^{-}\nu ^{+}\sigma ^{\mp };E)$. This Green function may
easily be derived from the generating functional $Z^0[j]$ represented in
Eqs. (96), (119), (138) and (150). The result is 
\begin{equation}
\begin{array}{c}
G(\lambda ^{+}\tau ^{-}\rho ^{-};\mu ^{-}\nu ^{+}\sigma ^{+};\tau _1-\tau
_2)=\frac 1{Z^0}\frac{\delta ^6Z^0[j\}}{\delta \eta _\lambda ^{*}(\tau
_1)\delta \overline{\eta }_\tau ^{*}(\tau _1)\delta \overline{\eta }_\mu
(\tau _2)\delta \eta _\nu (\tau _2)\delta \xi _\rho ^{*}(\tau _1)\delta \xi
_\sigma (\tau _2)}\mid _{j=0} \\ 
=\delta _{\lambda \nu }\Delta _q(\vec k_\lambda ,\tau _1-\tau _2)\delta
_{\tau \mu }\Delta _q(\vec k_\tau ,\tau _1-\tau _2)\delta _{\rho \sigma
}\Delta _g(\vec k_\rho ,\tau _1-\tau _2) \\ 
=G(\lambda ^{+}\tau ^{-}\rho ^{+};\mu ^{-}\nu ^{+}\sigma ^{-};\tau _1-\tau
_2)
\end{array}
\eqnum{249}
\end{equation}
The last equality in the above arises from the fact that $\Delta _g(\vec k%
_\rho ,\tau _1-\tau _2)$ is an even function of $\tau _1-\tau _2$. With the
expressions given in Eqs. (124) and (140), the Fourier transform of $%
G(\lambda ^{+}\tau ^{-}\rho ^{-};\mu ^{-}\nu ^{+}\sigma ^{+};\tau _1-\tau
_2) $ can be found to be 
\begin{equation}
G(\lambda ^{+}\tau ^{-}\rho ^{-};\mu ^{-}\nu ^{+}\sigma ^{+};\omega
_n)=\delta _{\lambda \nu }\delta _{\tau \mu }\delta _{\rho \sigma }\frac 18%
\Delta (\lambda \tau \rho )  \eqnum{250}
\end{equation}
where $\Delta (\lambda \tau \rho )$ will be specified soon later.

Substitution of Eqs. (169), (207) and (250) in Eq. (248) leads to 
\begin{equation}
\begin{array}{c}
K(\alpha ^{+}\beta ^{-};\gamma ^{+}\delta ^{-};E)=\frac 14%
[\sum\limits_{\lambda \rho }A(\alpha ^{+}\lambda ^{+}\rho ^{+})A(\lambda
^{+}\gamma ^{+}\rho ^{-})\Delta (\lambda \beta \rho )\delta _{\beta \delta }
\\ 
+\sum\limits_{\tau \rho }A(\tau ^{-}\beta ^{-}\rho ^{+})A(\delta ^{-}\tau
^{-}\rho ^{-})\Delta (\alpha \tau \rho )\delta _{\alpha \gamma
}-\sum\limits_\rho A(\alpha ^{+}\gamma ^{+}\rho ^{+})A(\delta ^{-}\beta
^{-}\rho ^{-})\Delta (\gamma \beta \rho ) \\ 
-\sum\limits_\rho A(\delta ^{-}\beta ^{-}\rho ^{+})A(\alpha ^{+}\gamma
^{+}\rho ^{-})\Delta (\alpha \delta \rho )]{\cal S}^{-1}(\gamma \delta ).
\end{array}
\eqnum{251}
\end{equation}
where the first two terms on the right hand side are unconnected and
represent the quark and antiquark self-energies, while the remaining two
terms precisely give the t-channel one-gluon exchange kernel. In view of the
expression in Eq. (80) and the definitions in Eqs. (67) and (71), we can
write 
\begin{equation}
A(\alpha ^{+}\gamma ^{+}\rho ^{\pm })=ig(2\pi )^3\delta ^3(\vec p_1-\vec q%
_1\mp \vec k)\overline{u}_{\sigma _1}(\vec p_1)T^c\gamma _\mu u_{\sigma
_1^{\prime }}(\vec q_1)(2\pi )^{-3/2}(2\omega (\vec k))^{-1/2}\varepsilon
_\mu ^\lambda (\vec k)  \eqnum{252}
\end{equation}
and 
\begin{equation}
\begin{array}{c}
A(\delta ^{-}\beta ^{-}\rho ^{\mp })=ig(2\pi )^3\delta ^3(\vec q_2-\vec p%
_2\pm \vec k)\overline{v}_{\sigma _2^{\prime }}(\vec q_2)T^c\gamma _\mu
v_{\sigma _2}(\vec p_2)(2\pi )^{-3/2}(2\omega (\vec k))^{-1/2}\varepsilon
_\mu ^\lambda (\vec k) \\ 
=-ig(2\pi )^3\delta ^3(\vec q_2-\vec p_2\pm \vec k)\overline{u}_{\sigma _2}(%
\overrightarrow{p_2})(-T^{c*})\gamma _\nu u_{\sigma _2^{\prime }}(\vec q%
_2)(2\pi )^{-3/2}(2\omega (\vec k))^{-1/2}\varepsilon _\nu ^{\lambda
^{\prime }}(\vec k)
\end{array}
\eqnum{253}
\end{equation}
where the last equality in Eq. (253) is obtained by the charge conjugation
transformation. In the above, we have set $\vec k_\alpha =\vec p_1$, $\vec k%
_\beta =\vec p_2$, $\vec k_\gamma =\vec q_1$, $\vec k_\delta =\vec q_2$ and $%
\vec k_\rho =$ $\vec k$. From Eqs. (249), (250), (124) and (140), it is easy
to get 
\begin{equation}
\begin{array}{c}
\Delta (\alpha \delta \rho )=\frac 12\int_{-\beta }^\beta d\tau e^{i\omega
_n\tau }\Delta _q(\vec p_1,\tau )\Delta _q(\vec q_2,\tau )\Delta _g(\vec k%
,\tau ) \\ 
=\overline{n}_f^\alpha \overline{n}_f^\delta \overline{n}_b^\rho \frac{%
e^{-\beta (\varepsilon _\alpha +\varepsilon _\delta +\omega _\rho )}-1}{%
i\omega _n-\varepsilon _\alpha -\varepsilon _\delta -\omega _\rho }-%
\overline{n}_f^\alpha \overline{n}_f^\delta n_b^\rho \frac{e^{-\beta
(\varepsilon _\alpha +\varepsilon _\delta -\omega _\rho )}-1}{i\omega
_n-\varepsilon _\alpha -\varepsilon _\delta +\omega _\rho } \\ 
-\overline{n}_f^\alpha n_f^\delta \overline{n}_b^\rho \frac{e^{-\beta
(\varepsilon _\alpha -\varepsilon _\delta +\omega _\rho )}-1}{i\omega
_n-\varepsilon _\alpha +\varepsilon _\delta -\omega _\rho }+\overline{n}%
_f^\alpha n_f^\delta n_b^\rho \frac{e^{-\beta (\varepsilon _\alpha
-\varepsilon _\delta -\omega _\rho )}-1}{i\omega _n-\varepsilon _\alpha
+\varepsilon _\delta +\omega _\rho } \\ 
-n_f^\alpha \overline{n}_f^\delta \overline{n}_b^\rho \frac{e^{\beta
(\varepsilon _\alpha -\varepsilon _\delta -\omega _\rho )}-1}{i\omega
_n+\varepsilon _\alpha -\varepsilon _\delta -\omega _\rho }+n_f^\alpha 
\overline{n}_f^\delta n_b^\rho \frac{e^{\beta (\varepsilon _\alpha
-\varepsilon _\delta +\omega _\rho )}-1}{i\omega _n+\varepsilon _\alpha
-\varepsilon _\delta +\omega _\rho } \\ 
+n_f^\alpha n_f^\delta \overline{n}_b^\rho \frac{e^{\beta (\varepsilon
_\alpha +\varepsilon _\delta -\omega _\rho )}-1}{i\omega _n+\varepsilon
_\alpha +\varepsilon _\delta -\omega _\rho }-n_f^\alpha n_f^\delta n_b^\rho 
\frac{e^{\beta (\varepsilon _\alpha +\varepsilon _\delta +\omega _\rho )}-1}{%
i\omega _n+\varepsilon _\alpha +\varepsilon _\delta +\omega _\rho } \\ 
\equiv \Delta (\overrightarrow{p_1},\overrightarrow{q_2},\overrightarrow{k;}%
\omega _n)
\end{array}
\eqnum{254}
\end{equation}
and 
\begin{equation}
\begin{array}{c}
\Delta (\gamma \beta \rho )=\frac 12\int_{-\beta }^\beta d\tau e^{i\omega
_n\tau }\Delta _q(\vec p_2,\tau )\Delta _q(\vec q_1,\tau )\Delta _g(\vec k%
,\tau ) \\ 
=\overline{n}_f^\gamma \overline{n}_f^\beta \overline{n}_b^\rho \frac{%
e^{-\beta (\varepsilon _\gamma +\varepsilon _\beta +\omega _\rho )}-1}{%
i\omega _n-\varepsilon _\gamma -\varepsilon _\beta -\omega _\rho }-\overline{%
n}_f^\gamma \overline{n}_f^\beta n_b^\rho \frac{e^{-\beta (\varepsilon
_\gamma +\varepsilon _\beta -\omega _\rho )}-1}{i\omega _n-\varepsilon
_\gamma -\varepsilon _\beta +\omega _\rho } \\ 
-\overline{n}_f^\gamma n_f^\beta \overline{n}_b^\rho \frac{e^{-\beta
(\varepsilon _\gamma -\varepsilon _\beta +\omega _\rho )}-1}{i\omega
_n-\varepsilon _\gamma +\varepsilon _\beta -\omega _\rho }+\overline{n}%
_f^\gamma n_f^\beta n_b^\rho \frac{e^{-\beta (\varepsilon _\gamma
-\varepsilon _\beta -\omega _\rho )}-1}{i\omega _n-\varepsilon _\gamma
+\varepsilon _\beta +\omega _\rho } \\ 
-n_f^\gamma \overline{n}_f^\beta \overline{n}_b^\rho \frac{e^{\beta
(\varepsilon _\gamma -\varepsilon _\beta -\omega _\rho )}-1}{i\omega
_n+\varepsilon _\gamma -\varepsilon _\beta -\omega _\rho }+n_f^\gamma 
\overline{n}_f^\beta n_b^\rho \frac{e^{\beta (\varepsilon _\gamma
-\varepsilon _\beta +\omega _\rho )}-1}{i\omega _n+\varepsilon _\gamma
-\varepsilon _\beta +\omega _\rho } \\ 
+n_f^\gamma n_f^\beta \overline{n}_b^\rho \frac{e^{\beta (\varepsilon
_\gamma +\varepsilon _\beta -\omega _\rho )}-1}{i\omega _n+\varepsilon
_\gamma +\varepsilon _\beta -\omega _\rho }-n_f^\gamma n_f^\beta n_b^\rho 
\frac{e^{\beta (\varepsilon _\gamma +\varepsilon _\beta +\omega _\rho )}-1}{%
i\omega _n+\varepsilon _\gamma +\varepsilon _\beta +\omega _\rho } \\ 
\equiv \Delta (\vec p_2,\vec q_1,\vec k;\omega _n)
\end{array}
\eqnum{255}
\end{equation}
where

\begin{equation}
\varepsilon _\alpha =\sqrt{\vec p_1^2+m_1},\text{ }\varepsilon _\beta =\sqrt{%
\vec p_2^2+m_2},\text{ }\varepsilon _\gamma =\sqrt{\vec q_1^2+m_1},\text{ }%
\varepsilon _\delta =\sqrt{\vec q_2^2+m_2},\text{ }\omega _\rho =\left| \vec 
k\right|  \eqnum{256}
\end{equation}
It is noted here that the chemical potential is not taken into account for
the bound state. Another expressions of the functions $\Delta (\alpha \delta
\rho )$ and $\Delta (\gamma \beta \rho )$ may be given by making use of the
expansions presented in Eqs. (A10) and (A19) in the Appendix. Since the
expressions contain infinite series, it might be not convenient for our
purpose. Upon inserting Eqs. (252)-(255) into the last two terms in Eq.
(251) and noticing the definition in Eq. (72), after completing the
integration over $\vec k$ and the summation over the polarization index, we
obtain 
\begin{equation}
\begin{array}{c}
K(\alpha ^{+}\beta ^{-};\gamma ^{+}\delta ^{-};E)=K(\vec p_1,\vec p_2;\vec q%
_1,\vec q_2;E) \\ 
=(2\pi )^{-3}\delta ^3(\vec p_1+\vec p_2-\vec q_1-\vec q_2)V(\vec p_{1,}\vec 
p_2;\vec q_1,\vec q_2;E)
\end{array}
\eqnum{257}
\end{equation}
where the $i\omega _n$ in Eqs. (254) and (255) has been replaced by $E$ and 
\begin{equation}
V(\vec p_{1,}\vec p_2;\vec q_1,\vec q_2;E)=\overline{u}_{\sigma _1}(\vec p%
_1)T^c\gamma _\mu u_{\sigma _1^{\prime }}(\vec q_1)\overline{u}_{\sigma _2}(%
\vec p_2)T^{c*}\gamma _\mu u_{\sigma _2^{\prime }}(\vec q_2)D(\vec p_{1,}%
\vec p_2;\vec q_1,\vec q_2)  \eqnum{258}
\end{equation}
in which

\begin{equation}
D(\vec p_{1,}\vec p_2;\vec q_1,\vec q_2)=\frac 1{8\omega (\vec p_1-\vec q_1)}%
[\Delta (\vec p_1,\vec q_2,\vec p_1-\vec q_1)+\Delta (\vec p_2,\vec q_1,\vec 
p_1-\vec q_1)]{\cal S}^{-1}(\vec q_1,\vec q_2)  \eqnum{259}
\end{equation}
With the kernel given above, the equation in Eq. (192) can be written as 
\begin{equation}
\lbrack E-\varepsilon (\vec p_1)-\varepsilon (\vec p_2)]\chi _{P\alpha }(%
\vec p_1,\vec p_2)=\int d^3q_1d^3q_2K(\vec p_1,\vec p_2;\vec q_1,\vec q%
_2)\chi _{P\alpha }(\vec q_1,\vec q_2)  \eqnum{260}
\end{equation}
When we introduce the cluster momenta 
\begin{equation}
\begin{array}{c}
\vec P=\vec p_1+\vec p_2,\text{ }\vec Q=\vec q_1+\vec q_2,\text{ }\vec q%
=\eta _2\vec p_1-\eta _1\vec p_2, \\ 
\vec k=\eta _2\vec q_1-\eta _1\vec q_2,\text{ }\eta _1=\frac{m_1}{m_1+m_2},%
\text{ }\eta _2=\frac{m_2}{m_1+m_2},
\end{array}
\eqnum{261}
\end{equation}
in the center of mass frame, Eq. (260) will be represented as 
\begin{equation}
\lbrack E-\varepsilon (\vec p_1)-\varepsilon (\vec p_2)]\chi _{P\alpha }(%
\vec q)=\int \frac{d^3k}{(2\pi )^3}V(\vec P,\vec q,\vec k)\chi _{P\alpha }(%
\vec k).  \eqnum{262}
\end{equation}
As we know, the spinor function in Eq. (258) can be written as $u_\sigma (%
\vec p)=U(\vec p)\varphi _\sigma $ where $\varphi _\sigma $ is the spin wave
function and $U(\vec p)$ is the ordinary Dirac spinor determined by the
Dirac equation. The spin wave functions may be absorbed into the amplitude $%
\psi (\alpha \beta ;E)$ appearing in Eq. (193). Thus, corresponding to the
equation in Eq. (262), the equation in Eq. (193) will be represented as 
\begin{equation}
\lbrack E-\varepsilon _1(\vec q)-\varepsilon _2(\vec q)]\psi _{P\alpha }(%
\vec q)=\int \frac{d^3k}{(2\pi )^3}\widehat{V}(\vec P;\vec q,\vec k)\psi
_{P\alpha }(\vec k)  \eqnum{263}
\end{equation}
where $\psi _{P\alpha }(\vec q)$ stands for the color singlet wave function
given in the Pauli spinor space and 
\begin{equation}
\widehat{V}(\vec P;\vec q,\vec k)=-\frac 43g^2\overline{U}(\vec p_1)\gamma
_\mu U(\vec q_1)\overline{U}(\vec p_2)\gamma ^\mu U(\vec q_2)D(\vec P,\vec q,%
\vec k)  \eqnum{264}
\end{equation}
represents the one-gluon exchange interaction Hamiltonian which formally is
the same as given in the case of zero-temperature [17]. In Eq. (264), we
have recovered the Minkowski metric for the $\gamma -$ matrix in order to
compare with the ordinary zero-temperature result and considered that the
expectation value of the $q\overline{q\text{ }}$color operator $%
T_1^a(-T_2^{a*})$ in the color singlet equals to $-\frac 43$.

\section{Concluding remarks}

In this paper, there are two new achievements. One is that the path-integral
formalism of the thermal QCD has been correctly established in the
coherent-state representation. The expression of the QCD generating
functional formulated in this paper not only gives an alternative
quantization of the QCD, but also provides a general method of calculating
the partition function, the thermal Green functions and thereby other
statistical quantities of QCD in the coherent-state representation. In
particular, the generating functional enables us to carry out analytical
calculations without being concerned with its discretized form. As one has
seen from Sect. 4, the analytical calculation of the zero-order generating
functional is more simple and direct than the previous calculations
performed in the discretized form given either in the coherent-state
representation or in the position space [18-22]. The coherent-state
path-integral formalism corresponds to the operator formalism represented in
terms of creation and annihilation operators. In comparison with the latter
formalism which was frequently applied in the many-body theory [21, 29], the
coherent-state path-integral formalism has a prominent advantage that in
calculations within this formalism, use of the operator commutators and the
Wick theorem is completely avoided. Therefore, it is more convenient for
practical applications. It should be noted that although the QCD generating
functional is derived in the Feynman gauge, the result is exact. This is
because QCD is a gauge-independent theory. As shown in Sect. 4, in the
partition function derived in the Feynman gauge, the unphysical part of the
partition function given by the unphysical degrees of freedom of gluons is
completely cancelled out by the partition function arising from the ghost
particles. Certainly, the generating functional formulated in the
coherent-state representation can be established in arbitrary gauges. But,
in this case, the gluon propagator would have a rather complicated form due
to that the longitudinal part of the propagator will involve the
polarization vector. Another point we would like to mention is that to
formulate the quantization of the thermal QCD in the coherent-state
representation, we limit ourself to work in the imaginary-time formalism. It
is no doubt that the theory can also be described in the real-time
formalism. We leave the discussion on this subject in the future.

The main achievement of this paper is the foundation of a rigorous
three-dimensional equation for the $q\overline{q}$ bound states at finite
temperature. Especially, the interaction kernel in the equation is given a
closed expression in the coherent-state representation. This kernel, as
shown in Eq. (226), contains only a few types of Green functions with some
definite coefficients. We also give the corresponding closed expression
represented in the position space. As shown in Eqs. (231)-(242), in this
expression only a few types of Green functions and commutators are involved.
Therefore, the kernel can not only easily be calculated by the perturbation
method, but also is suitable for nonperturbative investigations by using the
lattice gauge approach and some others. Since the kernel contains all the
interactions taking place in the bound state, obviously, the kernel and the
equation presented in the preceding sections are much suitable to study the
quark deconfinement at high temperature which is nowadays an important
theoretical problem in the high energy physics. It is expected that an
accurate nonperturbative calculation of the kernel could come up in the
future so as to give the problem of quark deconfinement a definitet solution

\section{Acknowledgment}

This work was supported by National Natural Science Foundation of China.

\section{Appendix: Derivation of the generating functional represented in
the position space}

To confirm the correctness of the results derived in Sect.4, in this
appendix, we plan to derive the familiar perturbative expansion of the
thermal QCD generating functional represented in the position space from the
corresponding one given in section 4. For this purpose, we need to derive
the generating functional represented in position space for the free system
which can be written as 
\begin{equation}
Z^0[J]=Z_g^0[J_\mu ^a]Z_q^0[I,\overline{I}]Z_c^0[K^a,\overline{K}^a] 
\eqnum{A1}
\end{equation}
where $Z_g^0[J_\mu ^a],$ $Z_q^0[I,\overline{I}]$ and $Z_c^0[K^a,\overline{K}%
^a]$ are the position space generating functionals arising respectively from
the free gluons, quarks and ghost particles and $J_\mu ^a$, $I$, $\overline{I%
}$, $K^a$ and $\overline{K}^a$ are the sources coupled to gluon, quark and
ghost particle fields respectively. In order to write out the $Z_q^0[I,%
\overline{I}]$, $Z_g^0[J_\mu ^a]$ and $Z_c^0[K^a,\overline{K}^a]$ from the
generating functionals given in Eqs. (119), (138) and (150), it is necessary
to establish relations between the sources introduced in the position space
and in the coherent-state representation. Let us separately discuss the
functionals $Z_q^0[I,\overline{I}]$, $Z_g^0[J_\mu ^a]$ and $Z_c^0[K^a,%
\overline{K}^a]$. First we focus our attention on the functioinal $Z_q^0[I,%
\overline{I}]$. Usually, the external source terms of fermions in the
generating functional given in the position space are of the form $%
\int_0^\beta d\tau \int d^3x[\overline{I}(\vec x,\tau )\psi (\vec x,\tau )+%
\overline{\psi }(\vec x,\tau )I(\vec x,\tau )]$ [22, 32]. Substituting in
this expression the Fourier expansions in Eqs. (58) and (59) for the quark
fields and in the following for the sources 
\begin{equation}
\begin{array}{c}
I(\vec x,\tau )=\int \frac{d^3k}{(2\pi )^{3/2}}I(\vec k,\tau )e^{i\vec k%
\cdot \vec x} \\ 
\overline{I}(\vec x,\tau )=\int \frac{d^3k}{(2\pi )^{3/2}}\overline{I}(\vec k%
,\tau )e^{-i\vec k\cdot \vec x}
\end{array}
\eqnum{A2}
\end{equation}
we have 
\begin{equation}
\begin{array}{c}
\int_0^\beta d\tau \int d^3x[\overline{I}(\vec x,\tau )\psi (\vec x,\tau )+%
\overline{\psi }(\vec x,\tau )I(\vec x,\tau )]=\int_0^\beta d\tau \int
d^3k[\eta _s^{*}(\vec k,\tau )b_s(\vec k,\tau ) \\ 
+b_s^{*}(\vec k,\tau )\eta _s(\vec k,\tau )+\overline{\eta }_s^{*}(\vec k%
,\tau )d_s(\vec k,\tau )+d_s^{*}(\vec k,\tau )\overline{\eta }_s(\vec k,\tau
)]
\end{array}
\eqnum{A3}
\end{equation}
where 
\begin{equation}
\begin{array}{c}
\eta _s(,\tau )=\overline{u}_s(\vec k)I(\vec k,\tau )\text{, }\eta _s^{*}(%
\vec k,\tau )=\overline{I}(\vec k,\tau )u_s(\vec k), \\ 
\overline{\eta }_s(\vec k,\tau )=-\overline{I}(-\vec k,\tau )v_s(\vec k)%
\text{, }\overline{\eta }_s^{*}(\vec k,\tau )=-\overline{v}_s(\vec k)I(-\vec 
k,\tau ).
\end{array}
\eqnum{A4}
\end{equation}
From these relations and the property of Dirac spinors, the relation in Eq.
(137) is easily proved. On substituting Eq. (A4) into Eq. (138), one can get 
\begin{equation}
Z_q^0[I,\overline{I}]=Z_q^0\exp \{\int_0^\beta d\tau _1\int_0^\beta d\tau
_2\int d^3k\overline{I}(\vec k,\tau _1)S_F(\vec k,\tau _1-\tau _2)I(\vec k%
,\tau _2)\}  \eqnum{A5}
\end{equation}
where 
\begin{equation}
S_F(\vec k,\tau _1-\tau _2)=[({\bf k}+m)/\varepsilon (\vec k)]\Delta _q(\vec 
k,\tau _1-\tau _2)  \eqnum{A6}
\end{equation}
with ${\bf k=}\gamma _\mu k^\mu $ and $k^\mu =(\vec k,i\varepsilon _n)$. By
making use of the inverse transformation of Eq. (A2), the generating
functional in Eq. (A5) is finally represented as [22, 32] 
\begin{equation}
Z_q^0[I,\overline{I}]=Z_q^0\exp \{\int_0^\beta d^4x_1\int_0^\beta d^4x_2%
\overline{I}(x_1)S_F(x_1-x_2)I(x_2)\}  \eqnum{A7}
\end{equation}
where $x=(\vec x,\tau )$, $d^4x=d\tau d^3x$ and 
\begin{equation}
S_F(x_1-x_2)=\int \frac{d^3k}{(2\pi )^3}S_F(\vec k,\tau _1-\tau _2)e^{i\vec k%
\cdot \vec x}  \eqnum{A8}
\end{equation}
It is well-known that the propagator $\Delta _q^{ss^{\prime }}(\vec k,\tau
_1-\tau _2)$ is antiperiodic, 
\begin{equation}
\begin{array}{c}
\Delta _q^{ss^{\prime }}(\vec k,\tau _1-\tau _2)=-\Delta _q^{ss^{\prime }}(%
\vec k,\tau _1-\tau _2-\beta ),\text{ }if\text{ }\tau _1\succ \tau _2 \\ 
\Delta _q^{ss^{\prime }}(\vec k,\tau _1-\tau _2)=-\Delta _q^{ss^{\prime }}(%
\vec k,\tau _1-\tau _2+\beta ),\text{ }if\text{ }\tau _1\prec \tau _2
\end{array}
\eqnum{A9}
\end{equation}
This can easily be proved from its representation in the operator formalism
as shown in Eq. (46) or Eq. (154) with the help of the translation
transformation $\widehat{b}_s(\tau )=e^{\beta \widehat{K}}$ $\widehat{b}%
_se^{-\beta \widehat{K}}$ and the trace property $Tr(AB)=Tr(BA)$. According
to the antiperiodic property of the propagator, we have the following
expansion 
\begin{equation}
\Delta _q(\vec k,\tau )=\frac 1\beta \sum_n\Delta _q(\vec k,\varepsilon
_n)e^{-i\varepsilon _n\tau }  \eqnum{A10}
\end{equation}
where $\tau =\tau _1-\tau _2$, $\varepsilon _n=\frac \pi \beta (2n+1)$ with $%
n$ being the integer and 
\begin{equation}
\Delta _q(\vec k,\varepsilon _n)=\int_0^\beta d\tau e^{i\varepsilon _n\tau
}\Delta _q(\vec k,\tau )=\frac{\varepsilon (\vec k)}{\varepsilon
_n^2+\varepsilon (\vec k)^2}  \eqnum{A11}
\end{equation}
Substituting Eq. (A10) into Eq. (A6) and noticing the above expression, it
can be found that [22, 32] 
\begin{equation}
S_F(x_1-x_2)=\frac 1\beta \sum_n\int \frac{d^3k}{(2\pi )^3}\frac{e^{i%
\overrightarrow{k}\cdot (\vec x_1-\vec x_2)-i\varepsilon _n(\tau _1-\tau _2)}%
}{\vec \gamma \cdot \vec k+m-i\varepsilon _n\gamma ^0}  \eqnum{A12}
\end{equation}
which is the familiar expression of the thermal fermion propagator in the
position space.

Next, we discuss the generating functional $Z_g^0[J_\mu ^a].$ The source
term of gluons in the generating functional given in the position space is
commonly taken as $\int_0^\beta d\tau \int d^3xJ_\mu ^a(\vec x,\tau )A^{a\mu
}(\vec x,\tau )$ [22, 32]$.$ Employing the expansions in Eq. (60) and in the
following 
\begin{equation}
J_\mu ^a(\vec x,\tau )=\int \frac{d^3k}{(2\pi )^{3/2}}J_\mu ^a(\vec k,\tau
)e^{i\vec k\cdot \vec x},  \eqnum{A13}
\end{equation}
we can write 
\begin{equation}
\int_0^\beta d\tau \int d^3xJ_\mu ^c(\vec x,\tau )A^{c\mu }(\vec x,\tau
)=\int_0^\beta d\tau \int d^3k[\xi _\lambda ^{c*}(\vec k,\tau )a_\lambda ^c(%
\vec k,\tau )+a_\lambda ^{c*}(\vec k,\tau )\xi _\lambda ^c(\vec k,\tau )] 
\eqnum{A14}
\end{equation}
where 
\begin{equation}
\xi _\lambda ^c(\vec k,\tau )=(2\omega (\vec k))^{-1/2}\epsilon _\lambda
^\mu (\vec k)J_\mu ^c(\vec k,\tau )=\xi _\lambda ^{c*}(-\vec k,\tau ) 
\eqnum{A15}
\end{equation}
in which the last equality follows from that the $J_\mu ^a(\vec x,\tau )$ is
a real function. Inserting the relations in Eq. (A15) and then the inverse
transformation of Eq. (A13) into Eq. (119) and considering completeness of
the polarization vectors, one may find the generating functional $%
Z_g^0[J_\mu ^a]$ such that [22, 32] 
\begin{equation}
Z_g^0[J_\mu ^a]=Z_g^0\exp \{\frac 12\int_0^\beta d^4x_1\int_0^\beta
d^4x_2J_\mu ^a(x_1)D_{\mu \nu }^{ab}(x_1-x_2)J_\nu ^b(x_2)\}  \eqnum{A16}
\end{equation}
where 
\begin{equation}
D_{\mu \nu }^{ab}(x_1-x_2)=\delta ^{ab}\delta _{\mu \nu }\int \frac{d^3k}{%
(2\pi )^3}\frac 1{\omega (\vec k)}\Delta _g(\vec k,\tau _1-\tau _2)e^{i\vec k%
\cdot (\vec x_1-\vec x_2)}  \eqnum{A17}
\end{equation}
By the same argument as mentioned for Eq. (A9), it can be proved that the
gluon propagator $\Delta _g(\vec k,\tau _1-\tau _2)$ is a periodic function 
\begin{equation}
\begin{array}{c}
\Delta _g(\vec k,\tau _1-\tau _2)=\Delta _g(\vec k,\tau _1-\tau _2-\beta ),%
\text{ }if\text{ }\tau _1\succ \tau _2; \\ 
\Delta _g(\vec k,\tau _1-\tau _2)=\Delta _g(\vec k,\tau _1-\tau _2+\beta ),%
\text{ }if\text{ }\tau _1\prec \tau _2.
\end{array}
\eqnum{A18}
\end{equation}
Therefore, we have the expansion 
\begin{equation}
\Delta _g(\vec k,\tau )=\frac 1\beta \sum_n\Delta _g(\vec k,\omega
_n)e^{-i\omega _n\tau }  \eqnum{A19}
\end{equation}
where $\omega _n=\frac{2\pi n}\beta $ and 
\begin{equation}
\Delta _g(\vec k,\omega _n)=\int_0^\beta d\tau e^{i\omega _n\tau }\Delta _g(%
\vec k,\tau )=\frac{\omega (\vec k)}{\varepsilon _n^2+\omega (\vec k)^2}. 
\eqnum{A20}
\end{equation}
Upon substituting Eqs. (A19) and (A20) in Eq. (A17), we arrive at [22,32] 
\begin{equation}
D_{\mu \nu }^{ab}(x_1-x_2)=\delta ^{ab}\frac 1\beta \sum_n\int \frac{d^3k}{%
(2\pi )^3}\frac{\delta _{\mu \nu }}{\varepsilon _n^2+\omega (\vec k)^2}e^{i%
\vec k\cdot (\vec x_1-\vec x_2)-i\omega _n(\tau _1-\tau _2)}  \eqnum{A21}
\end{equation}
which just is the gluon propagator given in the position space and in the
Feynman gauge.

Finally, we turn to the generating functional $Z_c^0[K^a,\overline{K}^a]$.
In accordance with the expansions in Eqs. (62) and (63) for the ghost
particle fields and those for the external sources: 
\begin{equation}
\begin{array}{c}
K^a(\vec x,\tau )=\int \frac{d^3k}{(2\pi )^{3/2}}K^a(\vec k,\tau )e^{i\vec k%
\cdot \vec x} \\ 
\overline{K}^a(\vec x,\tau )=\int \frac{d^3k}{(2\pi )^{3/2}}\overline{K}^a(%
\vec k,\tau )e^{-i\vec k\cdot \vec x}
\end{array}
\eqnum{A22}
\end{equation}
the relation between the sources in the position space and in the
coherent-state representation can be found to be 
\begin{equation}
\begin{array}{c}
\int_0^\beta d\tau \int d^3x[\overline{K}^a(\vec x,\tau )C^a(\vec x,\tau )+%
\overline{C}^a(\vec x,\tau )K(\vec x,\tau )]=\int_0^\beta d\tau \int
d^3k[\zeta _a^{*}(\vec k,\tau )c_a(\vec k,\tau ) \\ 
+c_a^{*}(\vec k,\tau )\zeta _a(\vec k,\tau )+\overline{\zeta }_a^{*}(\vec k%
,\tau )\overline{c}_a(\vec k,\tau )+\overline{c}_a^{*}(\vec k,\tau )%
\overline{\zeta }_a(\vec k,\tau )]
\end{array}
\eqnum{A23}
\end{equation}
where 
\begin{equation}
\begin{array}{c}
\zeta _a(\vec k,\tau )=(2\omega (\vec k))^{-1/2}K^a(\vec k,\tau ),\text{ }%
\zeta _a^{*}(\vec k,\tau )=(2\omega (\vec k))^{-1/2}\overline{K}^a(\vec k%
,\tau ), \\ 
\overline{\zeta }_a(\vec k,\tau )=-(2\omega (\vec k))^{-1/2}\overline{K}^a(-%
\vec k,\tau ),\text{ }\overline{\zeta }_a^{*}(\vec k,\tau )=-(2\omega (\vec k%
))^{-1/2}K^a(-\vec k,\tau ).
\end{array}
\eqnum{A24}
\end{equation}
from which the relations denoted in Eq. (149) directly follows. When the
above relations and the inverse transformations of Eq. (A22) are inserted
into Eq. (150), one can get 
\begin{equation}
Z_c^0[K^a,\overline{K}^a]=Z_c^0\exp \{-\int_0^\beta d^4x_1\int_0^\beta d^4x_2%
\overline{K}^a(x_1)\Delta _c^{ab}(x_1-x_2)K^b(x_2)\}  \eqnum{A25}
\end{equation}
where 
\begin{equation}
\begin{array}{c}
\Delta _c^{ab}(x_1-x_2)=\delta ^{ab}\int \frac{d^3k}{(2\pi )^3}\frac 1{%
\omega (\overrightarrow{k})}\Delta _g(\vec k,\tau _1-\tau _2)e^{i\vec k\cdot
(\vec x_1-\vec x_2)} \\ 
=\delta ^{ab}\frac 1\beta \sum\limits_n\int \frac{d^3k}{(2\pi )^3}\frac 1{%
\varepsilon _n^2+\omega (\overrightarrow{k})^2}e^{i\vec k\cdot (\vec x_1-%
\vec x_2)-i\omega _n(\tau _1-\tau _2)}
\end{array}
\eqnum{A26}
\end{equation}
which just is the free ghost particle propagator given in the position space
[22, 32]. In the last equality of Eq. (A26), the expansion given in Eqs.
(A19) and (A20) have been used.

With the generating functionals given in Eqs. (A7), (A16) and (A25), the
zeroth-order generating functional in Eq. (A1) is explicitly represented in
terms of the propagators and external sources. Clearly, the exact generating
functional can immediately be written out from Eq. (95) as shown in the
following 
\begin{equation}
Z[J]=\exp \{-\int_0^\beta d^4x{\cal H}_I(\frac \delta {\delta J(x)})\}Z^0[J]
\eqnum{A27}
\end{equation}
where $J$ stands for $I$, $\overline{I}$, $J_\mu ^a$ $K^a$ and $\overline{K}%
^a$, $\frac \delta {\delta J(x)}$ represents the differentials $\frac \delta
{\delta \overline{I}(x)}$, $-\frac \delta {\delta I(x)}$, $\frac \delta {%
\delta J_\mu ^a(x)}$, $\frac \delta {\delta \overline{K}^a(x)}$ and $-\frac 
\delta {\delta K^a(x)}$ and ${\cal H}_I(\frac \delta {\delta J(x)})$ can be
written out from Eq. (57) when the field functions in Eq. (57) are replaced
by the differentials with respect to the corresponding sources.

\section{\bf Reference}

[1] Miklos Gyulassy, Nucl. Phys. {\bf A}685, 432 (2001).

[2] D. Hardtke, Proceedings of XXI International Symposium on Lepton and
Photon Interactions at High Energies, p.329, at Fermilab, 2003.

[3] Reinhard Stock, J. Phys. G: Nucl. Part. Phys. 30, S633-S648 (2004).

[4] Y. Iwasaki, K. Kanaya, S. Kaya, S. Sakai and T. Yoshie, Phys. Rev. D69,
014507 (2004).

[5] M. D. Elia, A. Di Giacomo and E. Meggiolaro, Phys. Rev. D114504 (2003)

[6] A. M\`ocsy, F. Sannino and K. Tuominen, Phys. Rev. Lett. 92, 182302
(2004).

[7] Z. Yang and P. Zhuang, Phys. Rev. C69, 035203, (2004).

[8] H. Gies, Phys. Rev. D63, 025013 (2001).

[9] M. Asakawa, U. Heinz and B. M\"uller, Phys. Rev. Lett. 85, 2072 (2000).

[10] H. Li and C. M. Shakin, Phys. Rev. D66, 074016 (2002).

[11] S. Gupta, Phys. Rev. D64, 034507 (2001).

[12] S. Jesgarz, S. Lerma, P. O. Hess, O. Civitarese and M. Reboiro, Phys.
Rev. C67, 055210 (2003).

[13] G. F. Burgio, M. Baldo, P. K. Sahu and H.-J. Schulze, Phys. Rev. C66,
025802 (2002).

[14] B. Sheikholeslami-Sabzevari, Phys. Rev. C65, 054904 (2002).

[15] S. S. Wu, J. Phys. G: Nucl. Part. Phys. {\bf 16}, 1447 (1990); Commun.
Theor. Phys. {\bf 10} 181 (1988).

[16] J. C. Su and J. X. Chen, Phys. Rev. D {\bf 69}, 076002 (2004).

[17] J. C. Su, J. Phys. G: Nucl. Part. Phys. 30, 1309 ( 2004).

[18] A. Casher, D. Lurie and M. Revzen, J. Math. Phys. {\bf 9}, 1312 (1968).

[19] L. S. Schulman, Techniques and Applications of Path Integration, A
wiley-Interscience

Publication, New York (1981).

[20] M. Beccaria, B. All\'es and F. Farchioni, Phys. Rev. {\bf E} 55, 3870
(1997).

[21] J. W. Negele and H. Orland, Quantum Many-Particle Systems, Perseus
Books Publishing, L.L.C, 1998.

[22] M. Le Bellac, Thermal Field Theory, Cambridge University Press, 1996.

[23] J. C. Su, Phys. Lett. A {\bf 268}, 279 (2000).

[24] S. S. Schweber, J. Math, Phys. {\bf 3}, 831 (1962).

[25] L. D. Faddeev and A. A. Slavnov, Gauge Fields: Introduction to Quantum
Theory,

The Benjamin/Cummings Publishing Company, Inc. (1980).

[26] C. Itzykson and J-B. Zuber, Quantum Field Theory, McGraw-Hill Inc. New
York (1980).

[27] J. R. Klauder, Phys. Rev. D {\bf 19}, 2349 (1979).

[28] R. J. Glauber, Phys. Rev. {\bf 131}, 2766 (1963).

[39] A. L. Fetter and J. D. Walecka, Quantu Quantum Theory of Many-Particle
System, McGraw-Hill

(1971).

[30] D. Lurie, Particles and Fields, Interscience Publishers, a divison of
John Viley \& Sons, New York, 1968.

[31] H. Lehmann, Nuovo Cimento {\bf 11} (1954) 342.

[32] J. I. Kapusta, Finite-Temperature Field Theory, Cambridge University
Press, 1989.

\end{document}